\documentclass[twocolumn]{aastex701}

\usepackage{natbib}
\usepackage{rotating}
\usepackage{multirow}
\usepackage{hyperref}
\usepackage{booktabs} 
\usepackage{float}
\usepackage{amsmath}
\usepackage{color}
\usepackage{threeparttable}

\shorttitle{A Multi-Wavelength SED Analysis of Nine Open Clusters with Swift/UVOT}
\shortauthors{Cinar}

\graphicspath{{./}{Figures/}}
\begin{document}
\title {Hot Degenerate Components in Blue Stragglers: A Multi-Wavelength SED Analysis of Nine Open Clusters with Swift/UVOT}

\correspondingauthor{Deniz Cennet \c{C}\i nar}
\email{denizcennetcinar@gmail.com}

\author[0000-0001-7940-3731]{Deniz Cennet \c{C}\i nar} \affiliation{Istanbul University, Institute of Graduate Studies in Science, Programme of Astronomy and Space Sciences, 34116, Istanbul, Turkey} \email{denizcennetcinar@gmail.com}

\author[orcid=0000-0002-8988-8434, sname=Bisht, gname=D.]{D. Bisht} \affiliation{Indian Centre for Space Physics, 466, Barakhola, Singabari Road, Netai Nagar, Kolkata, West Bengal, 700099, India} \email{devendrabisht297@gmail.com}

\author[0000-0003-3510-1509]{Sel\c{c}uk Bilir} \affiliation{Istanbul University, Faculty of Science, Department of Astronomy and Space Sciences, 34119, Beyaz\i t, Istanbul, Turkey} \email{sbilir@istanbul.edu.tr}

\author[0000-0003-3713-2640]{Songmei Qin} \affiliation{Astrophysics Division, Shanghai Astronomical Observatory, Chinese Academy of Sciences, Shanghai 200030, China} \affiliation{School of Astronomy and Space Science, University of Chinese Academy of Sciences, No. 19A, Yuquan Road, Beijing 100049, PR China} \affiliation{Institut de Ci\`encies del Cosmos (ICCUB), Universitat de Barcelona (UB), Mart\'i i Franqu\`es 1, E-08028 Barcelona, Spain} \email{qinsongmei@shao.ac.cn}

\author[0000-0001-8031-1957]{Leila Saker} \affiliation{Observatorio Astron\'omico, Universidad Nacional de C\'ordoba, Laprida 854, 5000 C\'ordoba, Argentina} \email{leilasaker88@unc.edu.ar}

\begin{abstract}
We present a homogeneous multi-wavelength analysis of 35 blue straggler star (BSS) candidates in nine open clusters, combining \textit{Swift}/UVOT near-ultraviolet data with Gaia DR3 astrometry and optical-to-infrared photometry. We construct spectral energy distributions (SEDs) to search for signatures of hot companions associated with past mass transfer. Among the sample, 15 BSSs ($\sim43\%$) show ultraviolet excesses that are better described by two-component SED fits. The inferred companions are consistent with hot white dwarfs and pre-extremely low-mass (pre-ELM) white dwarf candidates, suggesting systems observed at different stages following mass transfer.  We examine the radial distribution of the BSSs and find evidence for mass segregation in dynamically evolved clusters, a result that is broadly consistent with the estimated half-mass relaxation timescales of the host systems. To place the clusters in a Galactic context, we compute their orbits using \texttt{galpy}, obtaining low eccentricities ($e \leq 0.1$) and disk-like trajectories. We also find a positive relation between the half-number radius of the BSS population ($r_{50}$) and the total number of BSSs.  Overall, our results are consistent with a scenario in which the BSS population in these clusters is dominated by binary evolution. The systems identified here provide observational constraints on post-mass-transfer evolutionary phases. While the number of robust detections is limited and intrinsic degeneracies remain in SED-based decomposition, these results provide a useful foundation for future spectroscopic confirmation.
\end{abstract}

\keywords{Blue straggler stars (168); Open star clusters (1160); Stellar evolution (1599), Ultraviolet astronomy (1736)}


\section{Introduction}
\label{sec:introduction}
Blue straggler stars (BSSs), first identified by \citet{Sandage1953}, are stars in stellar clusters that appear anomalously brighter and hotter than the main-sequence turn-off (MSTO), suggesting rejuvenation relative to the dominant cluster population. Their position in color--magnitude diagrams (CMDs) cannot be explained by standard single-star evolution at the cluster age, making BSSs valuable laboratories for investigating non-standard evolutionary pathways, particularly those involving binary interactions and stellar dynamics \citep[e.g.,][]{Bailyn1995, Ferraro2009F}.

\begin{table*}[!ht]
\centering
\caption{Literature survey of BSSs in OCs.}
\label{tab:bss_lit}
\resizebox{\linewidth}{!}{
\begin{tabular}{lllrrlll}
\toprule
Cluster & \multicolumn{1}{c}{$l$ ($^{\circ}$)} & \multicolumn{1}{c}{$b$ ($^{\circ}$)} & \multicolumn{1}{c}{$N_{\rm member}$} & $N_{\rm BSS}$ & BSS Type / Features & Data / Instrument & Reference \\
\midrule
Berkeley 39 & 223.5419 & +10.0879   & 729  & 17  & Single, Hot companion cand. & \textit{Swift}/UVOT    & \citet{2024AJ....168..278C} \\
King 2      & 122.8681 & $-$04.6843 & 107  & 25  & BSS+WD                      & \textit{Astrosat}/UVIT  & \citet{2021JApA...42...89J} \\
M67         & 215.6893 & +31.9256   & 1183 & 13  & BSS+WD, Barium BSS          & \textit{Astrosat}/UVIT  & \citet{2021MNRAS.507.2373P} \\
Melotte 66  & 259.5807 & $-$14.2625 & 1162 & 14  & BSS+WD, Eclipsing Binary    & \textit{Swift}/UVOT    & \citet{2022MNRAS.516.2444R} \\
NGC 188     & 122.8404 & +22.3804   & 3562 & ~1  & BSS+WD                      & UVIT, GALEX            & \citet{2021JApA...42...47R} \\
NGC 752     & 136.9070 & $-$23.3032 & 223  & 15  & BL, BSS+ELM WD              & UVIT, UVOT             & \citet{2024AA...688A.152J}  \\
NGC 2243    & 239.4799 & $-$18.0105 & 889  & 12  & BSS+WD, Single              & \textit{Swift}/UVOT    & \citet{2024AJ....168..274S} \\
NGC 2420    & 198.1087 & +19.6397   & 868  & ~3  & BSS+ELM WD                  & UVIT, UVOT             & \citet{2024ApJ...961..251Y} \\
NGC 2506    & 230.5626 & +09.9328   & 2175 & ~9  & BSS+WD (ELM/LM)             & \textit{Astrosat}/UVIT & \citet{2022MNRAS.516.5318P} \\
NGC 2627    & 251.5777 & +06.6531   & 422  & ~4  & BSS+ELM WD                  & \textit{Astrosat}/UVIT & \citet{2024AJ....168...97S} \\
NGC 6791    & 069.9606 & +10.9056   & 1654 & 47  & BSS+WD, BSS+ELM WD          & \textit{Astrosat}/UVIT & \citet{2023AA...676A..47J}  \\
NGC 6940    & 069.8306 & $-$07.1491 & 492  & ~1  & BSS+WD, BL                  & \textit{Astrosat}/UVIT & \citet{2024MNRAS.52710335P} \\
NGC 7142    & 105.3547 & +09.4835   & 546  & 10  & BSS+WD                      & UVIT, TESS             & \citet{2024MNRAS.527.8325P} \\
NGC 7789    & 115.5237 & $-$05.3675 & 2800 & 16  & BSS+ELM WD                  & \textit{Astrosat}/UVIT & \citet{2022MNRAS.511.2274V} \\
\bottomrule
\end{tabular}}
\end{table*}

Two principal formation channels are commonly invoked to explain the origin of BSSs. The first involves mass transfer or common-envelope evolution in binary systems, in which a star accretes mass from an evolved companion \citep{McCrea1964}. The second channel involves stellar collisions or mergers between low-mass stars \citep{Hills1976}. While collisional processes are expected to contribute significantly in the dense environments of globular clusters (GCs), numerous observational and theoretical studies indicate that mass transfer is the dominant mechanism in the relatively sparse environments of open clusters (OCs) \citep{Knigge2009, Mathieu2009}.

A key observational signature of the mass-transfer scenario is the presence of a hot white dwarf (WD) companion, which represents the remnant core of the original donor star. Identifying such companions is therefore central to constraining the evolutionary history of BSS systems. However, this task is challenging at optical wavelengths, where the luminous BSS primary overwhelms the faint WD's flux. In contrast, WDs emit strongly in the ultraviolet (UV), where their presence can be inferred through a measurable UV excess. As a result, UV photometry has become an essential tool for confirming mass-transfer-origin BSS systems.

The presence of other blue stellar populations further complicates the identification of genuine BSSs in OCs. In addition to classical BSSs, blue lurkers, which are mass-transfer products that have not yet migrated to the canonical BSS region, can overlap with the main sequence (MS) in optical CMD \citep{2019ApJ...881...47L, 2020JApA...41...45S}. Similarly, morphologically defined blue plume stars may occupy regions close to the BSS locus. In this context, UV observations, combined with multi-wavelength SED analysis, are particularly effective for distinguishing true BSSs from contaminating populations \citep{Gosnell2015, 2019ApJ...882...43S}. Hot components, such as WDs and hot subdwarfs (sdB/sdO), produce distinct UV excesses that are not detectable in optical bands alone. While contamination in UV-selected samples is generally reduced, it cannot be entirely excluded.

In recent years, UV observations have significantly advanced the study of BSS populations in open clusters, particularly with \textit{Astrosat}/UVIT \citep{Singh2014, Tandon2017}. As part of the UVIT Open Cluster Study (UOCS~I--XIV), several works have identified WD companions in clusters such as M67, NGC~188, NGC~752, and NGC~6791 (e.g., \citealt{2018MNRAS.481..226S, 2019ApJ...882...43S}; \citealt{2021JApA...42...89J, 2024AA...688A.152J}). These studies demonstrated that high spatial resolution and multi-filter UV coverage are crucial for constructing reliable SEDs that can isolate hot compact components, including low-mass and hot post-mass-transfer remnants.

Recent UV studies with \textit{Astrosat}/UVIT have therefore played a pivotal role in identifying hot companions to blue stragglers in several OCs \citep{Subramaniam2016, 2024MNRAS.52710335P}. However, these investigations are necessarily limited to targeted observations of a relatively small number of clusters. This naturally motivates the use of alternative UV facilities that provide broader sky coverage and more uniform datasets.

In this context, the \textit{Swift} Ultraviolet/Optical Telescope \citep[UVOT;][]{Roming2000, Roming2005} offers a complementary approach. \textit{Swift}/UVOT provides wide sky coverage, multiple UV filters, and a steadily growing archival database spanning many years. These characteristics make UVOT particularly well-suited for homogeneous studies across numerous clusters. Nevertheless, systematic assessments of BSS--WD systems based on UVOT photometry remain scarce in the literature. The current work represents one of the first systematic \textit{Swift}/UVOT-based SED analyses of BSSs across numerous OCs, complementing existing UVIT investigations and expanding the observational basis for understanding mass-transfer products in low-density cluster environments.

In this work, we use \textit{Swift}/UVOT archival data to systematically and homogeneously investigate the presence of hot companions to BSSs in OCs. By cross-matching a compilation of 502 known BSSs with 103 OCs observed by UVOT \citep{Siegel2019}, we identify 35 BSS candidates in nine OCs that have not previously been examined in the UV. For these targets, we construct multi-wavelength SEDs spanning the UV, optical, and infrared regimes, and we test for the presence of hot compact companions by comparing single-star models with binary composite models. We detect significant UV excesses in 15 of the 35 systems, consistent with WD companions, and derive their physical parameters, classifying them as low-mass or hot WDs.

\section{Data}\label{sec:data}

\begin{figure*}[!ht]
    \centering
    \includegraphics[width=1\linewidth]{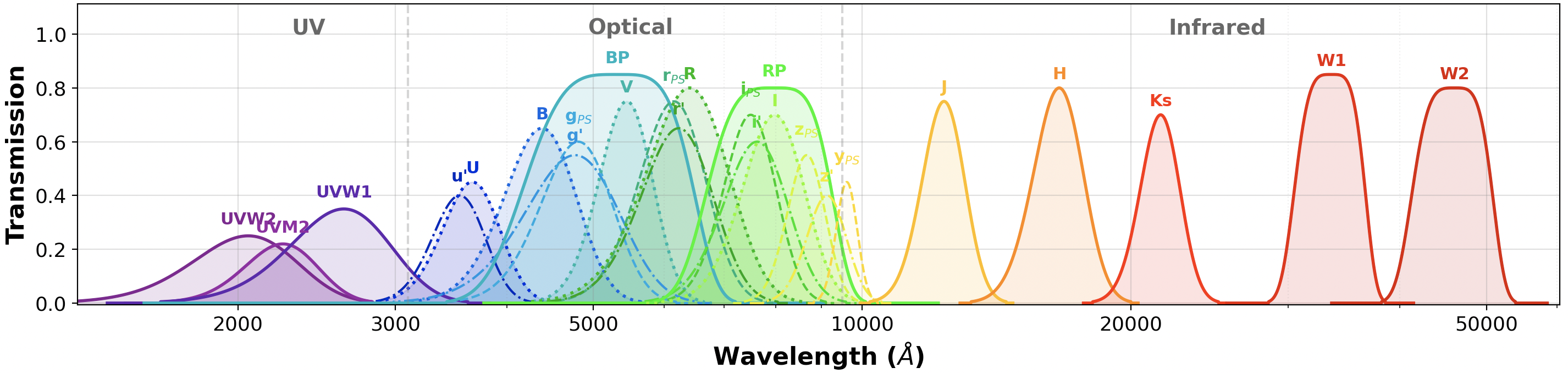}
    \caption{Normalized transmission curves for the photometric passbands adopted in the SED fitting. The coverage extends from the ultraviolet to the infrared regime.}
    \label{fig:filters}
\end{figure*}

To comprehensively characterize the physical properties of BSSs and reveal their potential hot companions, we compiled a multi-wavelength dataset spanning the ultraviolet (UV) to the mid-infrared (MIR). This extensive photometric coverage combines high-sensitivity UV observations from \textit{Swift}/UVOT with precision optical photometry and astrometry from Gaia DR3 \citep{GaiaCollaboration2023}, complemented by ground-based surveys and infrared data from Two Micron All Sky Survey \citep[2MASS;][]{2006AJ....131.1163S} and Wide-field Infrared Survey Explorer  \citep[WISE;][]{2010AJ....140.1868W}. By integrating these diverse data sources, we constructed robust SEDs essential for determining stellar parameters and detecting flux excesses indicative of binary systems. The normalized transmission curves and spectral coverage of the photometric passbands adopted in this analysis are shown in Figure \ref{fig:filters}.

\subsection{Ultraviolet (UV) Observations}
Detecting the faint thermal emission from hot, subluminous companions requires high-sensitivity observations in the ultraviolet regime, where these objects emit a substantial fraction of their flux. To achieve this, we utilized data from the \textit{Swift}/UVOT aboard the Neil Gehrels Swift Observatory \citep{Gehrels2004}. The UVOT, a modified 30 cm Ritchey-Chr\'{e}tien telescope with a $17^{\prime}\times17^{\prime}$ field of view, provides the critical Near-UV (NUV) and Far-UV (FUV) coverage needed for our analysis. Specifically, we focused on photometry from the $UVM2$ ($\lambda_{\rm eff} \approx 2246$ \AA) and $UVW2$ ($\lambda_{\rm eff} \approx 2086$ \AA) filters \citep{Breeveld2011}. Rather than performing a custom reduction of raw imagery, we leveraged the high-level science products available in the \textit{Swift}/UVOT Archive\footnote{\url{https://archive.stsci.edu/hlsp/uvot-oc}}. We adopted the calibrated flux values from the cluster point-source catalogs generated by \citet{Siegel2019}, thereby ensuring robust background subtraction and coincidence-loss corrections. A summary of the UVOT observations for the clusters in our sample is provided in Table \ref{tab:uvot_log}.

\begin{table}[ht]
\footnotesize
\centering
\caption{Log of \textit{Swift}/UVOT observations. The columns list the total exposure times for the filters.}
\label{tab:uvot_log}
\setlength{\tabcolsep}{4pt}
\renewcommand{\arraystretch}{1}
\begin{tabular}{c l c c c}
\hline \hline
ID & Cluster & \multicolumn{3}{c}{Exposure (s)} \\
\cline{3-5}
 & & $UVW2$ & $UVM2$ & $UVW1$ \\
\hline
1 & Berkeley 29 & 3151  & 2027  & 1665  \\
2 & Berkeley 31 & 1951  & 2083  & 2123  \\
3 & Berkeley 37 & 1644  & 1644  & 1511  \\
4 & Berkeley 75 & 5158  & ~~697 & ~~370 \\
5 & NGC 2192    & 1797  & 1797  & 2003  \\
6 & NGC 2204    & ~~974 & 1923  & --    \\
7 & NGC 2360    & 1881  & 1881  & 1905  \\
8 & NGC 2533    & 2132  & 1768  & 1718  \\
9 & NGC 6939    & 3992  & 3992  & 3968  \\
\hline
\end{tabular}
\end{table}

\begin{table*}[ht]
\scriptsize
\centering
\caption{Properties of the 35 candidates in the selected nine OCs. The columns list the Cluster name, Gaia DR3 Source ID, and the observed fluxes (in erg s$^{-1}$ cm$^{-2}$ \AA$^{-1}$) with uncertainties for the \textit{Swift}/UVOT $UVW2$, $UVM2$, and $UVW1$ filters.}
\label{tab:bss_fluxes_full}
\renewcommand{\arraystretch}{1.1} 
\setlength{\tabcolsep}{4pt} 
\begin{tabular}{l l c c c c}
\hline \hline
BSS ID & Cluster & Gaia DR3 ID & \multicolumn{3}{c}{Observed Flux (erg s$^{-1}$ cm$^{-2}$ \AA$^{-1}$)} \\ 
\cline{4-6}
&  & & \textit{UVW}2 & \textit{UVM}2 & \textit{UVW}1 \\ 
\hline
BSS 01  & Berkeley 29 & 3358218745306932224 & 1.80E-17 $\pm$ 3.24E-17 & 1.75E-17 $\pm$ 5.95E-17 & 2.10E-17 $\pm$ 5.41E-17 \\
BSS 02  & Berkeley 31 & 3157241012027949568 & 1.93E-16 $\pm$ 1.42E-17 & 2.00E-17 $\pm$ 1.66E-17 & 2.47E-16 $\pm$ 2.95E-17 \\
BSS 03  & Berkeley 31 & 3157241115107807872 & 5.12E-18 $\pm$ 8.49E-18 & 5.48E-18 $\pm$ 7.06E-18 & 6.63E-18 $\pm$ 7.32E-18 \\
BSS 04  & Berkeley 31 & 3157241282608500096 & 2.13E-17 $\pm$ 5.70E-18 & 2.46E-18 $\pm$ 5.45E-18 & 3.09E-17 $\pm$ 5.97E-18 \\
BSS 05  & Berkeley 31 & 3157241355625967744 & 1.55E-17 $\pm$ 1.57E-17 & 1.62E-17 $\pm$ 1.49E-17 & 1.88E-17 $\pm$ 2.08E-17 \\
BSS 06  & Berkeley 31 & 3157241523126689536 & 4.68E-17 $\pm$ 1.12E-17 & 4.91E-18 $\pm$ 9.49E-18 & 6.21E-17 $\pm$ 1.37E-17 \\
BSS 07  & Berkeley 31 & 3157241557486428160 & 3.86E-17 $\pm$ 9.59E-18 & 4.05E-18 $\pm$ 6.72E-18 & 4.73E-18 $\pm$ 7.85E-18 \\
BSS 08  & Berkeley 31 & 3157238263249488256 & 2.43E-17 $\pm$ 7.83E-18 & 3.24E-17 $\pm$ 6.96E-18 & 4.69E-17 $\pm$ 6.04E-17 \\
BSS 09  & Berkeley 31 & 3157241351327999104 & 1.09E-17 $\pm$ 3.60E-18 & 1.36E-18 $\pm$ 3.75E-18 & 1.78E-17 $\pm$ 4.75E-18 \\
BSS 10  & Berkeley 31 & 3157241527424009216 & 1.95E-17 $\pm$ 5.75E-18 & 2.18E-17 $\pm$ 5.22E-18 & 3.60E-17 $\pm$ 7.29E-18 \\
BSS 11  & Berkeley 31 & 3157241561783738496 & 1.74E-17 $\pm$ 4.98E-18 & 1.84E-17 $\pm$ 4.41E-18 & 3.45E-17 $\pm$ 4.69E-18 \\
BSS 12  & Berkeley 31 & 3157242283338251648 & 1.78E-17 $\pm$ 4.77E-18 & 2.08E-17 $\pm$ 3.65E-18 & 2.79E-17 $\pm$ 5.65E-18 \\
BSS 13  & Berkeley 37 & 3109976821085220736 & 3.34E-16 $\pm$ 4.31E-16 & 2.92E-16 $\pm$ 3.23E-16 & 2.48E-16 $\pm$ 2.28E-16 \\
BSS 14  & Berkeley 75 & 2922222806776944640 & 2.83E-17 $\pm$ 2.03E-17 & 2.02E-17 $\pm$ 1.36E-17 & --                      \\
BSS 15  & Berkeley 75 & 2922223253455323904 & 8.11E-17 $\pm$ 1.05E-17 & 7.53E-17 $\pm$ 1.87E-17 & 7.83E-17 $\pm$ 2.02E-17 \\
BSS 16  & Berkeley 75 & 2922222806776944640 & 1.91E-17 $\pm$ 1.46E-17 & --                      & --                      \\
BSS 17  & Berkeley 75 & 2922223249154669440 & 9.61E-17 $\pm$ 2.83E-17 & 9.37E-17 $\pm$ 7.42E-17 & 1.47E-16 $\pm$ 7.98E-17 \\
BSS 18  & NGC 2192    & 959419037352088960  & 2.38E-17 $\pm$ 2.41E-17 & 2.06E-17 $\pm$ 2.47E-17 & 2.84E-17 $\pm$ 2.36E-17 \\
BSS 19  & NGC 2192    & 959442852947148672  & 1.00E-17 $\pm$ 8.29E-18 & 9.90E-18 $\pm$ 8.21E-18 & 1.45E-17 $\pm$ 1.07E-17 \\
BSS 20  & NGC 2204    & 2941920110907962624 & 4.30E-17 $\pm$ 1.70E-17 & --                      & --                      \\
BSS 21  & NGC 2204    & 2941922928406484864 & 4.42E-17 $\pm$ 1.26E-17 & 4.75E-17 $\pm$ 1.23E-17 & --                      \\
BSS 22  & NGC 2204    & 2941936259984685312 & 1.45E-16 $\pm$ 2.14E-16 & 1.35E-16 $\pm$ 1.49E-16 & --                      \\
BSS 23  & NGC 2204    & 2941936741021002240 & 8.39E-17 $\pm$ 8.50E-17 & 7.79E-17 $\pm$ 7.89E-17 & --                      \\
BSS 24  & NGC 2204    & 2941939283641661056 & 9.80E-16 $\pm$ 1.54E-16 & 9.49E-16 $\pm$ 1.40E-16 & --                      \\
BSS 25  & NGC 2204    & 2941939283641661568 & 1.21E-16 $\pm$ 2.56E-16 & 1.20E-16 $\pm$ 1.88E-16 & --                      \\
BSS 26  & NGC 2204    & 2941935847667840128 & 3.23E-17 $\pm$ 1.04E-17 & 3.54E-17 $\pm$ 1.44E-17 & 8.47E-17 $\pm$ 8.58E-17 \\
BSS 27  & NGC 2204    & 2941936191265206016 & 3.26E-17 $\pm$ 3.90E-17 & 3.07E-17 $\pm$ 3.11E-17 & 3.60E-17 $\pm$ 3.32E-17 \\
BSS 28  & NGC 2204    & 2941939386720881536 & 5.15E-17 $\pm$ 1.80E-17 & 5.91E-17 $\pm$ 1.52E-17 & 1.14E-17 $\pm$ 1.47E-17 \\
BSS 29  & NGC 2360    & 3031253334726488576 & 7.29E-17 $\pm$ 5.37E-17 & 6.32E-17 $\pm$ 5.24E-17 & 9.34E-17 $\pm$ 5.16E-17 \\
BSS 30  & NGC 2533    & 5596625961920878848 & 5.23E-17 $\pm$ 3.37E-17 & 4.17E-17 $\pm$ 3.46E-17 & 7.63E-17 $\pm$ 4.22E-17 \\
BSS 31  & NGC 6939    & 2194727330484552960 & 1.12E-16 $\pm$ 1.03E-16 & 8.24E-17 $\pm$ 1.14E-17 & 1.48E-16 $\pm$ 1.23E-16 \\
BSS 32  & NGC 6939    & 2194728120758530944 & 5.58E-16 $\pm$ 6.68E-18 & 3.84E-16 $\pm$ 5.30E-18 & 9.94E-16 $\pm$ 1.01E-17 \\
BSS 33  & NGC 6939    & 2194818899185893760 & 2.11E-16 $\pm$ 5.04E-17 & 1.21E-16 $\pm$ 3.35E-17 & 3.98E-16 $\pm$ 5.13E-17 \\
BSS 34  & NGC 6939    & 2194724856583515776 & 2.11E-16 $\pm$ 5.04E-17 & 1.21E-16 $\pm$ 3.35E-17 & 3.98E-16 $\pm$ 5.13E-17 \\
BSS 35  & NGC 6939    & 2194724959662728064 & 1.12E-16 $\pm$ 1.03E-16 & 8.24E-17 $\pm$ 1.14E-17 & 1.48E-16 $\pm$ 1.23E-16 \\ 
\hline
\end{tabular}
\end{table*}

\subsection{Optical Photometry and Astrometry}
The foundational optical data for this study were derived from the Gaia mission's third Data Release \citep{GaiaCollaboration2023}. {\it Gaia} DR3 database provides the precise astrometric parameters ($\alpha$, $\delta$, $\varpi$, $\mu_{\alpha} \cos\delta$, $\mu_{\delta}$) essential for establishing robust cluster membership. We also utilized its high-precision photometry in the $G$ ($\lambda_{\rm eff} \approx 673$ nm), $G_{\rm BP}$ ($\lambda_{\rm eff} \approx 532$ nm), and $G_{\rm RP}$ ($\lambda_{\rm eff} \approx 797$ nm) passbands. To construct comprehensive SEDs covering the full optical window, we complemented the {\it Gaia} data with ground-based surveys. We retrieved broadband photometry ($g, r, i, z, y$) from the Pan-STARRS1 Data Release 2 \citep[DR2;][]{2020ApJS..251....7F}. For southern targets, we extended our coverage using the SkyMapper Southern Survey \citep[SMSS;][]{2007PASA...24....1K} Data Release 4 ($u, v, g, r, i, z$) \citep{2024PASA...41...61O}.

\subsection{Infrared (IR) Archival Data}
To constrain the Rayleigh-Jeans tail of the stellar SEDs, we incorporated archival data from near infrared (NIR) and MIR surveys. NIR magnitudes in the $J$ (1.24 $\mu$m), $H$ (1.66 $\mu$m), and $K_{\rm s}$ (2.16 $\mu$m) bands were sourced from the 2MASS \citep{2006AJ....131.1163S}. For the MIR regime, we utilized photometry from the AllWISE data release of the WISE \citep{2014yCat.2328....0C}. Our analysis prioritized the $W1$ (3.35 $\mu$m) and $W2$ (4.60 $\mu$m) bands because of their superior signal-to-noise ratios relative to the longer-wavelength channels. Finally, all multi-wavelength photometric data were ingested into the Virtual Observatory SED Analyzer\footnote{\url{https://svo2.cab.inta-csic.es/theory/vosa}}  \citep[VOSA;][]{Bayo2008} to perform the SED fitting procedures described in the subsequent sections.

\section{Method}
\subsection{Sample Selection} \label{sec:sample_selection}

\begin{figure}[h]
    \centering
\includegraphics[width=1\linewidth]{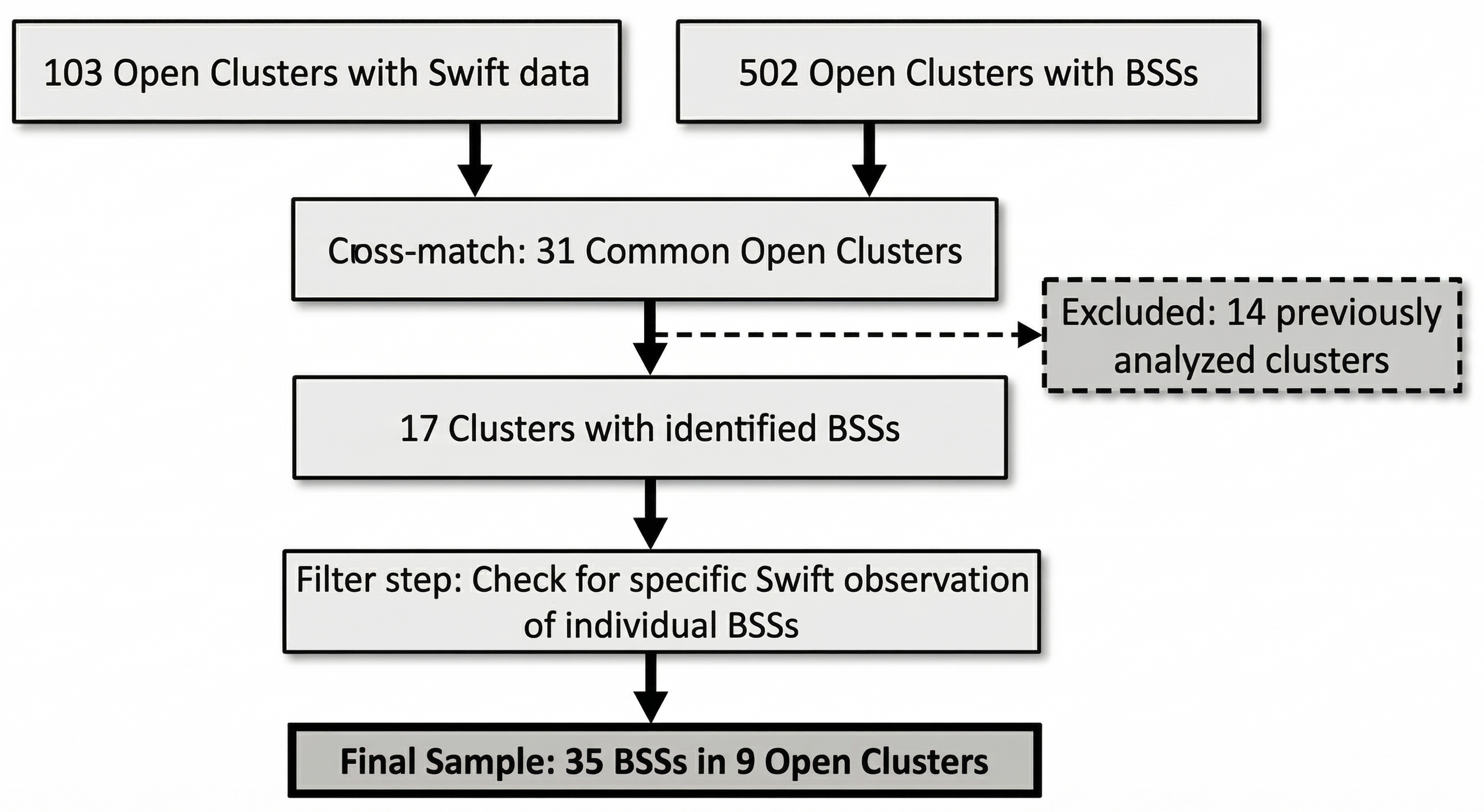}
    \caption{Flowchart of the sample selection procedure.}
    \label{fig:flowchart}
\end{figure}

To construct a comprehensive sample of BSSs suitable for UV analysis, we utilized the archival data from the \textit{Swift}/UVOT. The sample selection process is summarized in Figure~\ref{fig:flowchart}. We began by cross-correlating two primary datasets: a catalog of 502 OCs known to host BSS populations \citep[e.g.,][]{Qin2026, Li2023, Rain2024, Rain2021, Jadhav2021, Ahumada2007} and the list of 103 OCs available in the \textit{Swift} archive \citep{Siegel2019}. This initial cross-match yielded 31 common OCs that possess both identified BSS candidates and available archival UV observations.

To ensure the novelty of our analysis and to avoid duplicating results, we conducted a literature review of these 34 OCs. We excluded 14 OCs that had been previously analyzed using \textit{Swift}/UVOT or \textit{Astrosat}/UVIT data in recent studies. This filtering step yielded a subset of 17 candidate clusters, containing 47 identified BSS candidates.

Finally, we visually inspected the available \textit{Swift} images and checked the exposure times for each of the 47 candidates to ensure sufficient data quality for photometric analysis. In several cases, the BSS candidates fell outside the instrument's field of view or lacked observations in the required filters. Consequently, our final sample comprises 35, distributed across nine OCs.

\begin{figure*}[!ht]
    \centering
    \includegraphics[width=0.24\linewidth]{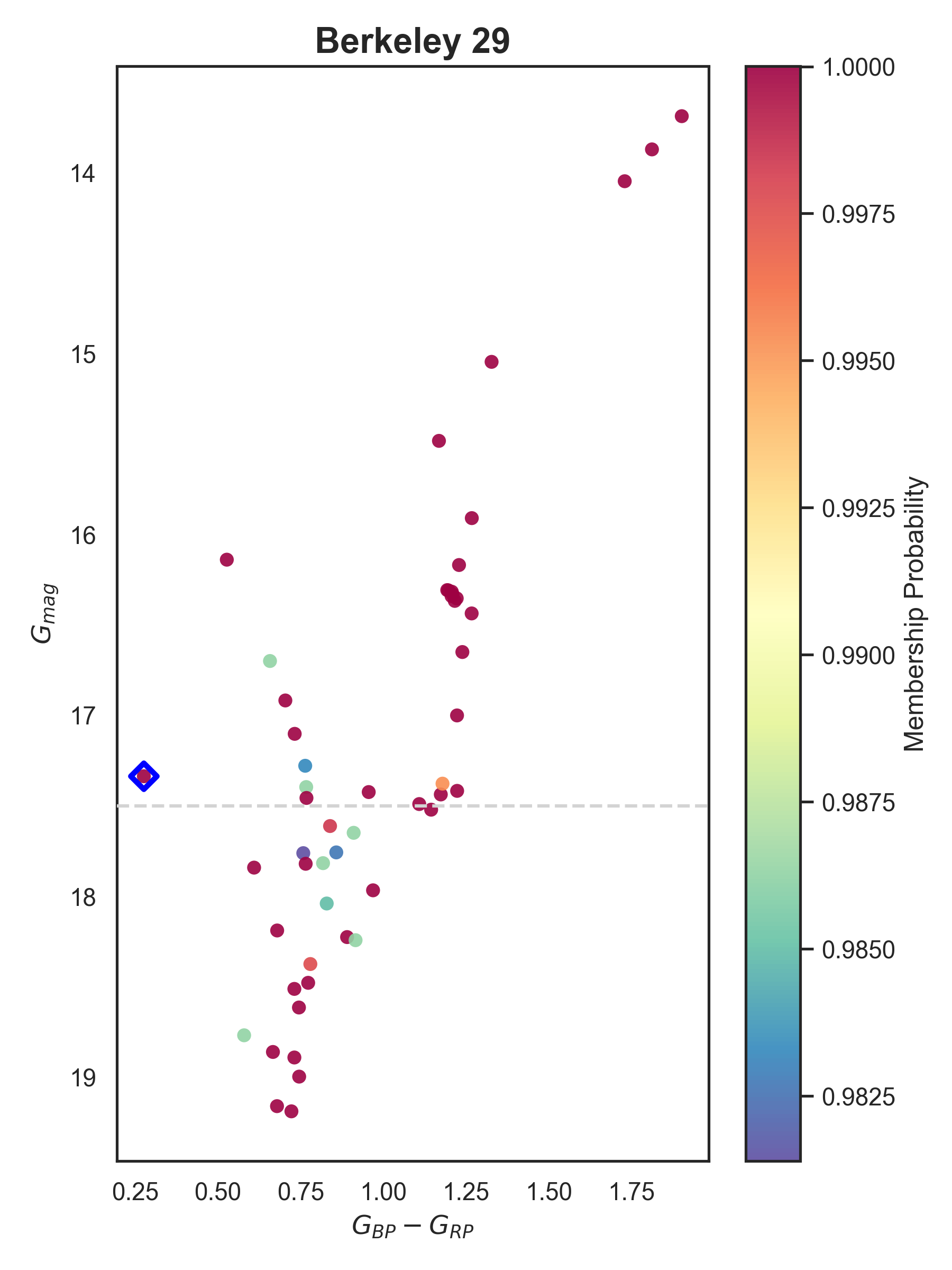}
    \includegraphics[width=0.24\linewidth]{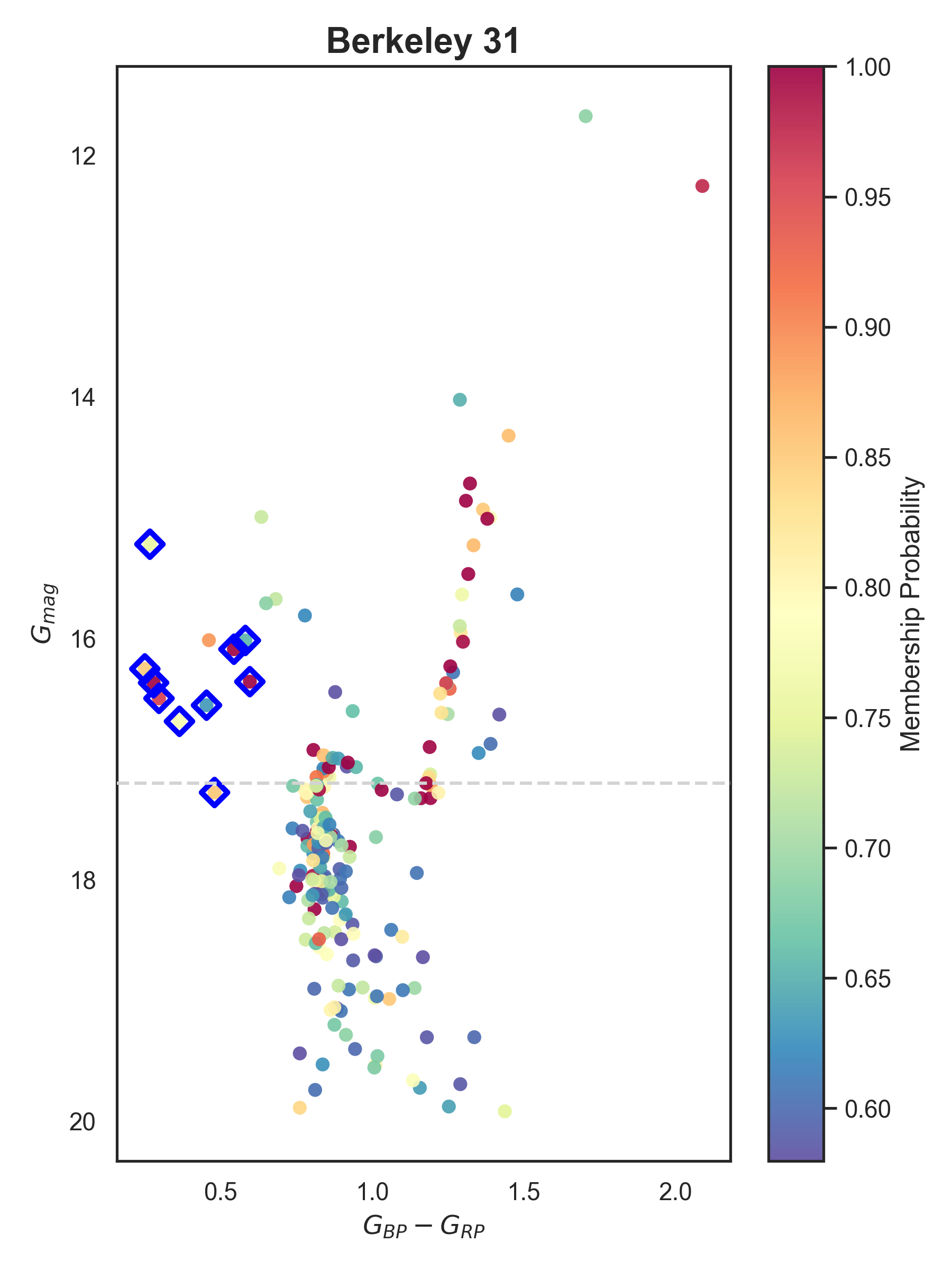}
    \includegraphics[width=0.24\linewidth]{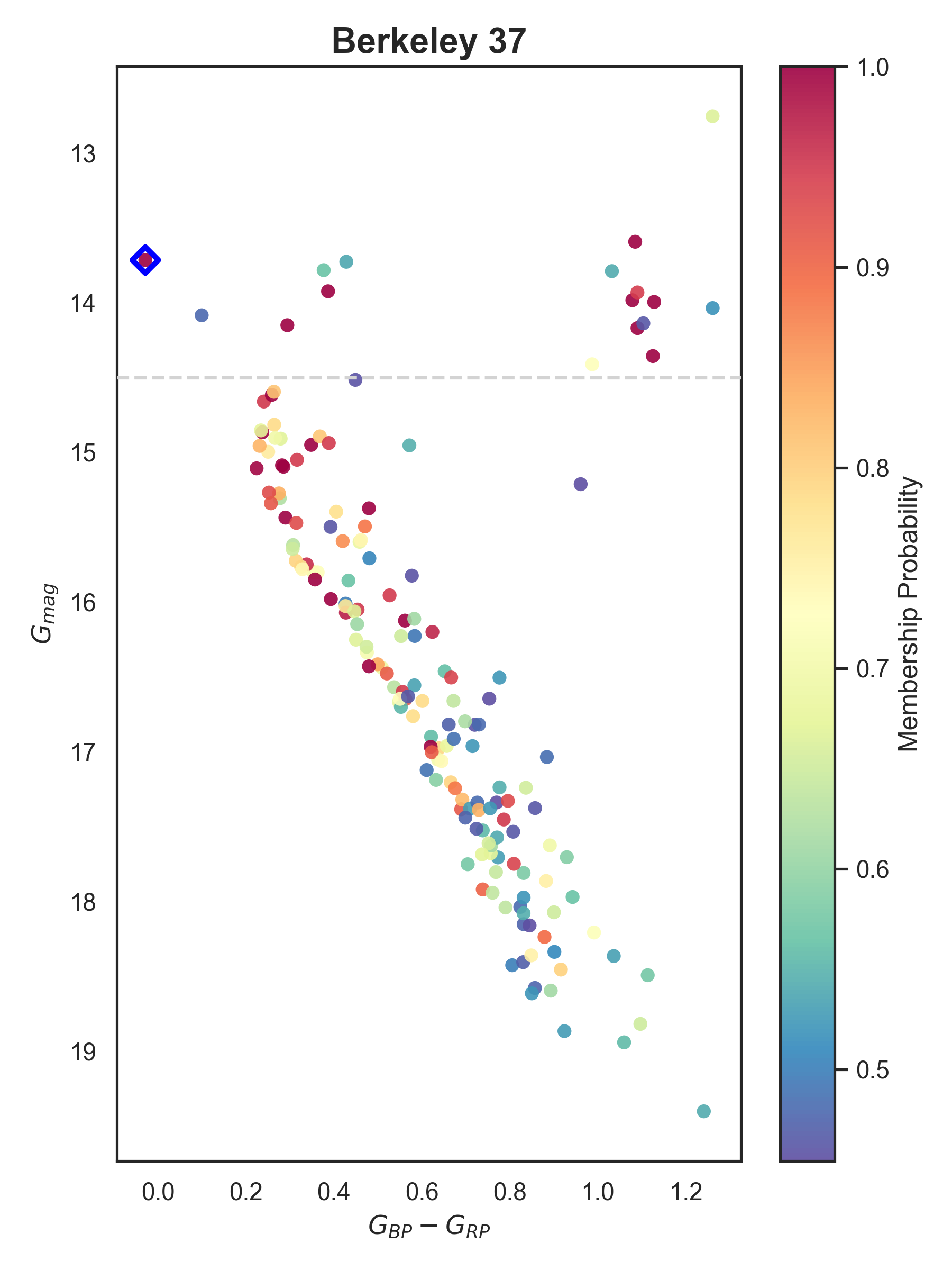}\\
    \includegraphics[width=0.24\linewidth]{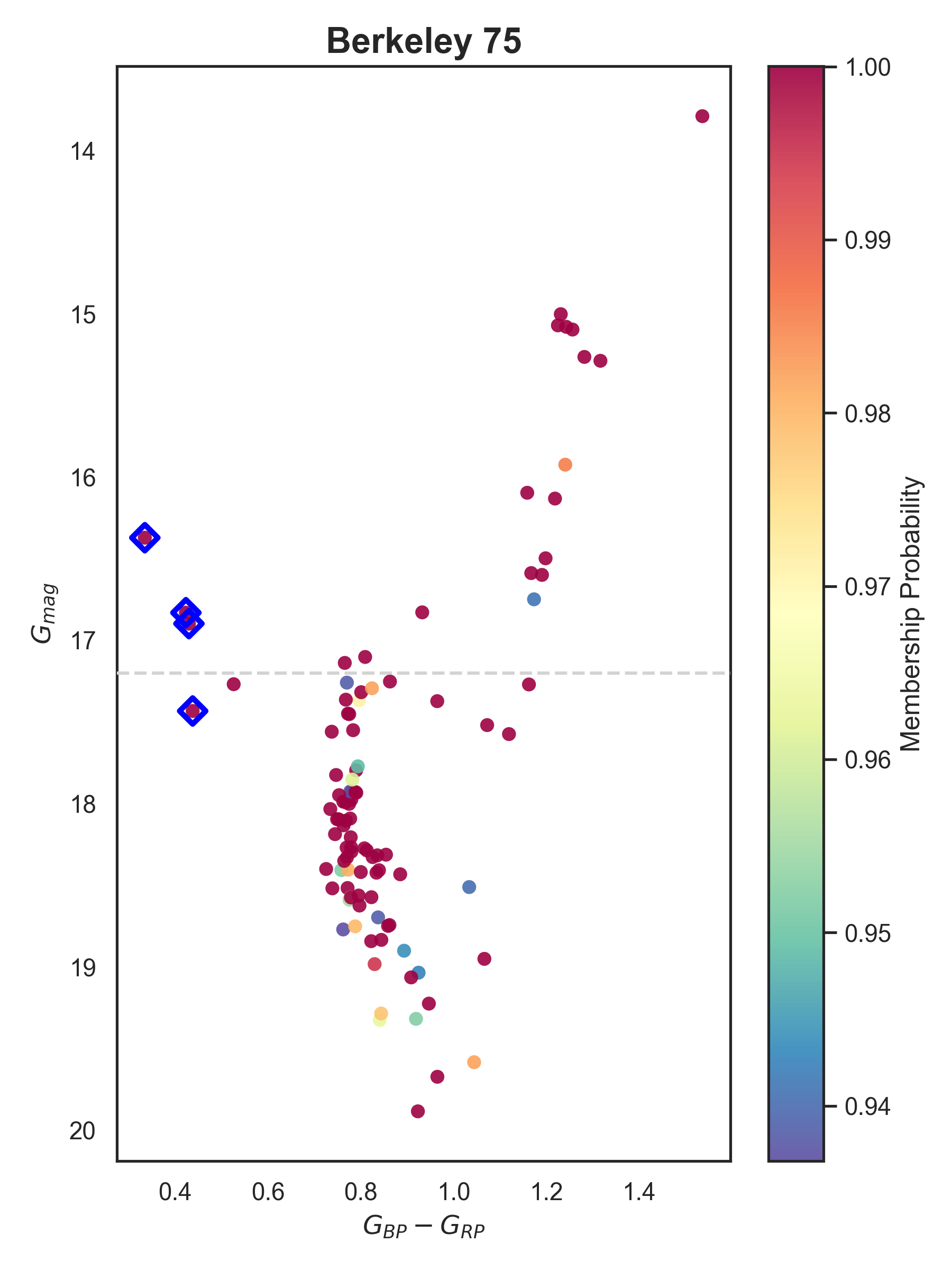}
    \includegraphics[width=0.24\linewidth]{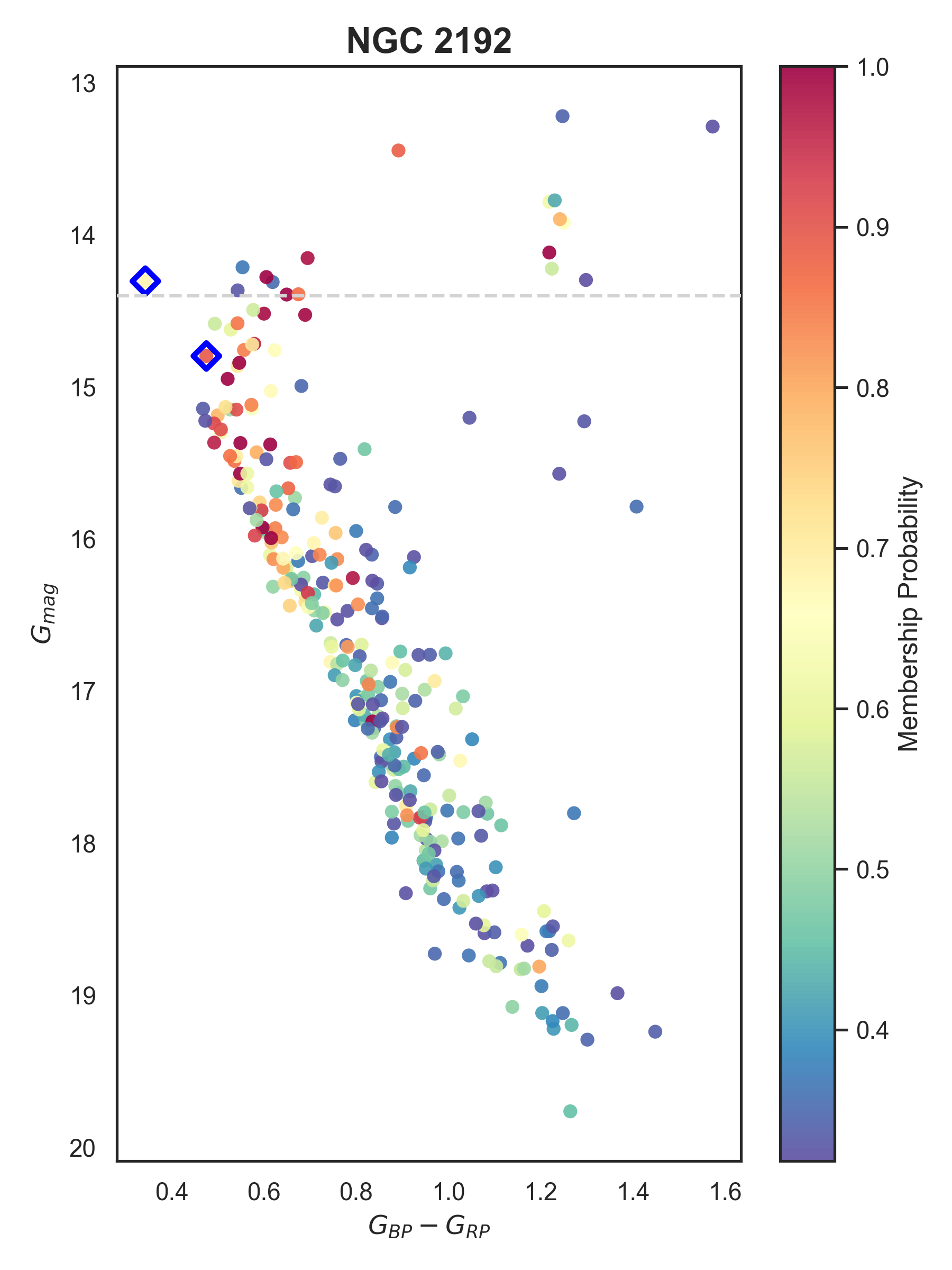}
    \includegraphics[width=0.24\linewidth]{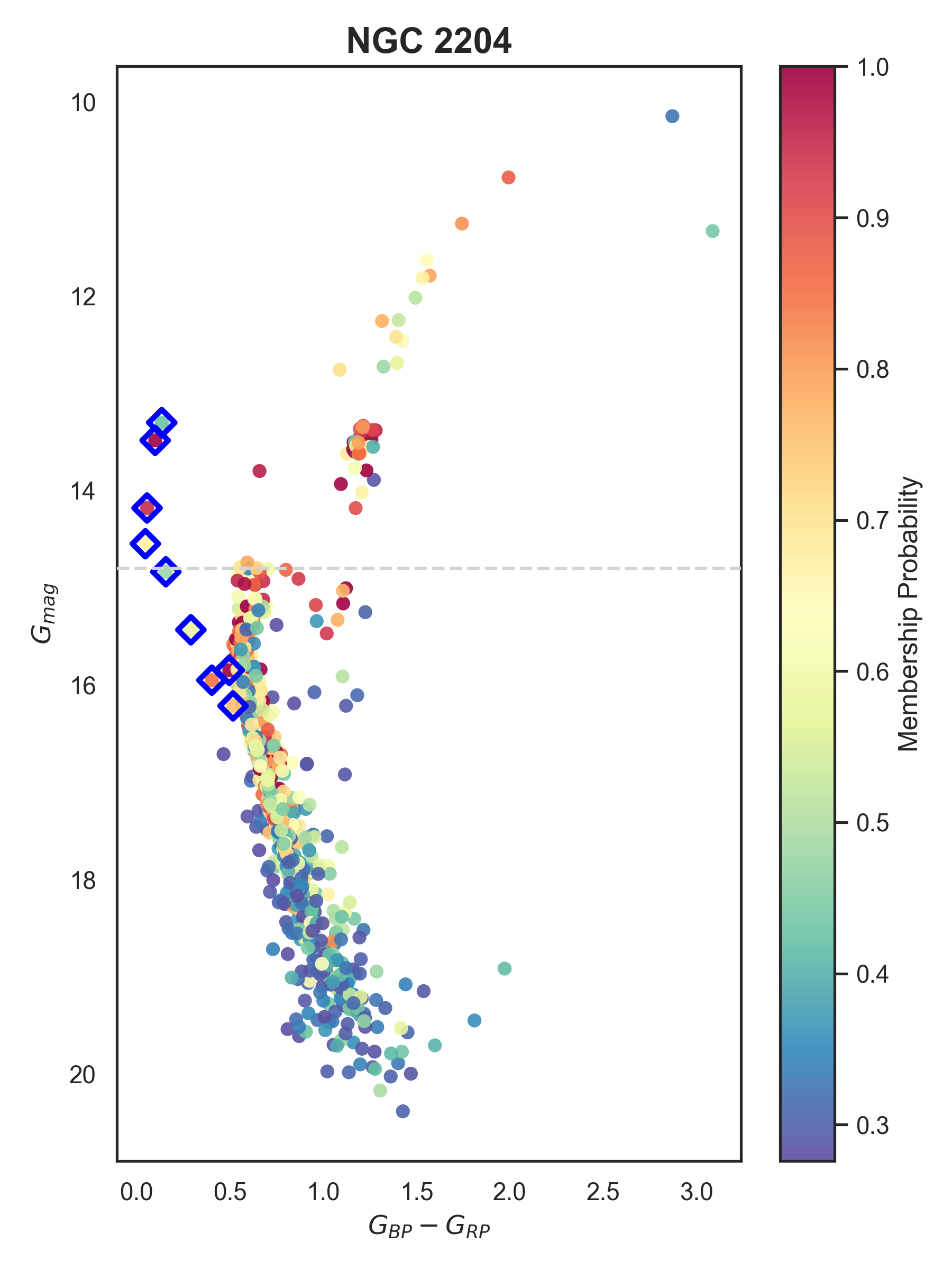}\\
    \includegraphics[width=0.24\linewidth]{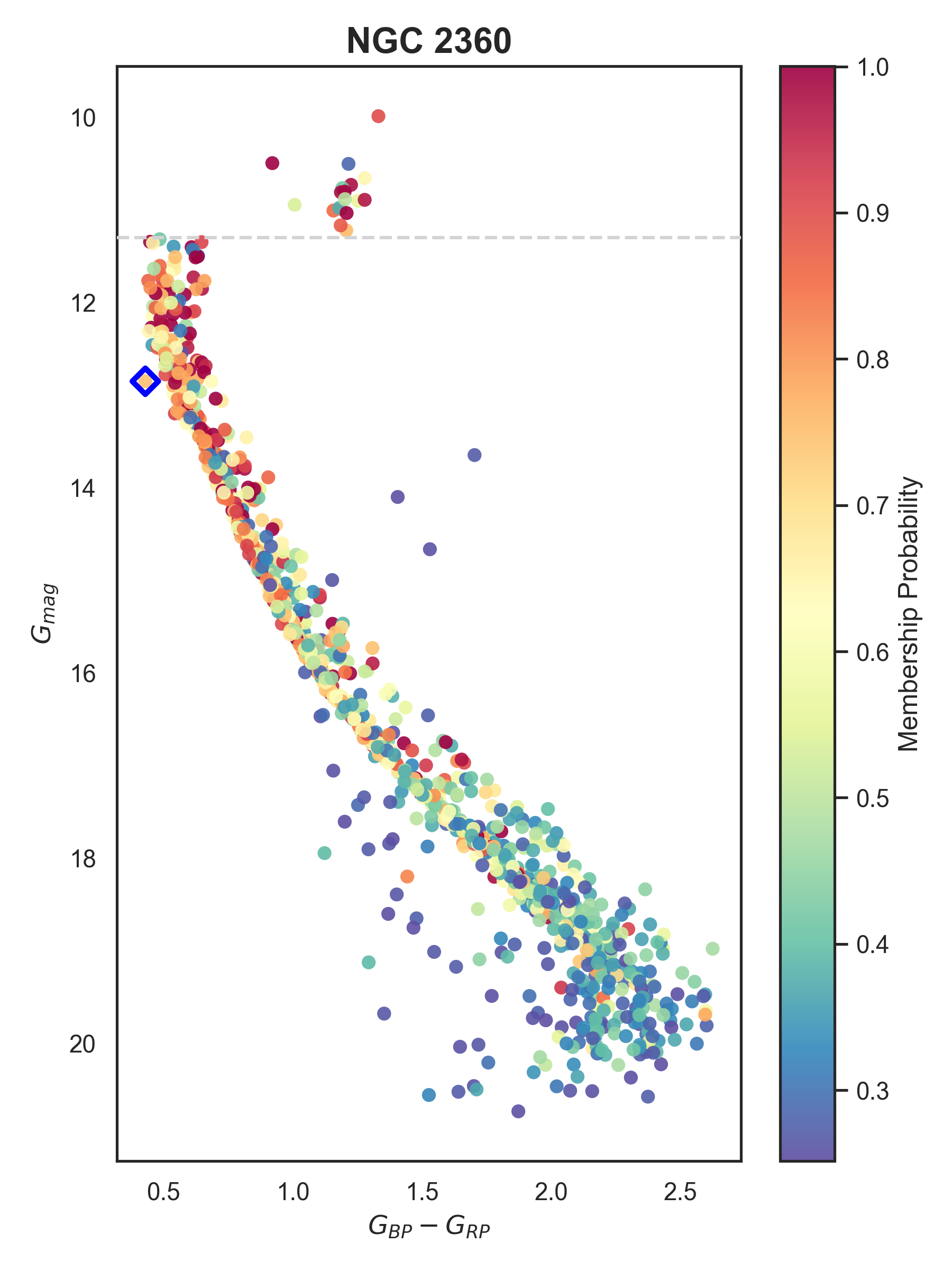}
    \includegraphics[width=0.24\linewidth]{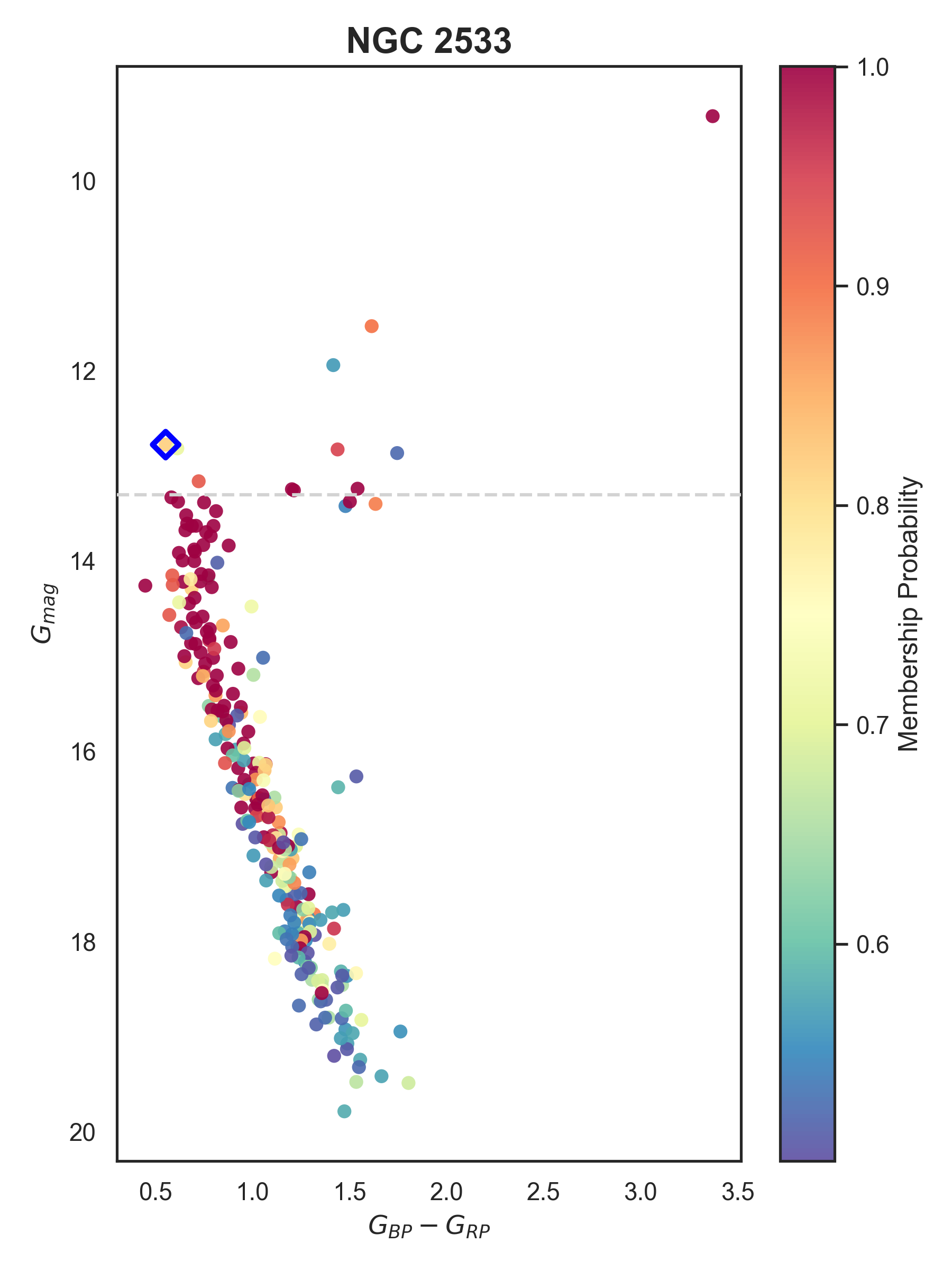}
    \includegraphics[width=0.24\linewidth]{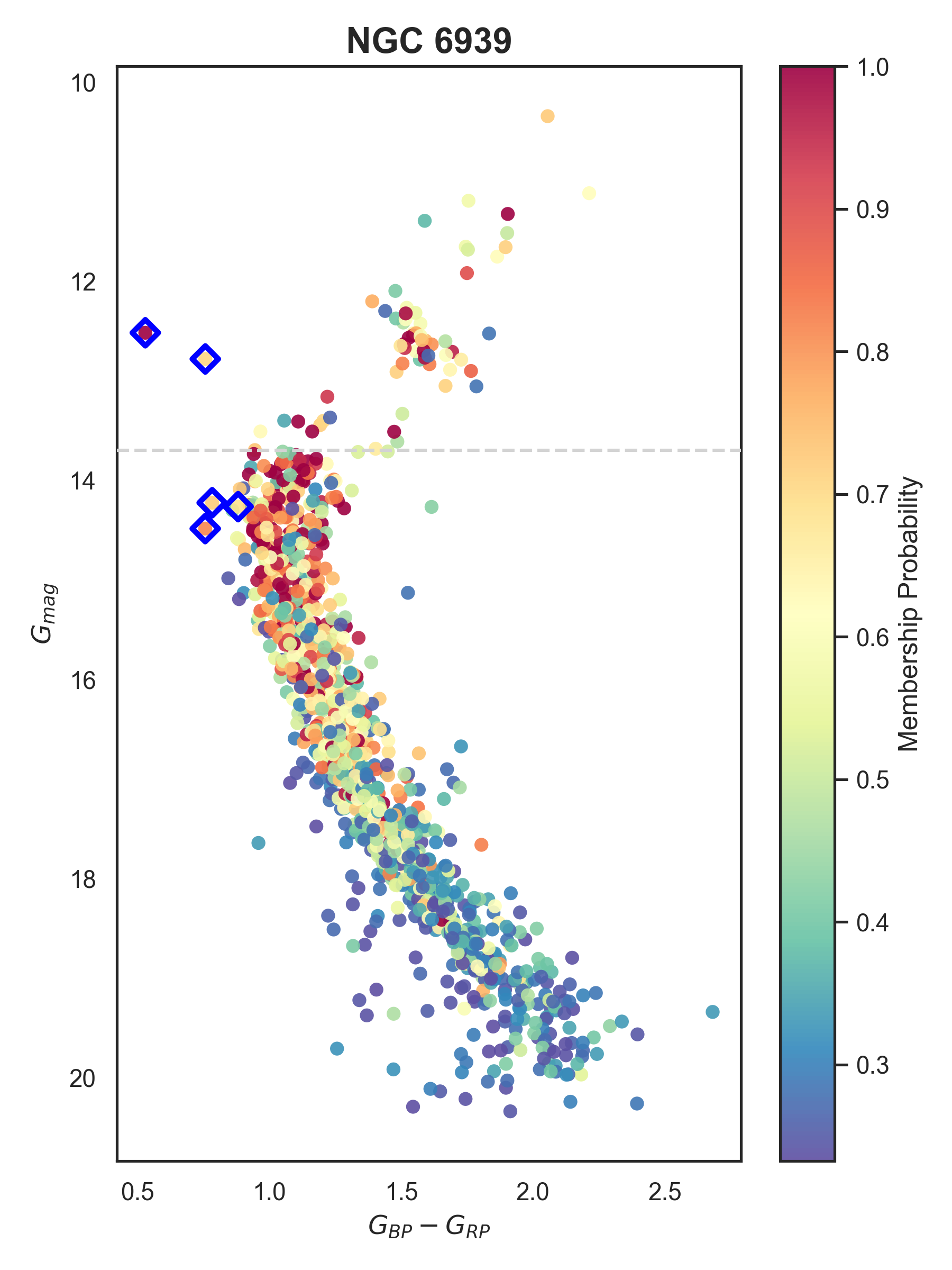}
    \caption{CMDs for the studied nine OCs, showing the the $G$-band apparent magnitude versus $G_{\rm BP}-G_{\rm RP}$ color. The plotted stars correspond to cluster members adopted from \citet{Hunt2024}, and the color scale indicates their associated membership probabilities, with redder colors denoting higher membership likelihood. Blue diamond symbols mark the identified BSS candidates with available \textit{Swift}/UVOT observations in each cluster. The horizontal dashed line represents the adopted MSTO magnitude in this study.}
    \label{fig:CMDs}
\end{figure*}

To further validate the membership status and evolutionary state of our final sample, we cross-matched the 35 candidates and their host clusters against the comprehensive OCs catalog of \citet{Hunt2024}. This catalog provides high-precision astrometric data and robust membership probabilities derived from Gaia DR3. Using these membership probabilities, we constructed the CMDs presented in Figure~\ref{fig:CMDs}. We note that the horizontal dashed line marks an approximate MSTO magnitude used to separate MS and evolved stars, and should not be taken as a strict boundary for BSS identification.

\begin{table*}[ht]
    \centering
    \footnotesize
    \caption{Fundamental parameters of the nine OCs adopted from \citet{Hunt2024}. 
    The columns list equatorial coordinates, proper-motion components, and trigonometric parallax, radial velocity, logarithmic age, $V$-band extinction, distance modulus, and half-mass relaxation time. The relaxation times are computed as described in Section~\ref{sec:radial}.}
    \label{tab:cluster_params}
    \setlength{\tabcolsep}{1.5pt}
    \renewcommand{\arraystretch}{1.1}
    \begin{tabular}{c c c c c c c c c c c c} 
        \hline \hline
        No. & Cluster & $\alpha_{2000}$ & $\delta_{2000}$ & $\mu_{\alpha} \cos\delta$ 
            & $\mu_{\delta}$ & $\varpi$ & RV & $\log t$ & $A_{\rm V}$ & $(m-M)$ 
            & $t_{\rm relax}$ \\
         & & (hh:mm:ss) & (dd:mm:ss) & (mas\,yr$^{-1}$) & (mas\,yr$^{-1}$) & (mas) 
           & (km\,s$^{-1}$) & (yr) & (mag) & (mag) & (Myr) \\
        \hline
        1 & Berkeley~29 & 06:53:03.81 & $+$16:55:33.16 & $+0.12 \pm 0.10$ & $-1.06 \pm 0.05$ 
          & $0.05 \pm 0.05$ & $25.39 \pm 0.18$ & $8.41 \pm 0.62$ & $1.49 \pm 0.78$ 
          & $15.60 \pm 0.10$ & ~~4 \\
        2 & Berkeley~31 & 06:57:37.61 & $+$08:17:17.29 & $+0.11 \pm 0.13$ & $-0.92 \pm 0.09$ 
          & $0.12 \pm 0.08$ & $65.27 \pm 5.06$ & $9.08 \pm 0.23$ & $0.43 \pm 0.21$ 
          & $13.90 \pm 0.19$ & 17 \\
        3 & Berkeley~37 & 07:20:16.54 & $-$00:59:53.55 & $-1.05 \pm 0.06$ & $+0.71 \pm 0.07$ 
          & $0.18 \pm 0.05$ & $81.75 \pm 3.79$ & $9.01 \pm 0.19$ & $0.07 \pm 0.06$ 
          & $13.51 \pm 0.22$ & 13 \\
        4 & Berkeley~75 & 06:48:59.75 & $-$23:59:45.57 & $-0.24 \pm 0.07$ & $+1.18 \pm 0.11$ 
          & $0.11 \pm 0.07$ & $125.07 \pm 1.81$ & $9.27 \pm 0.21$ & $0.20 \pm 0.16$ 
          & $14.36 \pm 0.22$ & 17 \\
        5 & NGC~2192 & 06:15:17.39 & $+$39:50:33.94 & $+0.19 \pm 0.05$ & $-1.97 \pm 0.06$ 
          & $0.23 \pm 0.05$ & $20.75 \pm 5.53$ & $9.03 \pm 0.19$ & $0.19 \pm 0.11$ 
          & $13.08 \pm 0.19$ & 13 \\
        6 & NGC~2204 & 06:15:32.45 & $-$18:40:17.37 & $-0.58 \pm 0.06$ & $+1.96 \pm 0.07$ 
          & $0.20 \pm 0.05$ & $93.89 \pm 1.42$ & $9.26 \pm 0.17$ & $0.05 \pm 0.05$ 
          & $13.29 \pm 0.22$ & 76 \\
        7 & NGC~2360 & 07:17:45.75 & $-$15:38:39.91 & $+0.39 \pm 0.14$ & $+5.63 \pm 0.13$ 
          & $0.92 \pm 0.05$ & $27.25 \pm 0.73$ & $9.16 \pm 0.18$ & $0.12 \pm 0.08$ 
          & $10.02 \pm 0.17$ & 36 \\
        8 & NGC~2533 & 08:07:04.48 & $-$29:51:37.53 & $-3.18 \pm 0.09$ & $+5.07 \pm 0.08$ 
          & $0.35 \pm 0.05$ & $40.58 \pm 5.16$ & $8.86 \pm 0.18$ & $0.83 \pm 0.19$ 
          & $12.04 \pm 0.16$ & 27 \\
        9 & NGC~6939 & 20:31:36.36 & $+$60:39:04.76 & $-1.82 \pm 0.10$ & $-5.46 \pm 0.10$ 
          & $0.53 \pm 0.03$ & $-18.81 \pm 1.12$ & $9.02 \pm 0.15$ & $1.33 \pm 0.18$ 
          & $11.32 \pm 0.14$ & 28 \\
        \hline
    \end{tabular}
\end{table*}

\subsection{SED Analysis}
\label{sec:section3.2}
To determine the fundamental physical properties ($T_{\rm eff}$, $R$, $L$) of the BSS candidates, and to screen for potential unresolved hot companions, we performed a detailed SED analysis. We employed the VOSA to construct and fit SEDs using a broad photometric baseline. The dataset covers the ultraviolet ($Swift$/UVOT), optical ({\it Gaia}~DR3, Pan-STARRS, SkyMapper, TESS), and infrared (2MASS, WISE) regimes, as detailed in Section~\ref{sec:data}.

For the fitting process, we provided VOSA with the equatorial coordinates of the sources, along with cluster distances ($d_{\rm geo}$) adopted from the Gaia EDR3-based geometric distance estimates of \citet{BailerJones2021}. These distances account for the non-linearity of the parallax distance transformation and incorporate a physically motivated Galactic prior, making them more robust than simple parallax inversions, particularly for distant and low-parallax cluster members. VOSA then retrieved the available multiwavelength photometry within a $3\arcsec$ search radius around each source. A similar approach has been adopted in previous VOSA-based SED studies of BSS populations in OCs \citep{ 2024AJ....168..278C, Chand2024, 2024AJ....168..274S, Sheikh2024}. To ensure data integrity, we visually inspected each source using Aladin\footnote{\url{https://aladin.u-strasbg.fr/}} and confirmed that the photometry was free from contamination by nearby sources. We applied extinction corrections using the $R_{\rm V} = 3.1$ reddening laws of \cite{Fitzpatrick1999} and \cite{Indebetouw2005}.

\begin{figure}
    \centering
    \includegraphics[width=1\linewidth]{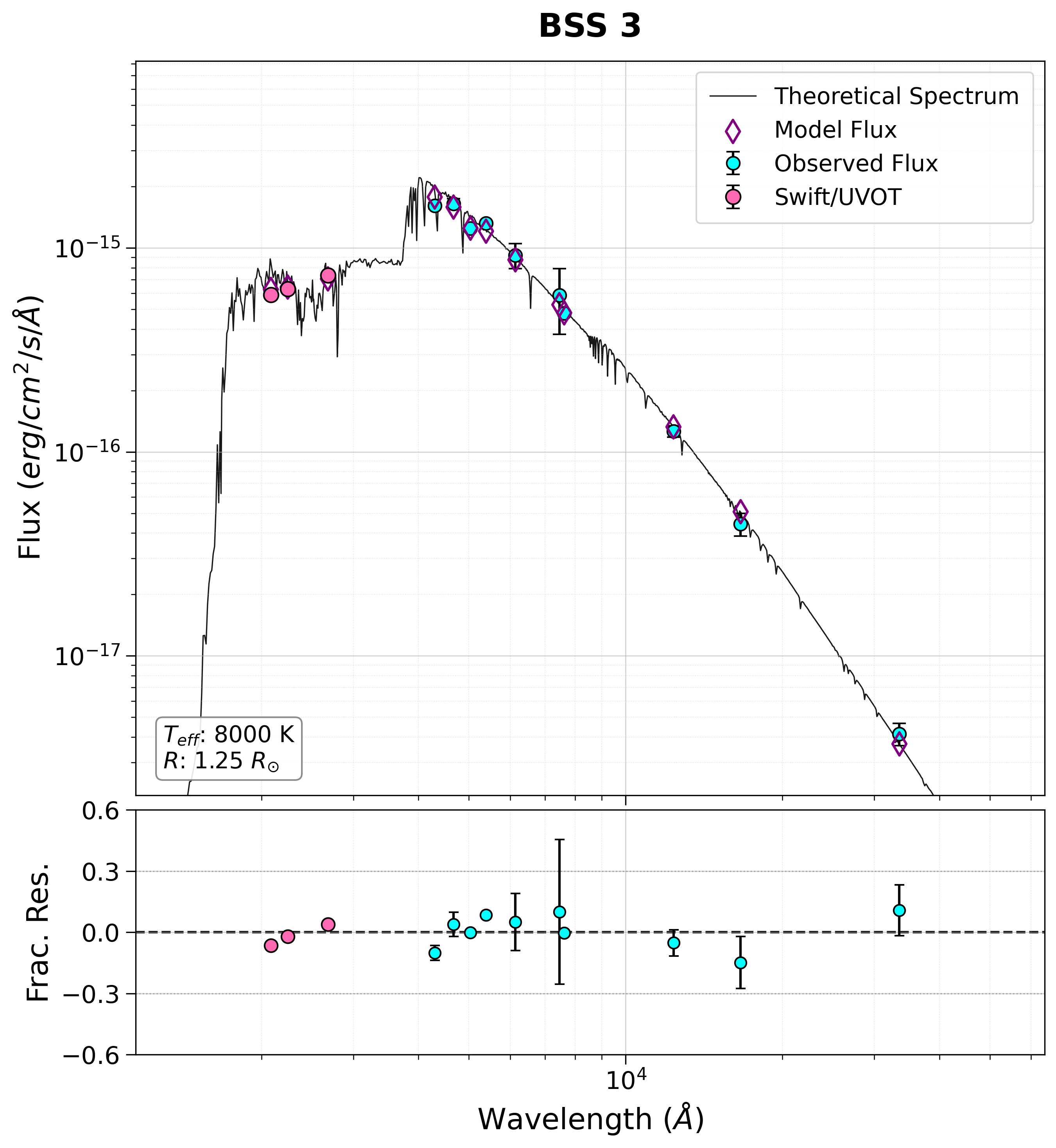}
    \caption{Representative SED fitting analysis for a single-component BSS candidate in Berkeley 31. \textbf{Top panel:} Cyan circles with error bars denote extinction-corrected observed multi-wavelength photometric fluxes. The solid black curve represents the best-fitting theoretical spectrum derived from \citet{Castelli1997} models. Purple diamonds indicate the synthetic fluxes integrated over the corresponding pass-bands. \textbf{Bottom panel:} The fractional residuals between the observed data and the model fluxes, demonstrating the quality of the fit. The SED plots for the remaining single-component BSS candidates are presented in Appendix~\ref{AppendixA}.}
    \label{fig:single_sed}
\end{figure}

We initially modeled the SEDs using single-star atmospheres from the Kurucz ODFNEW/NOVER grid \citep{Castelli1997, Castelli2003}. The parameter space was restricted to temperatures between $5000$--$15000$~K and surface gravities of $\log g = 2.5$--$5.0$~dex, with metallicities fixed to literature-based cluster values  \citep[e.g.,][]{Casamiquela2019, Donor2020, Dias2021, Randich2022, Fu2022,  Netopil2022}. The quality of the fit was assessed using the visual goodness-of-fit parameter, $V_{\rm gfb}$ \citep{Bayo2008}. To account for potentially underestimated photometric errors, VOSA calculates a modified reduced $\chi^2$ where flux errors are enforced to be at least 10\% of the observed flux. Following established conventions \citep[e.g.,][]{Jimenez-Esteban2021, Zeng2025}, we considered fits with $V_{\rm gfb} < 10$--$15$ to be reliable representations of single stars \citep{Rebassa-Mansergas2021}. Figure~\ref{fig:single_sed} displays the SED analysis for a representative single BSS in NGC~2204. The corresponding fits for all other single stars in our sample can be found in Appendix~\ref{AppendixA}.

The uncertainties in the derived radii and luminosities were estimated through standard error propagation within the VOSA framework. The stellar radius is obtained from the scaling factor $M_{\rm d} = (R/d)^2$, where $d$ is the cluster distance, giving $R = d\,\sqrt{M_{\rm d}}$. Its uncertainty is:
\begin{equation}
    \sigma_{\rm R} = R \cdot \sqrt{\left(\frac{\sigma_{\rm d}}{d}\right)^2 + \frac{1}{4}\left(\frac{\sigma_{M_{\rm d}}}{M_{\rm d}}\right)^2},
\end{equation}
where $\sigma_{\rm d}$ and $\sigma_{M_{\rm d}}$ are the uncertainties in distance \citep{BailerJones2021} and scaling factor, respectively. The luminosity is computed from $L = 4\pi R^2 \sigma T_{\rm eff}^4$, with uncertainty \citep{Eker2015, Eker2018, Eker2024}:
\begin{equation}
    \frac{\sigma_{\rm L}}{L} = \sqrt{\left(\frac{2\sigma_{\rm R}}{R}\right)^2 + \left(\frac{4\sigma_{T_{\rm eff}}}{T_{\rm eff}}\right)^2}.
\end{equation}
The temperature uncertainties were estimated via Monte Carlo simulations, in which the observed photometric fluxes were perturbed within their measurement errors, and the SED fitting was repeated over multiple realizations; the dispersion in $T_{\rm eff}$ was adopted as the $1\sigma$ uncertainty. As a result, the temperature uncertainties vary between sources, largely reflecting differences in photometric quality and wavelength coverage. The similar uncertainty levels in radius and luminosity are mainly driven by the shared fractional distance error, which is expected to contribute significantly to the radius uncertainty for members of the same cluster.

A critical component of our analysis was the search for UV excesses, which often betray the presence of a WD companion \citep{Gosnell2015, Rao2022}. We analyzed the fractional residuals between the observed and model fluxes, specifically in the \textit{Swift}/UVOT filters. Sources exhibiting a residual excess greater than $\sim30\%$ in at least two UV bands were flagged as binary candidates \citep{Chand2024}. A UV excess of $\geq$30\% in at least two UV bands is required to ensure that the detected excesses are robust against photometric uncertainties and small extinction mismatches. This threshold is chosen to minimize contamination from photometric uncertainties and extinction mismatches while ensuring robust identification of genuine excess emission. Similar multi-band UV-excess criteria have been employed in previous UV studies of blue stragglers and post-mass-transfer systems, where they have reliably identified hot, compact companions \citep{Gosnell2015, 2022MNRAS.511.2274V, Chand2024}. 

For candidates showing a significant UV excess, we used the VOSA \texttt{Binary Fit} module. These systems were modeled as a combination of a cool BSS component \citep{Castelli2003} and a hot companion, represented by WD models \citep{Koester2010}, with parameters ranging from $T_{\rm eff, WD} = 10,000$ to $80,000$~K and $\log g_{\rm WD} = 6.5$ to $9.5$~dex. The parameters for both components were optimized simultaneously.

In addition to these cases, we identify BSS~28 as a source exhibiting a noticeable IR excess. Such excess emission may be indicative of circumstellar material, including possible dust or debris disk structures \citep{Patel14}. However, mid-infrared excesses derived from broadband photometry can be affected by background contamination, blending, or photometric uncertainties, especially in crowded cluster fields. Therefore, the nature of the excess in BSS 28 cannot be conclusively established with the current data. Therefore, we consider BSS~28 a tentative candidate, and its detailed characterization will require higher-quality, higher-resolution infrared observations in future studies.

We note that SED fitting of composite systems is subject to intrinsic degeneracies between effective temperature and radius, especially for hot compact companions. The inclusion of UV data, in particular, reduces these degeneracies, as the UV flux is highly sensitive to the temperature of the hot component while contributing negligibly at optical wavelengths. As a result, the derived parameters for the hot companions, particularly in systems with strong UV excesses, are reasonably well constrained; however, they remain subject to degeneracies and should be interpreted with caution in the absence of spectroscopic constraints.

We note that the derived parameters for BSS 15, BSS 30, and BSS 32 remain comparatively uncertain because these systems lack strong FUV constraints. In such cases, the decomposition of the composite SED is more strongly affected by degeneracies between the temperature and radius of the hot component. Therefore, the inferred parameters for these systems should be interpreted with caution until deeper ultraviolet observations or spectroscopic follow-up data become available.

\section{Results and Discussion}\label{sec:results}

In this section, we present the results of the multi-wavelength SED analysis performed on 35 candidates across nine OCs. One of the key outcomes of this study is the identification of candidate hot WD companions in several systems within our sample. These hot, compact components provide important observational constraints on the mass-transfer formation channel and are typically obscured at optical wavelengths by the BSS primary's dominant luminosity. However, the high UV sensitivity and broad spectral coverage provided by the \textit{Swift}/UVOT observations enabled the detection and characterization of these faint companions.

Our analysis shows that the majority of the BSS sample (20 out of 35 targets) consists of systems whose additional components cannot be resolved or detected within the sensitivity limits of the current dataset, rather than being conclusively single stars. Among the remaining systems, 15 objects ($\sim43\%$) exhibit ultraviolet excesses indicative of potential hot companions. We further examine whether the fraction of binary BSSs identified in each cluster depends on global cluster properties, particularly the overall binary fraction. However, homogeneous estimates of cluster-wide binary fractions are not available for all targets in our sample, which prevents a direct quantitative comparison. Instead, we perform a qualitative assessment of the relative occurrence of UV-excess BSSs across clusters. Clusters with larger BSS populations tend to exhibit a higher number of candidate binary systems, primarily reflecting sample-size effects rather than an intrinsic physical correlation. Therefore, while our results are consistent with a binary-driven formation scenario, a robust test of any correlation between the cluster binary fraction and BSS binarity will require homogeneous measurements of binary fractions from spectroscopic or variability studies. In the following subsections, we present the fundamental physical parameters ($T_{\rm eff}$, $R$, and $L$) derived for both single and binary systems and discuss the implications of these newly discovered binaries for the evolutionary history of BSS.

\begin{figure*}[ht]
    \centering
    
    \includegraphics[width=0.3\linewidth]{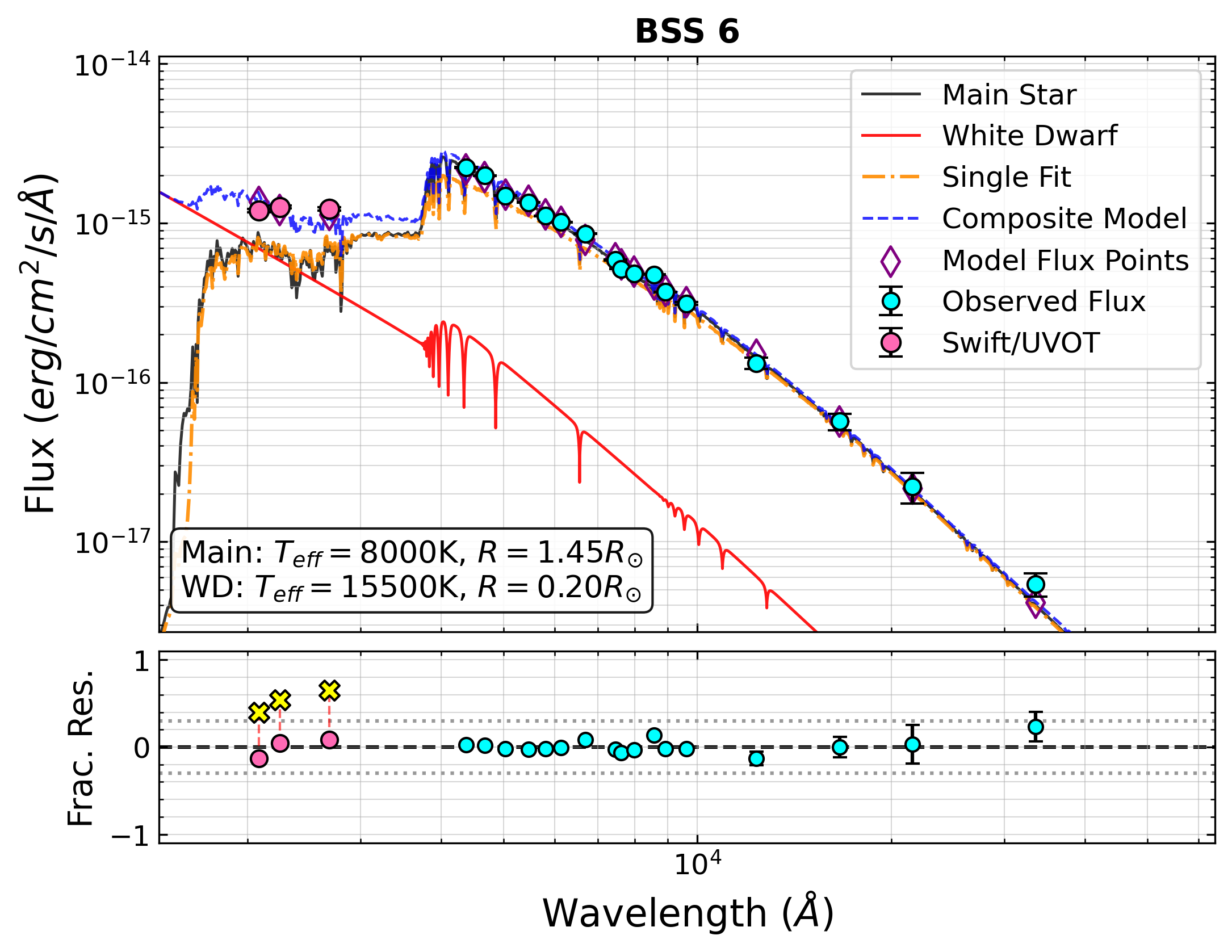}
    \includegraphics[width=0.3\linewidth]{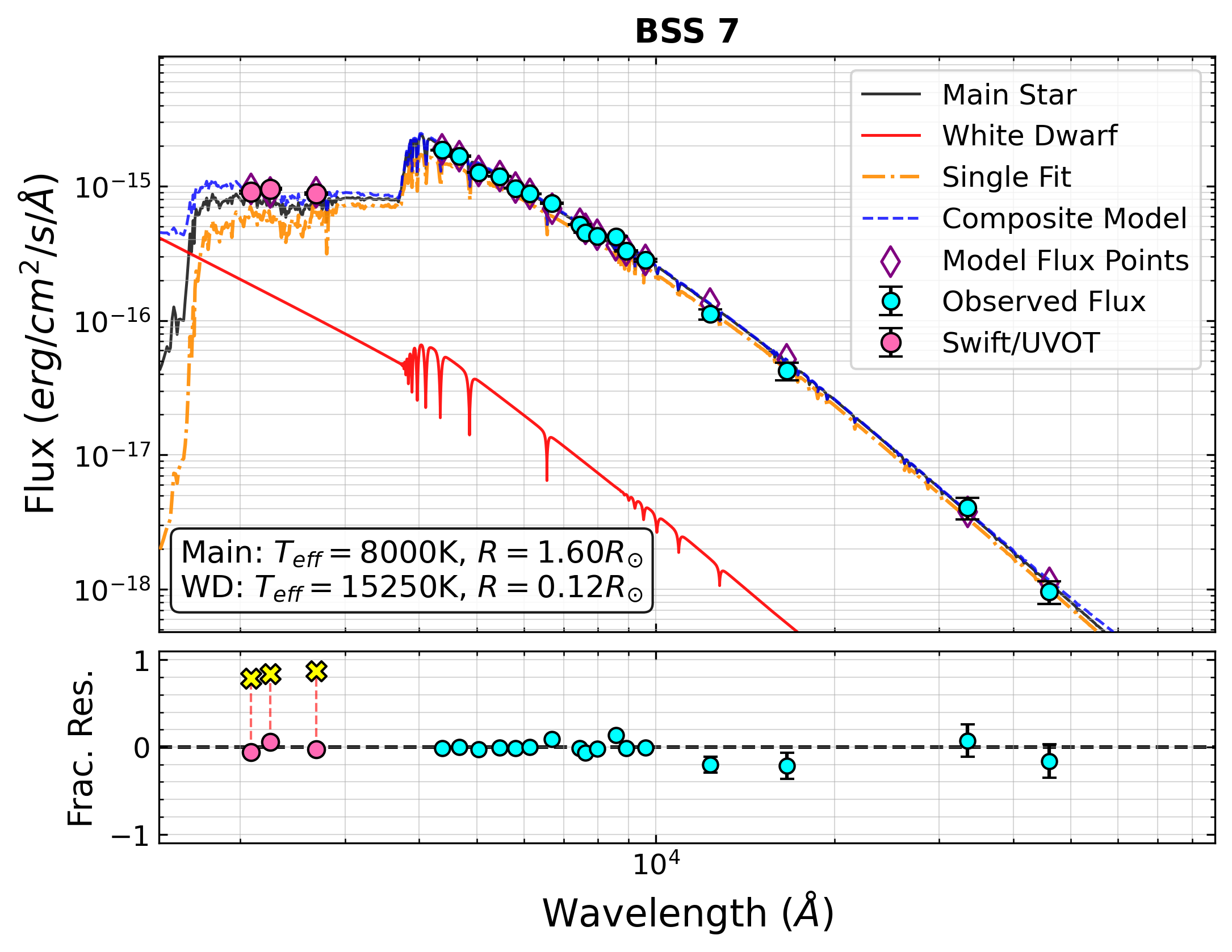}
    \includegraphics[width=0.3\linewidth]{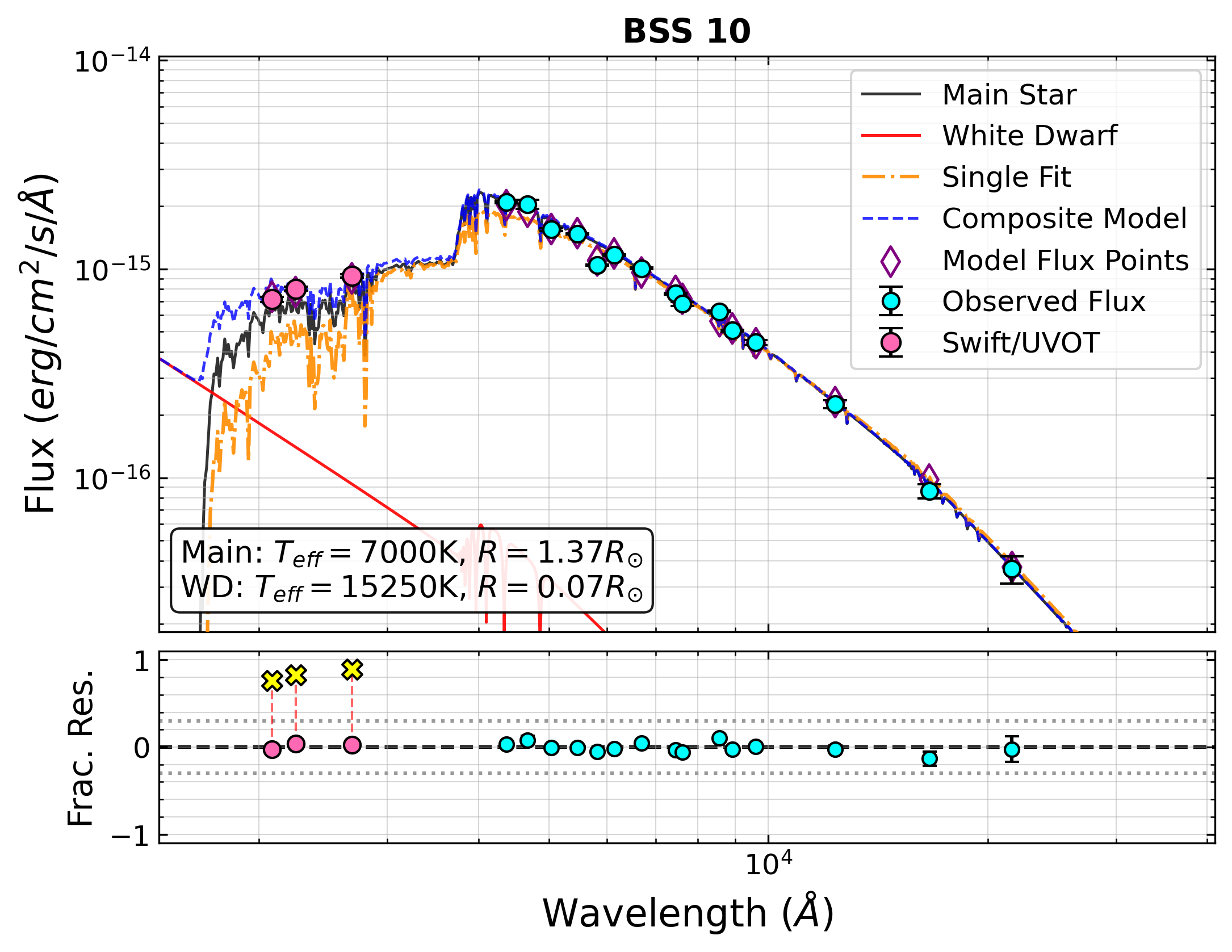}\\
    \includegraphics[width=0.3\linewidth]{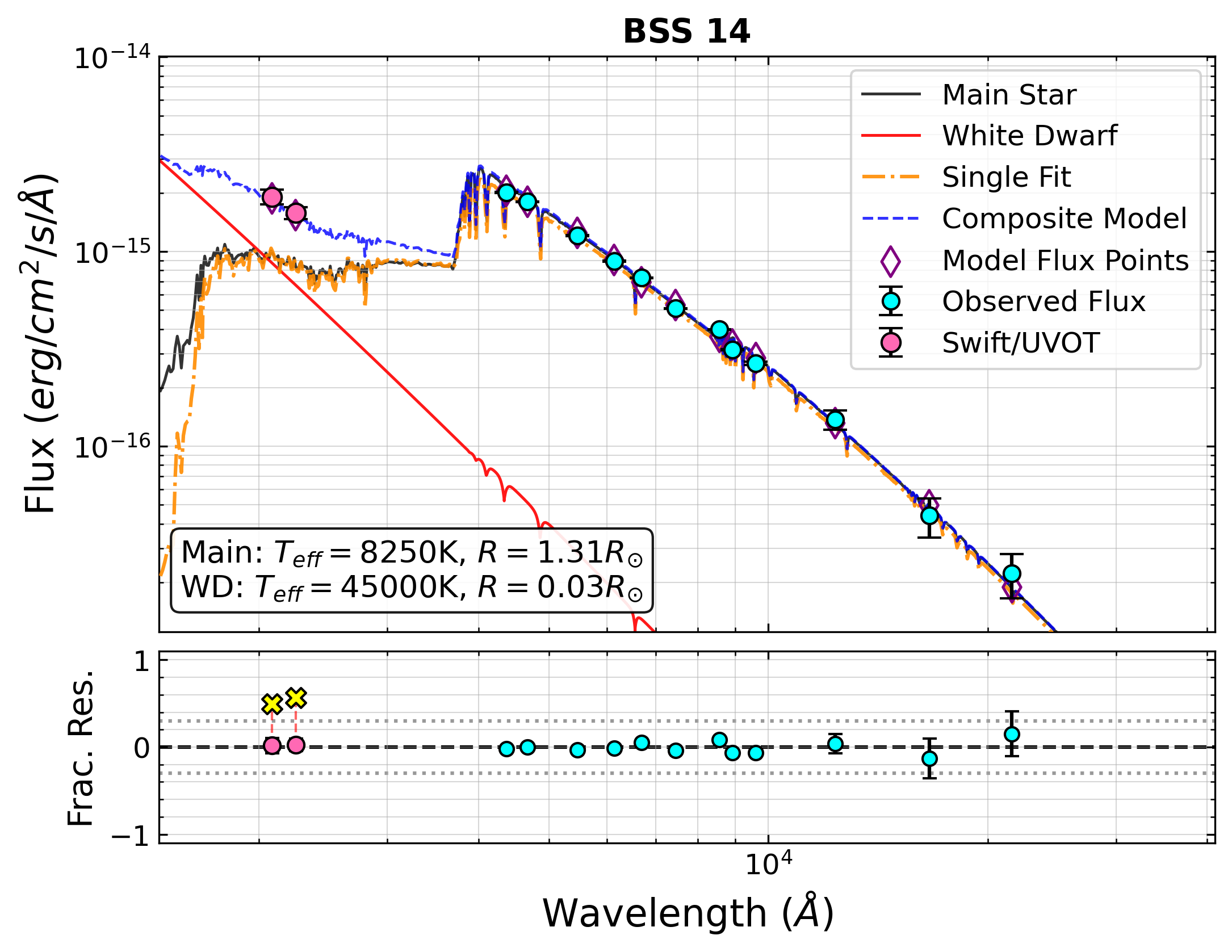}
    \includegraphics[width=0.3\linewidth]{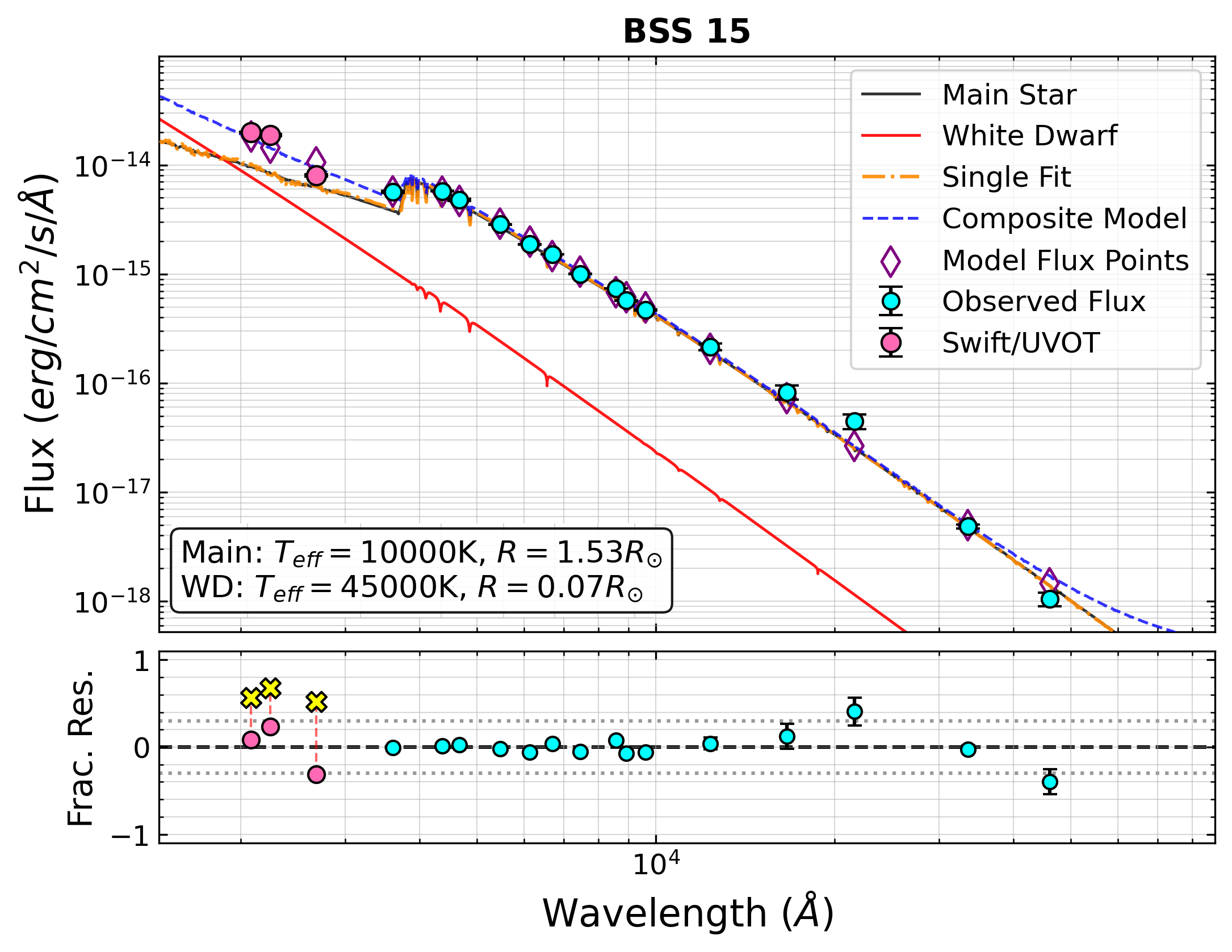}
    \includegraphics[width=0.3\linewidth]{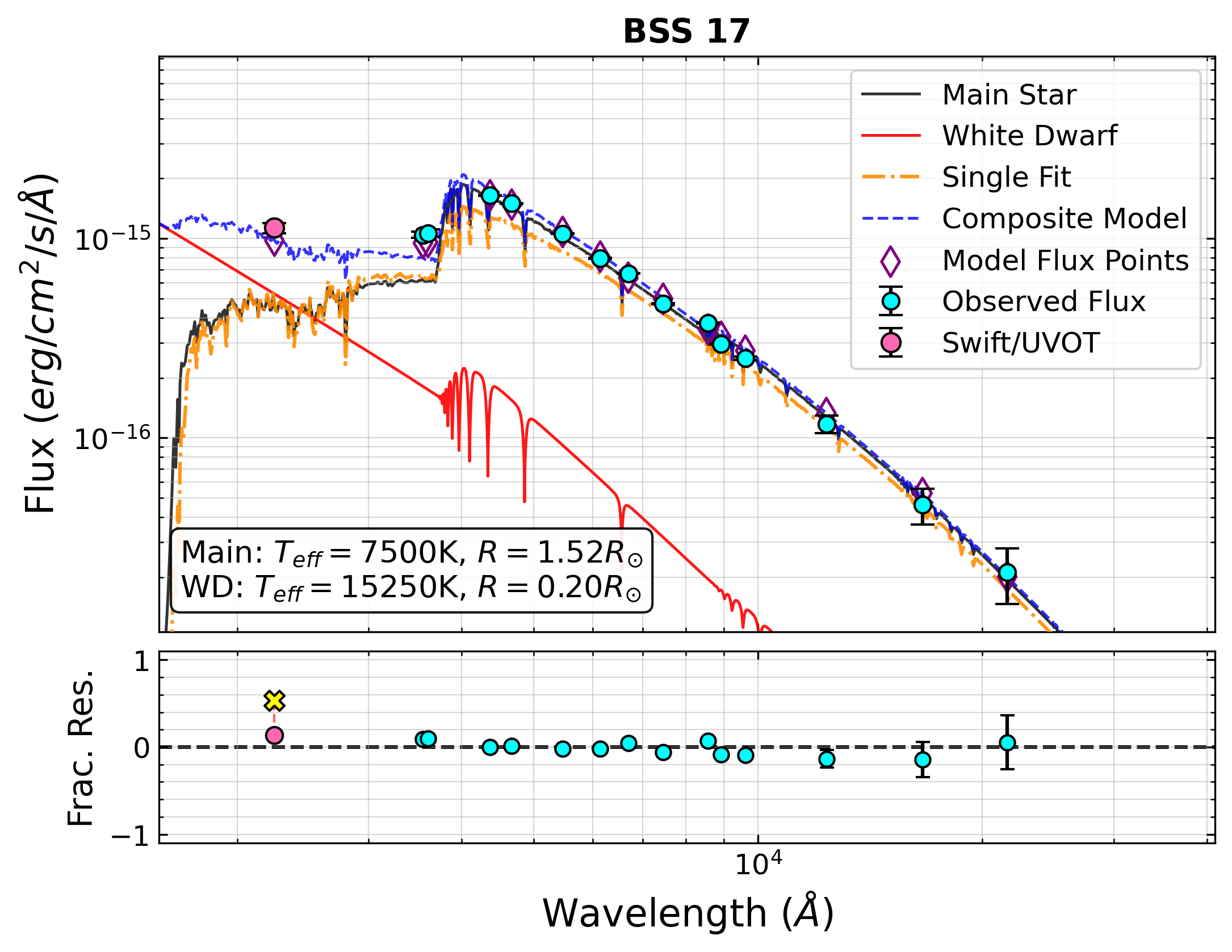}\\
    \includegraphics[width=0.3\linewidth]{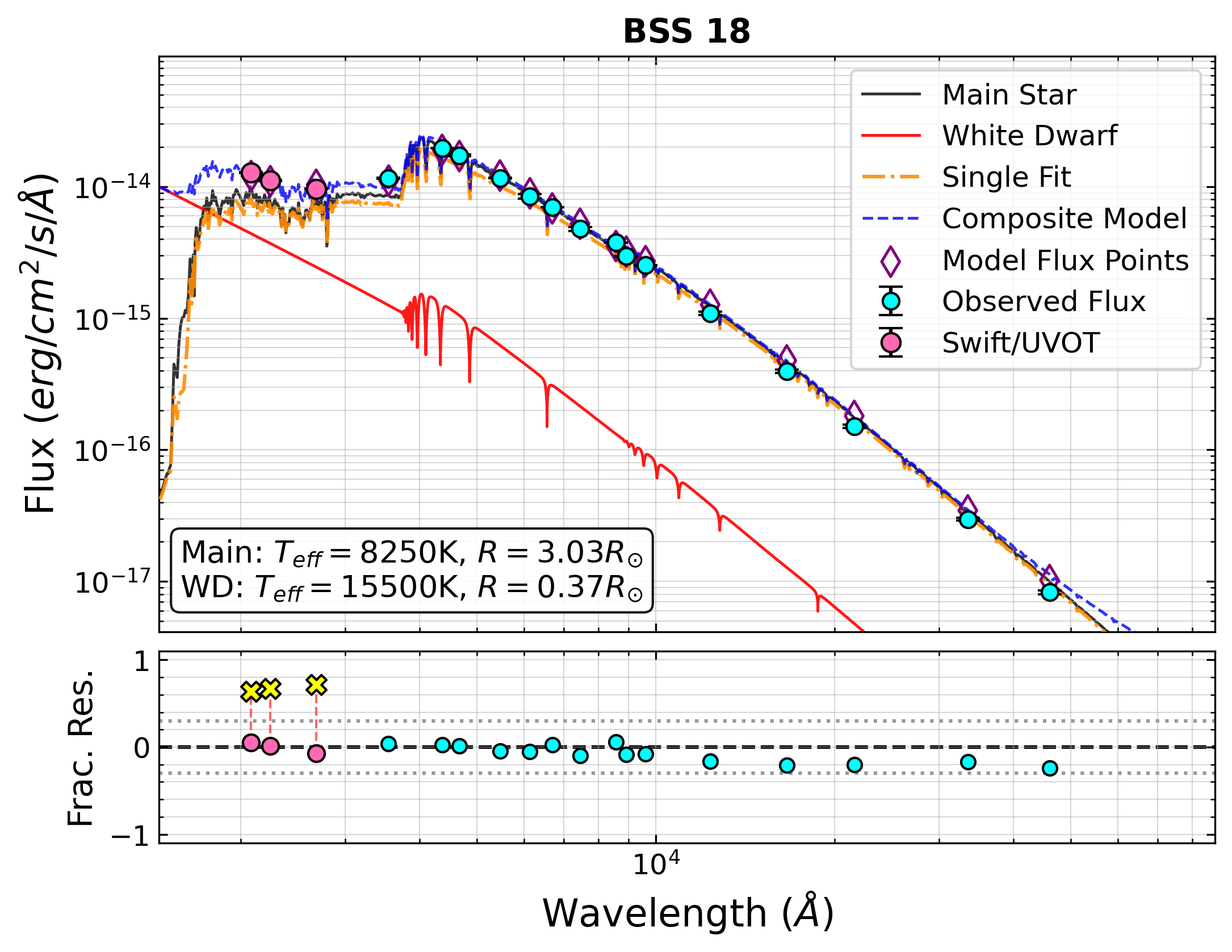}
    \includegraphics[width=0.3\linewidth]{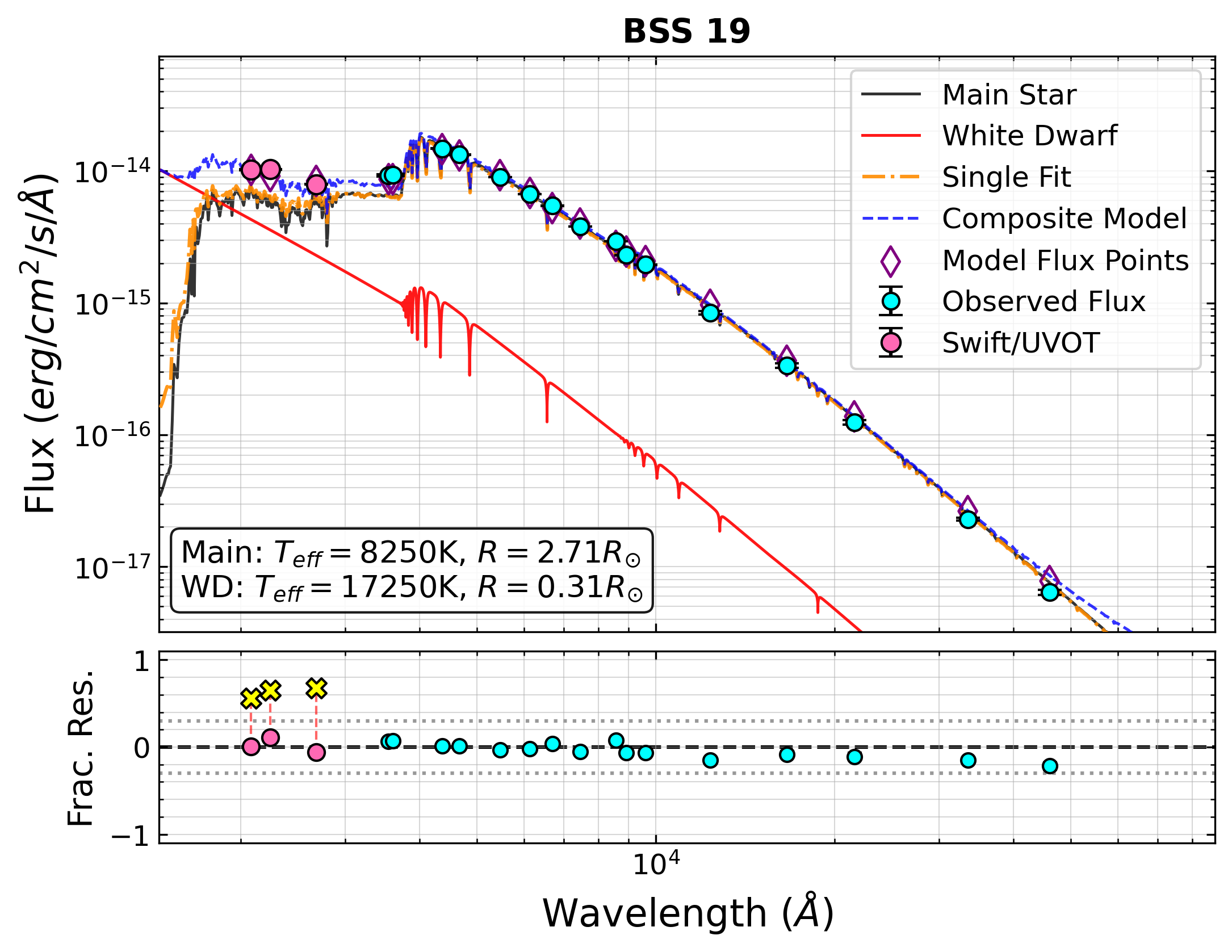}
    \includegraphics[width=0.3\linewidth]{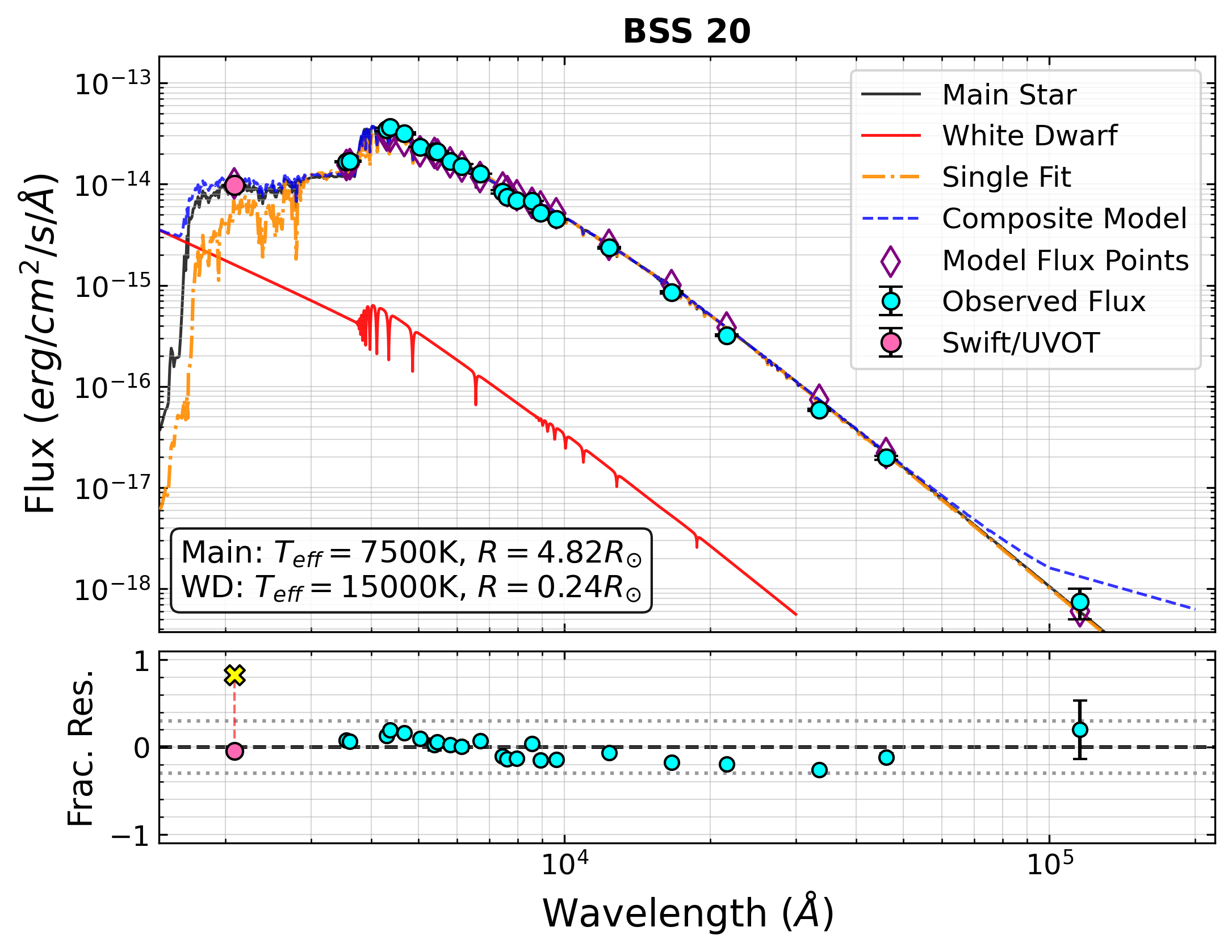}\\
    \includegraphics[width=0.3\linewidth]{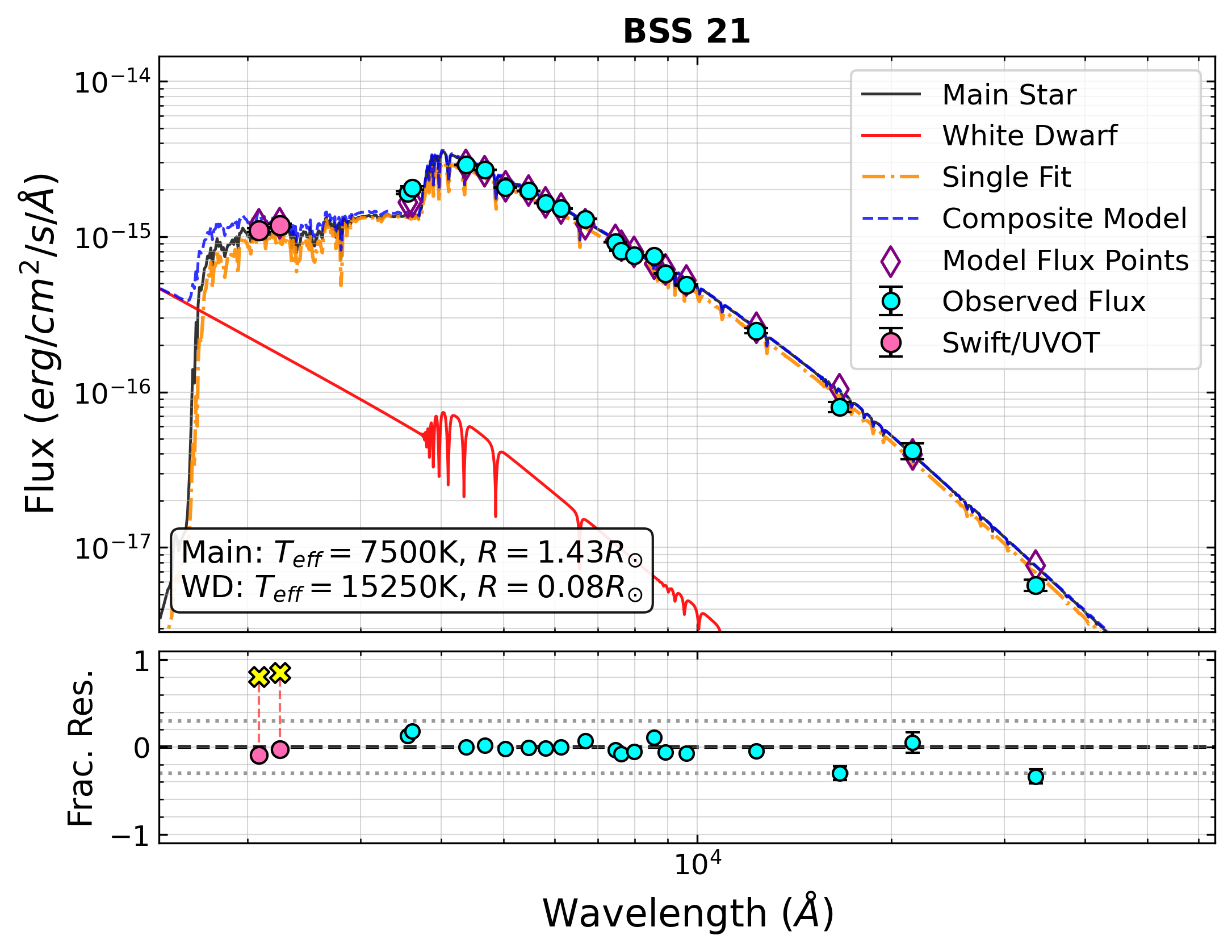}
    \includegraphics[width=0.3\linewidth]{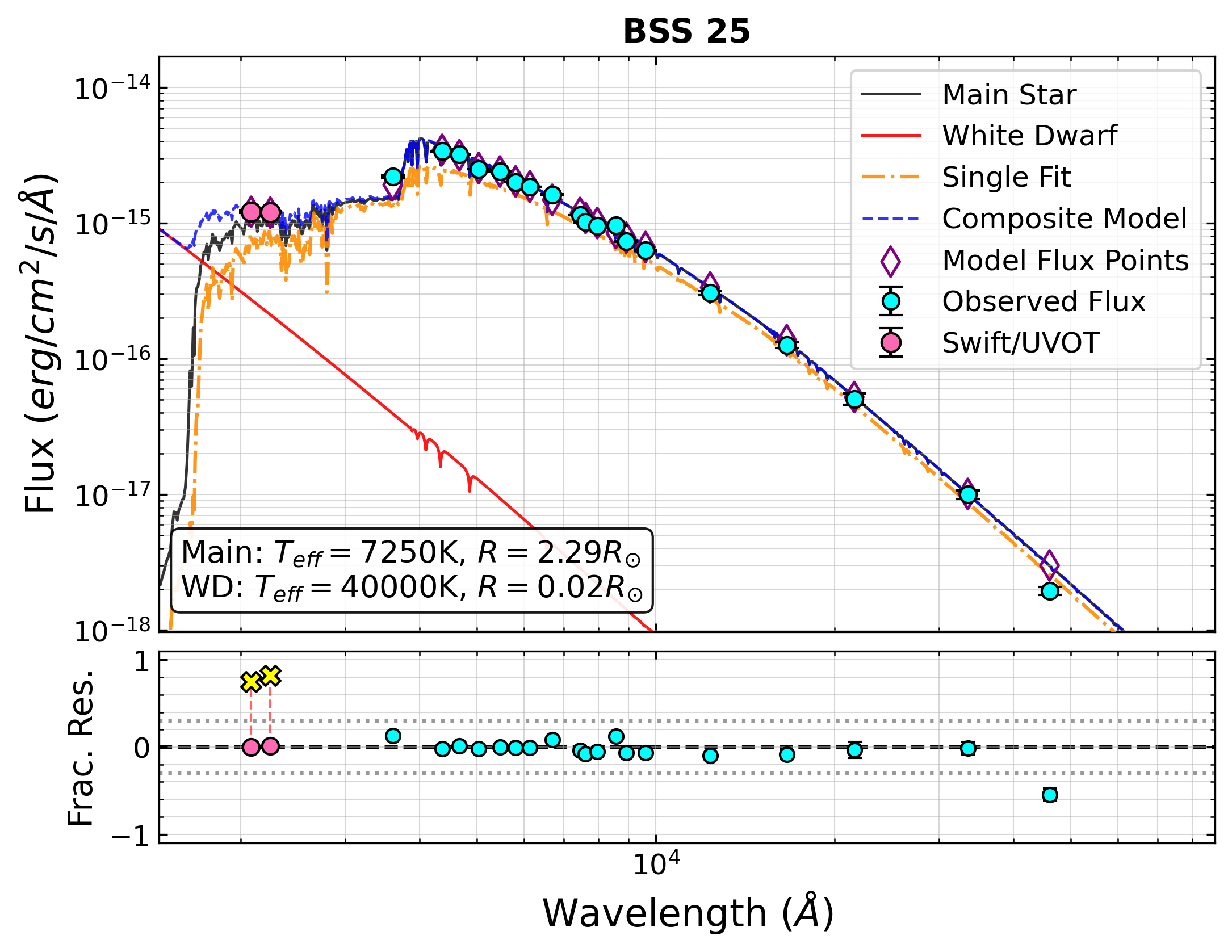}
    \includegraphics[width=0.3\linewidth]{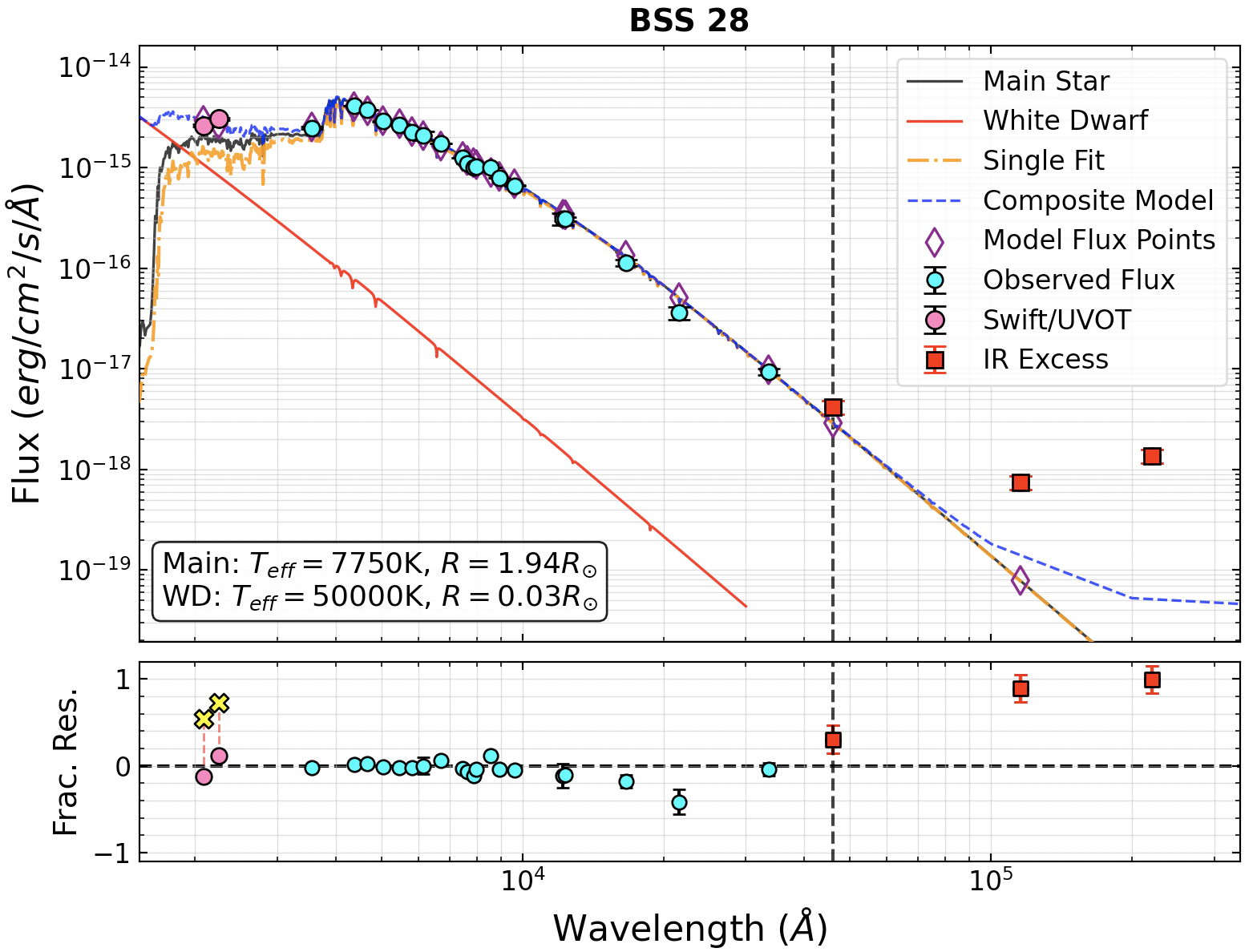}\\
    \includegraphics[width=0.3\linewidth]{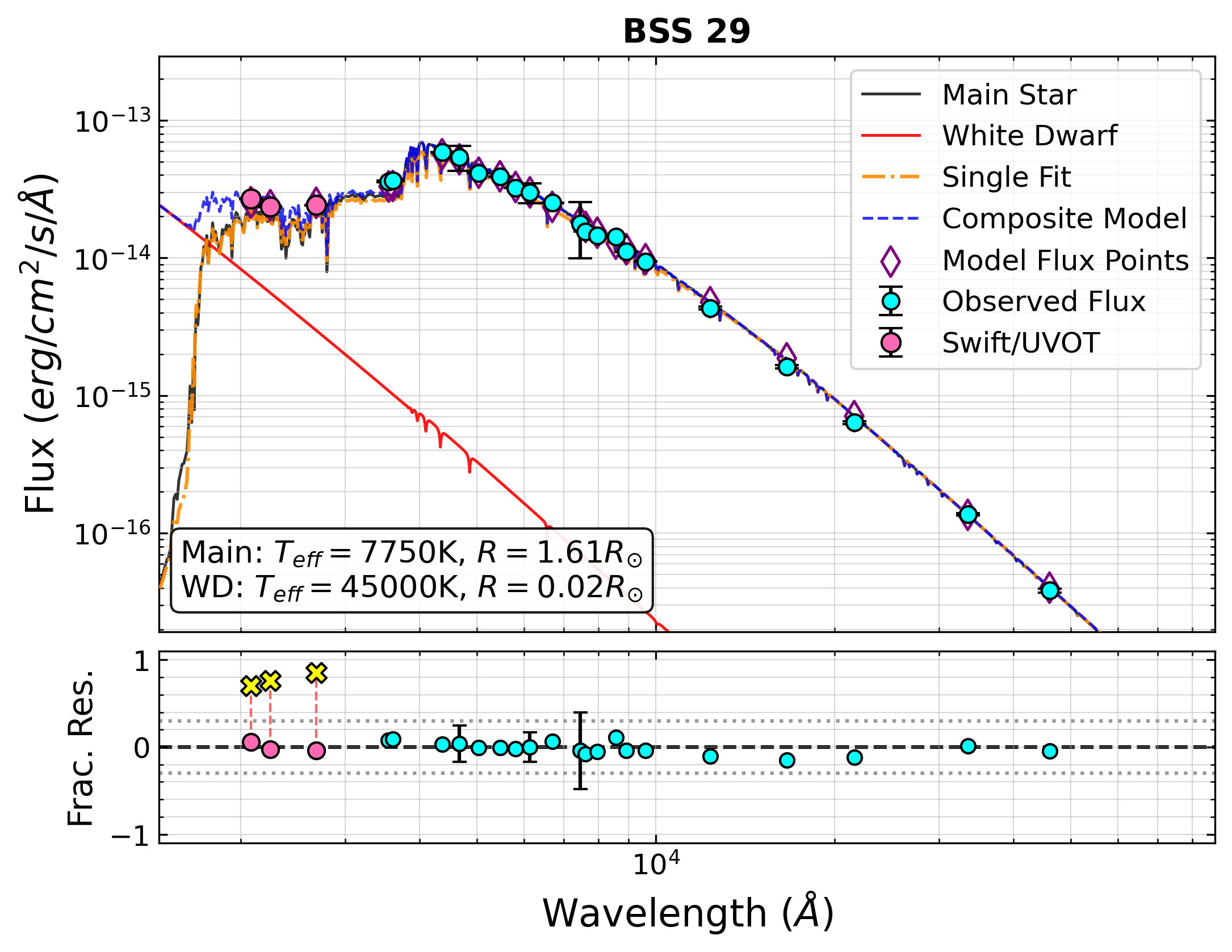}
    \includegraphics[width=0.3\linewidth]{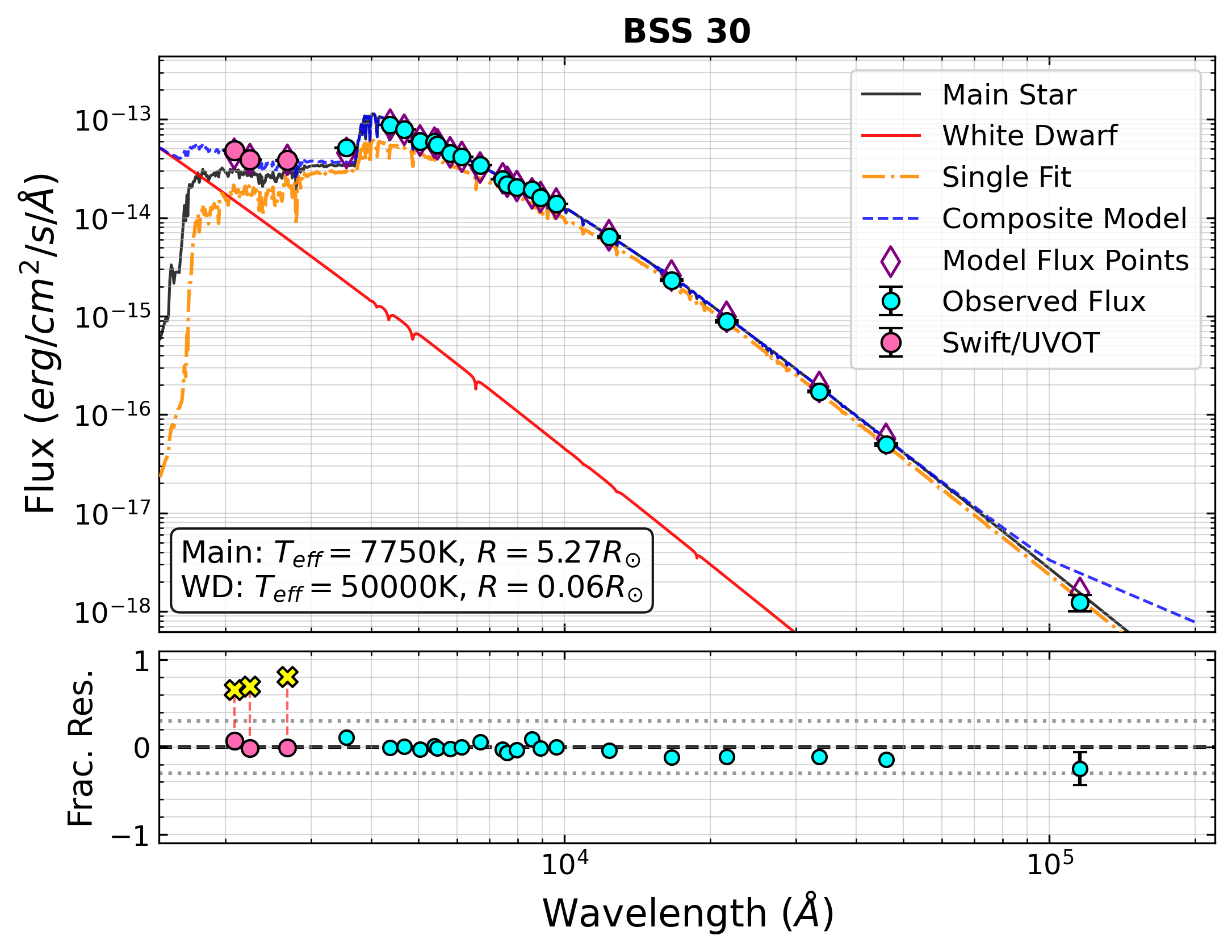}
    \includegraphics[width=0.3\linewidth]{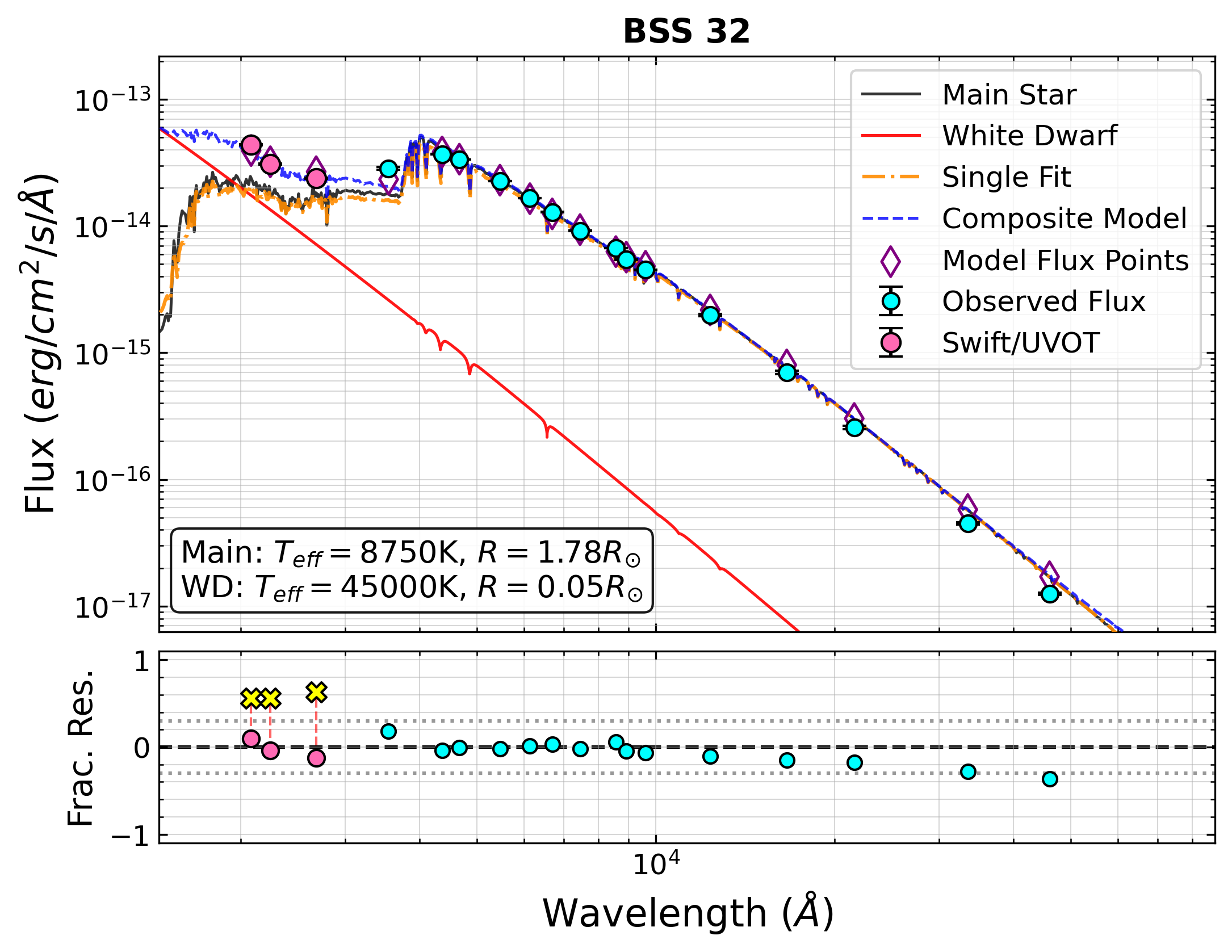}
        
    \caption{Two-component SED fitting analysis for the binary BSS candidates. \textbf{Top panels:} Cyan circles with error bars represent the observed multi-wavelength photometric fluxes. The solid black and red curves illustrate the best-fit model spectra of the cool BSS primary \citep{Castelli1997} and the hot WD secondary \citep{Koester2010}, respectively. The blue-dashed curve shows the total composite model spectrum (primary + secondary). The single-component fit is shown in orange. Purple diamonds denote the synthetic fluxes computed for the corresponding photometric passbands. \textbf{Bottom panels:} The fractional residuals between the observed fluxes and the best-fitting composite model. Yellow crosses show the \textit{Swift}/UVOT residuals relative to the single-star Kurucz model of the primary, clearly revealing the UV excess.}
    \label{fig:binary_seds}
\end{figure*}

We present the fundamental physical parameters ($T_{\rm eff}$, $R$, and $L$) derived for the 35 candidates in our sample. As outlined in Section~\ref{sec:section3.2}, we constructed comprehensive multi-wavelength SEDs by anchoring our dataset with \textit{Swift}/UVOT photometry. To rigorously constrain the stellar properties, observed fluxes were fitted against theoretical atmosphere models using the VOSA framework. We employed Kurucz ODFNEW/NOVER grids \citep{Castelli1997} to model the BSS primaries. For targets exhibiting significant UV excesses, a strong indicator of binarity, we adopted a two-component fitting strategy, combining the cool stellar model with hot WD spectra from \citet{Koester2010}. The final adopted parameters, along with uncertainties estimated via Monte Carlo simulations to account for photometric and distance errors, are listed in Table \ref{tab:sed_results}. Additionally, representative SEDs for four binary BSS systems are shown in Figure \ref{fig:binary_seds}.

\begin{table*}[ht]
\centering
\footnotesize
\caption{Results of the two-component SED fitting analysis, showing the derived parameters for the binary BSS candidates. The columns list the host cluster, Gaia DR3 source identifier, and equatorial coordinates ($\alpha$, $\delta$). For each binary component (designated A = primary, B = secondary companion), the derived fundamental parameters are provided: effective temperature ($T_{\rm eff}$), radius ($R$), and luminosity ($L$). Finally, the scaling factors, the number of photometric points used in the fit ($N_{\rm fit}$), reduced chi-square ($\chi^2_\nu$), the goodness-of-fit metric (vgf and vgf$_b$) are presented.} \label{tab:sed_results}
\renewcommand{\arraystretch}{1.1}
\setlength{\tabcolsep}{2pt}
\begin{tabular}{llccccccccccc}
\hline
BSS ID & Cluster & $\alpha$ & $\delta$ & Comp. & $T_{\rm eff}$ & $R$ & $L$ & Scaling Factor & $N_{\rm fit}$ & $\chi^{2}_\nu$ & vgf & vgf$_b$ \\
\hline

BSS 06 & Berkeley 31 & 06:57:37.44 & +08:17:16.8 & A & 8000$\pm$121 & 1.45$\pm$0.09 & 7.71$\pm$0.79 & 3.69E-23 & 20 & 4.35 & 9.46 & 0.65 \\
       &              &             &             & B & 15500$\pm$125 & 0.20$\pm$0.01 & 2.08$\pm$0.21 & 7.08E-25 &    &      &      &      \\

BSS 07 & Berkeley 31 & 06:57:40.56 & +08:17:45.6 & A & 8000$\pm$150 & 1.60$\pm$0.10 & 9.47$\pm$0.97 & 3.49E-23 & 20 & 3.09 & 7.56 & 1.18 \\
       &              &             &             & B & 15250$\pm$125 & 0.12$\pm$0.01 & 0.72$\pm$0.07 & 2.01E-25 &    &      &      &      \\

BSS 10 & Berkeley 31 & 06:57:36.24 & +08:17:24.0 & A & 7000$\pm$177 & 1.37$\pm$0.09 & 4.06$\pm$0.42 & 7.63E-23 & 18 & 1.74 & 9.27 & 2.51 \\
       &              &             &             & B & 15250$\pm$125 & 0.07$\pm$0.01 & 0.22$\pm$0.02 & 1.81E-25 &    &      &      &      \\

BSS 14 & Berkeley 75 & 06:49:00.24 & -23:59:52.8 & A & 8250$\pm$221 & 1.31$\pm$0.08 & 7.13$\pm$0.73 & 3.29E-23 & 14 & 5.51 & 6.8 & 0.77 \\
       &              &             &             & B & 45000$\pm$2500 & 0.03$\pm$0.01 & 3.67$\pm$0.37 & 3.86E-26 &    &      &      &      \\

BSS 15 & Berkeley 75 & 06:49:00.72 & -24:00:50.4 & A & 10000$\pm$178 & 1.53$\pm$0.05 & 21.07$\pm$1.44 & 3.31E-23 & 18 & 6.98 & 14.21 & 0.99\\
       &              &             &             & B & 45000$\pm$2500 & 0.06$\pm$0.01 & 16.76$\pm$1.12 & 1.34E-25 &    &      &      &      \\

BSS 17 & Berkeley 75 & 06:49:00.00 & -23:59:56.4 & A & 7500$\pm$148 & 1.52$\pm$0.09 & 6.58$\pm$0.68 & 3.74E-23 & 15 & 7.09 & 21.08 & 2.09 \\
       &              &             &             & B & 15250$\pm$125 & 0.20$\pm$0.01 & 2.01$\pm$0.21 & 6.79E-25 &    &      &      &      \\

BSS 18 & NGC 2192 & 06:15:19.94 & +39:48:56.9 & A & 8250$\pm$204 & 3.03$\pm$0.18 & 38.44$\pm$3.93 & 3.01E-22 & 19 & 15.25 & 34.27 & 2.24 \\
       &           &             &             & B & 15500$\pm$125 & 0.37$\pm$0.02 & 7.22$\pm$0.73 & 4.50E-24 &    &      &      &      \\

BSS 19 & NGC 2192 & 06:15:30.07 & +39:52:57.4 & A & 8250$\pm$185 & 2.71$\pm$0.16 & 30.58$\pm$3.08 & 2.29E-22 & 19 & 18.09 & 37.41 & 2.01 \\
       &           &             &             & B & 17250$\pm$125 & 0.31$\pm$0.02 & 7.80$\pm$0.78 & 3.07E-24 &    &      &      &      \\

BSS 20 & NGC 2204 & 06:15:52.32 & -18:45:21.6 & A & 7500$\pm$125 & 4.82$\pm$0.41 & 66.30$\pm$10.41 & 7.36E-22 & 24 & 19.12 & 18.98 & 3.42 \\
       &           &             &             & B & 15000$\pm$125 & 0.24$\pm$0.02 & 2.71$\pm$0.42 & 1.93E-24 &    &      &      &      \\

BSS 21 & NGC 2204 & 06:15:30.24 & -18:38:45.6 & A & 7500$\pm$123 & 1.43$\pm$0.09 & 5.86$\pm$0.60 & 7.56E-23 & 23 & 8.28 & 16.27 & 3.36 \\
       &           &             &             & B & 15250$\pm$125 & 0.08$\pm$0.01 & 0.30$\pm$0.03 & 2.25E-25 &    &      &      &      \\

BSS 25 & NGC 2204 & 06:15:25.68 & -18:38:38.4 & A & 7250$\pm$135 & 2.29$\pm$0.14 & 13.01$\pm$1.33 & 1.04E-22 & 21 & 14.9 & 14.01 & 2.34 \\
       &           &             &             & B & 40000$\pm$1750 & 0.02$\pm$0.01 & 1.08$\pm$0.11 & 1.44E-26 &    &      &      &      \\

BSS 28 & NGC 2204 & 06:15:54.72 & -18:50:45.6 & A & 7750$\pm$125 & 1.94$\pm$0.19 & 12.29$\pm$2.23 & 9.37E-23 & 25 & 6.00 & 19.05 & 1.58 \\
       &           &             &             & B & 50000$\pm$3750 & 0.03$\pm$0.01 & 3.64$\pm$0.65 & 4.17E-26 &    &      &      &      \\

BSS 29 & NGC 2360 & 07:17:48.00 & -15:42:10.8 & A & 7750$\pm$160 & 1.61$\pm$0.10 & 8.38$\pm$0.93 & 1.28E-21 & 23 & 26.25 & 22.01 & 1.02 \\
       &           &             &             & B & 45000$\pm$2500 & 0.02$\pm$0.01 & 1.21$\pm$0.12 & 3.20E-25 &    &      &      &      \\

BSS 30 & NGC 2533 & 08:07:01.44 & -29:57:10.8 &  A & 7750$\pm$150 & 5.27$\pm$0.32 & 90.28$\pm$9.09 & 5.78E-25 & 26 & 27.53 & 18.74 & 1.09\\
       &              &             &             & B & 50000$\pm$3750 & 0.06$\pm$0.01 & 18.13$\pm$1.81 & 1.83E-21 &    &      &      &      \\

BSS 32 & NGC 6939 & 20:31:54.24 & +60:40:22.8 &  A & 8750$\pm$125 & 1.78$\pm$0.10 & 16.77$\pm$1.67 & 4.83E-22 & 19 & 14.22 & 17.05 & 3.73\\
       &              &             &             & B & 45000$\pm$2500 & 0.06$\pm$0.01 & 8.93$\pm$0.89 & 7.68E-25 &    &      &      &      \\ \hline
\end{tabular}%
\end{table*}

Our methodology aligns with and expands upon recent studies by \citet{Sheikh2024} and \citet{Zeng2025}, confirming the efficacy of combining Gaia DR3 astrometry with broad-band photometry for characterizing BSS populations. Despite differences in cluster environments, ranging from the old, metal-poor NGC~2243 to the younger NGC~6134, both studies independently concluded that binary mass transfer is a dominant formation pathway. While \citet{Sheikh2024} successfully identified post-mass-transfer systems hosting hot WD companions, \citet{Zeng2025} detected active mass-transfer phases (EA-type binaries). Building on this robust multi-wavelength framework, our analysis extends to a diverse sample of nine OCs, leveraging the sensitivity of \textit{Swift}/UVOT to unveil hot companions that remain elusive in optical bands.

Our analysis of the 35 BSS targets identifies 15 systems exhibiting ultraviolet excess. For the remaining 20 systems, the observed SEDs are well reproduced by single-component models; their best-fit parameters and corresponding SEDs are presented in Appendix~\ref{AppendixA}. These BSSs span a broad range of effective temperatures ($T_{\rm eff} \approx 7000$-12,750~K), with radii of $R \approx 0.9$-4.5 $R_{\odot}$ and luminosities between $L \approx 2$ and 195 $L_{\odot}$. Most of the effective temperature values are broadly consistent with those reported for BSSs in similar intermediate-age OCs, typically covering $6500$--$10{,}250$~K \citep{ 2022MNRAS.511.2274V, 2024MNRAS.52710335P}. The higher end of our range ($T_{\rm eff} \sim 12{,}000$--$12{,}750$~K) is also in line with what is expected for more massive BSSs, as found in younger or less evolved clusters where the MSTO occurs at higher temperatures \citep{Sindhu2019, 2021MNRAS.507.2373P}. Such temperatures can be produced through efficient mass transfer or merger processes, which place these objects above the cluster turn-off on the H-R diagram \citep{Bailyn1995, Ferraro2009F}. The generally low goodness-of-fit values (vgf$_{\rm b} \approx 0.45$-3.42) indicate excellent agreement between the models and the observed photometry. The properties of the confirmed binary systems and their evolutionary implications are discussed in detail in the following section.

\subsection{Evolutionary Status of BSSs}
\label{sec:evolutionary_status}

The physical properties inferred from the SED fitting offer valuable clues to the evolutionary status of the BSS population. For systems currently consistent with single-component SEDs, the inferred properties suggest that the more luminous objects may correspond to BSSs at relatively advanced evolutionary stages, possibly evolving toward the subgiant branch, as expected from post-MS evolution in this temperature range \citep[e.g.,][]{Bressan2012, Chen2015, Marigo2017}, whereas the cooler and less luminous targets are compatible with BSSs near the MS turn-off. However, these interpretations should be treated with caution, as the presence of unresolved companions cannot be ruled out. In contrast, the most astrophysically informative results arise from the 15 binary systems identified across seven OCs. The detected hot secondary components are classified into two distinct evolutionary groups based on their temperatures and radii: pre-ELM WDs (stripped-core candidates) and hot WDs.

\subsubsection{Pre-ELM WD Candidates} 

This group comprises a subset of systems in which the hot companions exhibit moderate effective temperatures ($T_{\rm eff}\simeq 15{,}000$--$17{,}250$~K) and relatively large radii ($R\simeq 0.12$--$0.37\,R_{\odot}$). Such radii are substantially larger than those expected for hot WDs ($R\sim0.01\,R_{\odot}$) and instead are consistent with the predicted properties of proto-WDs or pre-ELM WDs. These objects are interpreted as stripped stellar cores that are still undergoing gravitational contraction and have not yet entered the final WD cooling sequence. In total, six systems in our sample fall into this category (BSS 06, 07, 17, 18, 19, and 20), located in four OCs: Berkeley 31, Berkeley 75, NGC 2192, and NGC 2204 (see Table~\ref{tab:evo_track}).

\begin{figure}[ht]
        \centering
        \includegraphics[width=1\linewidth]{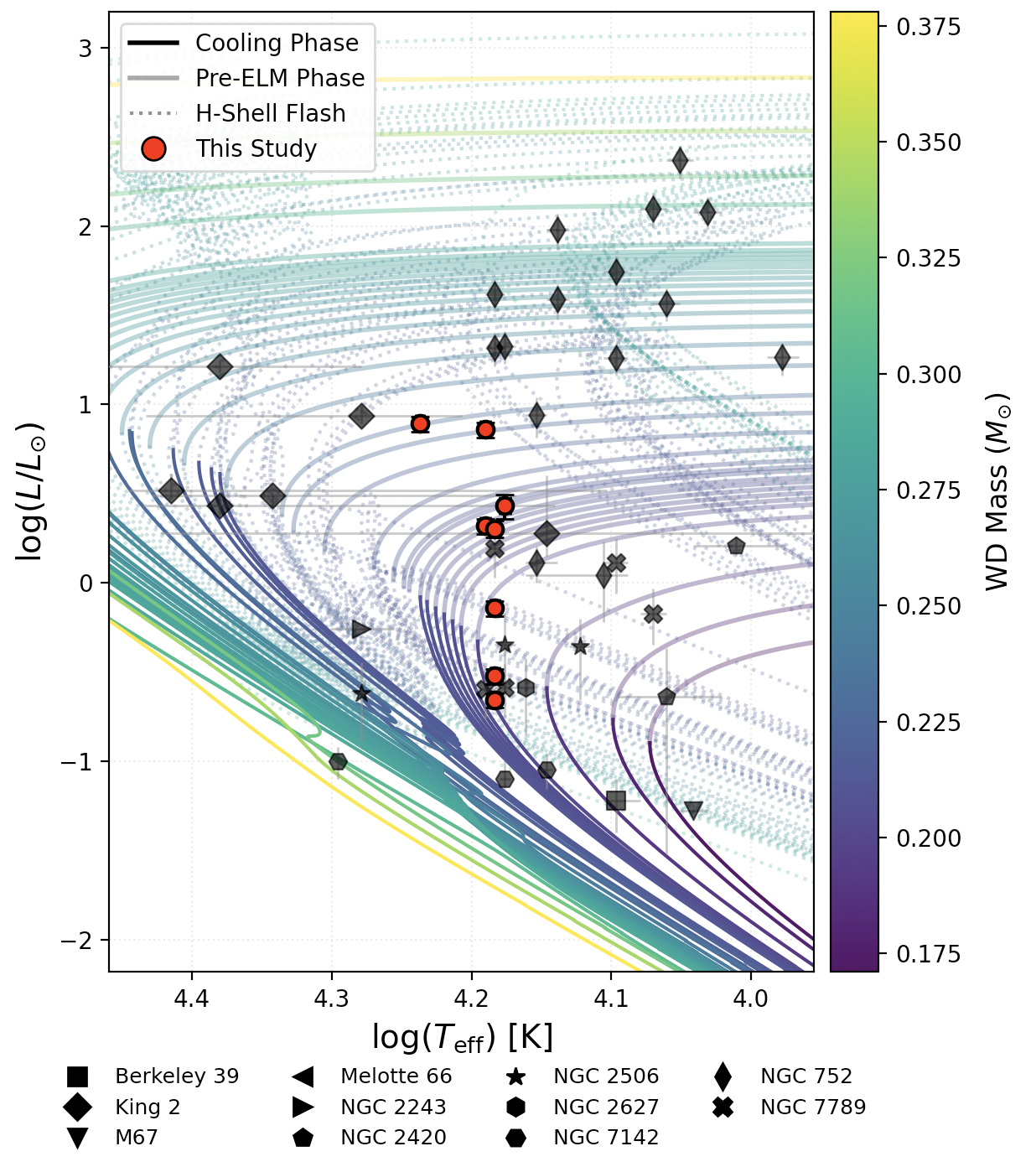}
        \caption{The $\log(L/L_{\odot})$ versus $\log(T_{\text{eff}})$ diagram showing the evolutionary status of the pre-ELM candidates derived from SED analysis. Red circles indicate the candidates from this study, labeled with the last six digits of their Gaia DR3 IDs. Triangles represent comparison objects from various OCs found in the literature. Theoretical evolutionary tracks are taken from \citet{Istrate2016}. The tracks are color-coded by stellar mass and segmented into different evolutionary phases: pre-ELM phase (dashed lines), H-shell flash episodes (dotted lines), and the final cooling phase (solid lines).}
	\label{fig:hr_diagram_elm}
\end{figure}

To assess their evolutionary status, we placed these companions on the Hertzsprung-Russell (H-R) diagram together with low-mass helium-core evolutionary tracks from \citet{Istrate2016} computed for solar metallicity, and shown in Figure~\ref{fig:hr_diagram_elm}. A complementary compilation of previously studied hot component stars, used as a comparison sample in the H-R diagram analysis, is provided in Appendix~\ref{AppendixB}. Most of the objects lie close to tracks corresponding to masses of approximately $M \sim 0.21$--$0.26\,M_{\odot}$, consistent with He-core WD evolutionary models \citep{Althaus2013}. Their locations coincide with the pre-ELM contraction phase (thick gray segments in Figure~\ref{fig:hr_diagram_elm}), indicating that these objects have not yet reached the temperatures associated with CNO-driven hydrogen-shell flashes nor settled onto the final WD cooling sequence. This strongly suggests that the companions are stripped stellar cores currently undergoing gravitational contraction following recent mass-transfer events \citep{Istrate2016}.

Following the methodology of \citet{Dattatrey2023}, we interpolated the evolutionary tracks of \citet{Istrate2016} to assess the evolutionary status of the stripped-core candidates. The reference point is the maximum radius, corresponding to the onset of the pre-ELM contraction phase. Within this framework, most systems occupy similar positions along the contraction sequence, indicating that they are in comparable pre-ELM evolutionary stages \citep{Maxted2013, El-Badry2021}. Two systems (BSS 10 and BSS 21) are found beyond this phase, lying on the early WD cooling track. In this sense, they can be interpreted as slightly more evolved counterparts of the pre-ELM objects, representing the immediate post–pre-ELM stage within the same evolutionary sequence, rather than a distinct companion class. The relatively large radius inferred for BSS 18 (0.37 $R_\odot$) remains consistent with the expected range ($\sim 0.1$–$0.6\,R_\odot$) for proto–ELM objects during the early contraction phase \citep{Istrate2016}.

The host OCs in our sample span a broad range of intermediate ages, extending from a few hundred Myr to nearly 2~Gyr. At these ages, the corresponding MSTO masses are typically of order $\sim 2$--$2.5\,M_{\odot}$, significantly higher than those characteristic of old GCs. This implies that the progenitors of the identified pre-ELM candidates were intermediate-mass stars. Consequently, our results indicate that in intermediate-age OC environments, binary evolution involving relatively massive progenitors can lead to the formation of low-mass helium-core pre-WD dwarfs. These evolutionary pathways are likely distinct from those commonly inferred in GC populations, where the progenitor masses are substantially lower.

\subsubsection{Hot WD Candidates} 

In contrast to the pre-ELM or stripped-core candidates, a distinct subset of systems in our sample hosts extremely hot and compact companions that are fully consistent with canonical WDs. These objects are characterized by very high effective temperatures ($T_{\rm eff} \gtrsim 40{,}000$~K and reaching up to $\sim 80{,}000$~K) and small radii ($R \sim 0.02$--$0.03\,R_{\odot}$), with surface gravities omitted since they cannot be robustly constrained from SED analysis. Such parameters place them firmly on the WD cooling sequence and clearly separate them from the pre-ELM population discussed above.

The adopted radius range for the hot WD companions is consistent with theoretical mass--radius relations for carbon--oxygen (CO) and oxygen--neon (ONe) core WDs in the mass range $0.4 \lesssim M/M_{\odot} \lesssim 1.3$ \citep{Nauenberg1972, Althaus2010}. At these masses, degenerate equation-of-state models predict radii of approximately $0.008$--$0.018\,R_{\odot}$ for the most massive WDs and up to $\sim 0.04\,R_{\odot}$ for lower-mass CO-core remnants \citep{Koester2010}. The inferred radii in our sample fall within this theoretically expected range and are furthermore consistent with values reported for hot WD companions in similar SED-based studies of OC BSSs \citep[e.g.,][]{2021MNRAS.507.2373P,Rao2022, 2022cosp...44.2216V, Sheikh2024, 2026arXiv260412494B}.

Seven systems in our sample fall into this category (BSS 14, 15, 25, 28, 29, 30, and 32; see  Table~\ref{tab:evo_track}). The hottest companions identified in our sample reach effective temperatures of $T_{\rm eff} \approx 50{,}000$~K. These properties are broadly consistent with theoretical mass-radius relations for hot WDs \citep[e.g.,][]{Althaus2010, Koester2010}, indicating fully degenerate remnants formed after core helium ignition in the progenitor.

The detection of such hot and compact remnants provides strong evidence for a post-mass-transfer evolutionary origin. Given their extreme temperatures, these WDs are expected to be very young objects on the cooling sequence, with cooling ages of at most a few Myr according to modern WD evolutionary models \citep{Bedard2020}. This implies that mass transfer in these systems must have ceased very recently on evolutionary timescales, and that the companions have only just settled onto the WD cooling track following envelope ejection, possibly via a common-envelope phase.

\begin{figure}
    \centering
    \includegraphics[width=1\linewidth]{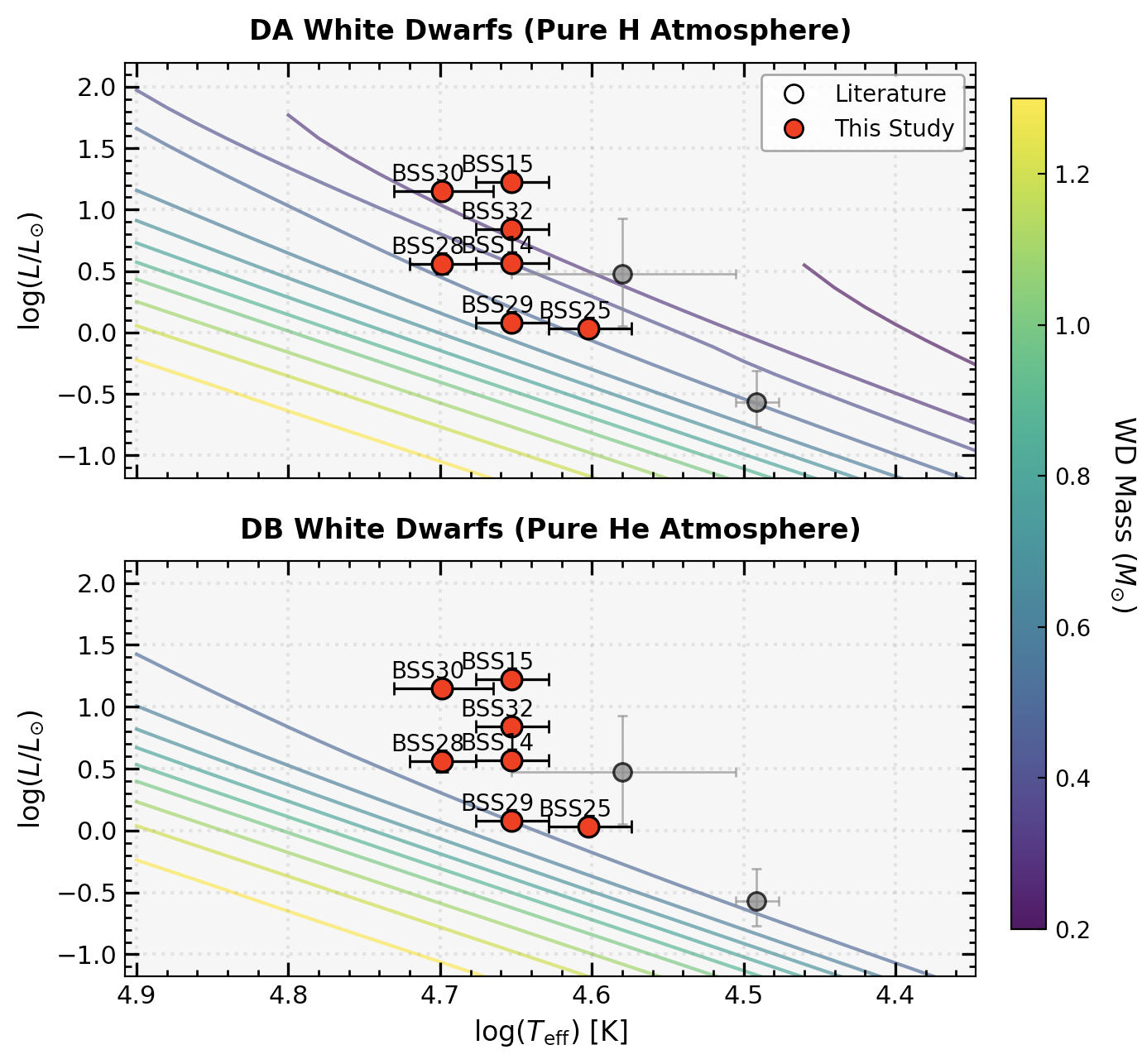}
    \caption{Comparison of theoretical cooling tracks with the observational positions of the WD candidates in the H-R diagram. The upper and lower panels show models for DA and DB WDs, respectively. The solid colored curves represent evolutionary cooling tracks for WDs with masses in the range $0.2 \leq M/M_{\odot} \leq 1.3$. The color of each track corresponds to the WD mass, as indicated by the color bar on the right. The evolutionary models are adopted from \citet{2025AN....34640118C}. Red circles mark the hot WD candidates analyzed in this work, while gray circles show comparison systems from the literature in Melotte~66 and NGC~6940. Error bars indicate the uncertainties in the derived effective temperature and luminosity.}
    \label{fig:CO_WD_mass}
\end{figure}

Unlike the pre-ELM candidates, which represent stars that have undergone interrupted stellar evolution before core helium ignition, these hot, luminous companions follow the standard cooling sequences of fully evolved WDs. By interpolating their positions across the recent DA (pure hydrogen atmosphere) and DB (pure helium atmosphere) model grids \citep[e.g.,][]{2025AN....34640118C}, we systematically derived their current masses, cooling ages, and corresponding progenitor masses. Their presence in blue straggler systems underscores the remarkable diversity of mass-transfer pathways operating in OCs, producing a wide spectrum of stellar remnants that ranges from stripped, low-mass helium cores to massive CO and ONe WDs.

\begin{table*}
\centering
\footnotesize
\caption{Properties of BSS systems with masses and radii estimated from evolutionary-track interpolation. The cool component corresponds to the BSS, while the hot component represents the WD companion. For hot WD systems, values are given as H/He. Missing values are indicated with `--'.}
\label{tab:evo_track}
\begin{tabular}{l c ll ll}
\hline
 &  Evolutionary & \multicolumn{2}{c}{Cool Component (A)} & \multicolumn{2}{c}{Hot Component (B)} \\
\cline{3-6}
BSS ID & Stage & Mass ($M_\odot$) & Radius ($R_\odot$) & Mass ($M_\odot$) & Radius ($R_\odot$) \\
\hline

BSS 06 & 1 & 1.35$\pm$0.02 & 1.45$\pm$0.09 & 0.263$\pm$0.002 & 0.20$\pm$0.01 \\
BSS 07 & 1 & 1.36$\pm$0.03 & 1.60$\pm$0.10 & 0.260$\pm$0.008 & 0.12$\pm$0.01 \\
BSS 10 & 2 & 1.17$\pm$0.04 & 1.37$\pm$0.10 & 0.256$\pm$0.030 & 0.07$\pm$0.01 \\
BSS 14 & 3 & 1.37$\pm$0.03 & 1.31$\pm$0.10 & 0.399$\pm$0.067/-- & 0.03$\pm$0.01/-- \\
BSS 15 & 3 & 1.51$\pm$0.01 & 1.53$\pm$0.08 & 0.258$\pm$0.012/-- & 0.07$\pm$0.01/-- \\
BSS 17 & 1 & 1.28$\pm$0.03 & 1.52$\pm$0.10 & 0.210$\pm$0.001 & 0.20$\pm$0.01 \\
BSS 18 & 1 & 1.85$\pm$0.02 & 3.04$\pm$0.21 & 0.228$\pm$0.003 & 0.37$\pm$0.02 \\
BSS 19 & 1 & 1.83$\pm$0.02 & 2.71$\pm$0.18 & 0.233$\pm$0.001 & 0.31$\pm$0.02 \\
BSS 20 & 1 & 1.53$\pm$0.00 & 4.84$\pm$0.41 & 0.213$\pm$0.003 & 0.24$\pm$0.02 \\
BSS 21 & 2 & 1.27$\pm$0.03 & 1.44$\pm$0.09 & 0.216$\pm$0.008 & 0.08$\pm$0.01 \\
BSS 25 & 3 & 1.52$\pm$0.01 & 2.29$\pm$0.15 & 0.474$\pm$0.044/0.500$\pm$0.010 & 0.02$\pm$0.01/0.02$\pm$0.01 \\
BSS 28 & 3 & 1.45$\pm$0.03 & 1.95$\pm$0.19 & 0.469$\pm$0.041/-- & 0.02$\pm$0.01/-- \\
BSS 29 & 3 & 1.32$\pm$0.03 & 1.61$\pm$0.11 & 0.543$\pm$0.053/0.500$\pm$0.049 & 0.02$\pm$0.01/0.02$\pm$0.01 \\
BSS 30 & 3 & 2.14$\pm$0.03 & 5.28$\pm$0.33 & 0.288$\pm$0.038/-- & 0.05$\pm$0.01/-- \\
BSS 32 & 3 & 1.60$\pm$0.06 & 1.78$\pm$0.10 & 0.292$\pm$0.035/-- & 0.04$\pm$0.01/-- \\
\hline
\hline
\end{tabular}
\noindent \\
(1) pre-ELM, (2) cooling, (3) hot WD.
\end{table*}

\subsection{Dynamical State of Host Clusters and Implications for BSS Formation}

The presence and distribution of BSSs within clusters provide important insights into their dynamical evolution. In low-density OCs, stellar collisions are expected to be rare, and binary evolution is the dominant mechanism responsible for BSS formation \citep{Mathieu2009, 2011MNRAS.415.3771L}. The identification of numerous BSS systems hosting hot degenerate companions in our sample strongly supports this scenario. The ages of the studied clusters range from approximately 0.7 to 2 Gyr, corresponding to MSTO masses of $\sim1.5$–$2.5\,M_\odot$. At these ages, binary mass transfer involving intermediate-mass progenitors is expected to efficiently produce BSSs and low-mass WD companions. The presence of both pre-ELM and fully formed WDs in our sample indicates that binary evolution is ongoing and spans multiple evolutionary stages. 

\subsubsection{Radial Distribution and Mass Segregation}\label{sec:radial}

We analyzed the radial distributions of the BSS candidates relative to other cluster members to assess their dynamical states. Because BSSs are generally more massive than mean cluster members, they are expected to undergo mass segregation and migrate toward the cluster core via dynamical friction over time. We calculated the projected distance of each star from the cluster center using Gaia DR3 coordinates \citep{GaiaCollaboration2023}. For comparison, cluster members were divided into two reference populations using a cluster-specific magnitude limit ($G_{\rm lim}$) to separate evolved turn-off and giant stars from the lower MS.

\begin{figure*}[ht]
    \centering
    \includegraphics[width=0.9\linewidth]{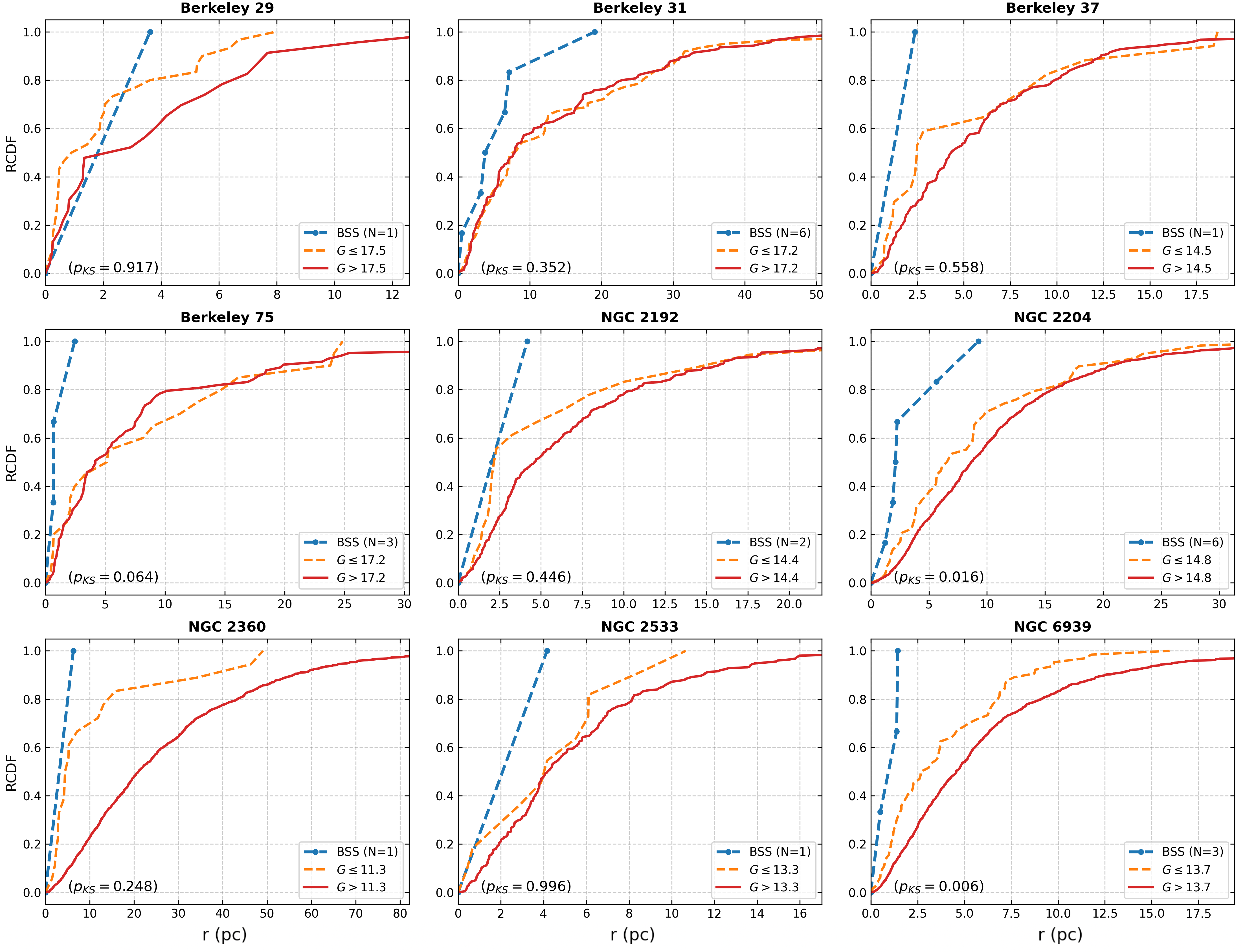}
    \caption{RCDFs of BSSs (blue dashed lines), evolved stars (orange dashed lines), and MS stars (red solid lines) for each cluster. The projected distance from the cluster center is shown on the horizontal axis. The MS and evolved samples are defined relative to the MSTO lines indicated in the corresponding CMDs. The number of BSSs in each cluster is indicated in the legend. The KS test probabilities ($p_{\rm KS}$) comparing the radial distributions of BSSs and the reference populations are reported in each panel.}
    \label{fig:RCDFs}
\end{figure*}

Figure~\ref{fig:RCDFs} shows the resulting radial cumulative distribution functions (RCDFs), comparing BSS candidates to the bright ($G \le G_{\rm lim}$) and faint ($G > G_{\rm lim}$) reference populations. We performed a two-sample Kolmogorov-Smirnov (KS) test to compare the radial distributions of the BSS candidates and the unsegregated faint population. The resulting $p$-values ($p_{\rm KS}$) quantify the probability that both samples are drawn from the same parent distribution; $p_{\rm KS} < 0.05$ indicates statistically significant mass segregation.

The RCDFs reflect the expected mass hierarchy within the clusters. Evolved stars (giants; orange dashed lines) are more centrally concentrated than the lower-mass MS stars (dwarfs; red solid lines), which dominate the cluster outskirts (typically at $r>10$~pc). The BSS and evolved RCDFs rise steeply and reach unity at smaller radii, demonstrating spatial stratification and confirming mass segregation among the more massive stellar populations.

In clusters such as NGC~2204 and NGC~6939, BSS candidates exhibit a strong central concentration relative to the faint population ($p_{\rm KS} = 0.016$ and $0.006$, respectively). This indicates that these clusters are dynamically relaxed and their BSS populations are mass-segregated \citep{Ferraro2012}. In OCs with few BSSs (e.g., Berkeley~29, $N=1$), statistical significance is inherently limited, though the candidates still preferentially reside in the inner regions. Overall, the BSS spatial distributions closely track those of the massive, bright members \citep{Lanzoni2016}.

In the low-density environments of OCs, BSS formation is primarily driven by mass transfer in binary systems, where the progenitor accretes mass from a donor that later evolves into a WD \citep{2011Natur.478..356G}. Our identification of BSS candidates with degenerate companions at various evolutionary stages (pre-ELM, cooling, and hot WDs; see Table~\ref{tab:evo_track}) supports this channel. 

The dynamical analysis highlights varying degrees of mass segregation among these binary systems. The highly segregated BSS population in NGC~6939 is consistent with massive binaries (BSS+WD) that have migrated to the core, aligning with the detection of a $\sim0.6\,M_\odot$ hot WD companion. NGC~2204 similarly shows clear segregation and hosts diverse companions. Conversely, the radial distributions of BSSs in NGC~2533 ($p_{\rm KS} = 0.996$) and NGC~2192 ($p_{\rm KS} = 0.446$) are statistically indistinguishable from the reference populations. This lack of segregation may reflect a younger dynamical age or a less dynamically evolved cluster environment. However, the presence of pre-ELM companions (e.g., BSSs~18 and 19 in NGC~2192) indicates that the mass-transfer mechanism operates actively even before significant dynamical relaxation occurs \citep{Gosnell2015, Bisht2026}.

To provide a more quantitative dynamical context for the KS test results, we estimated the relaxation time $t_{\rm relax}$ for each cluster following the standard approximation \citep{Binney2008}:
\begin{equation}
    t_{\rm relax} = \frac{0.138\,N^{1/2}\,R_{\rm h}^{3/2}}
    {G^{1/2}\,\bar{m}^{1/2}\,\ln(0.4\,N)},
    \label{eq:trelax}
\end{equation}
\noindent where $N$ is the number of cluster members, $R_{\rm h}$ is the  half-mass radius, $\bar{m}$ is the mean stellar mass, and $G$ is the  gravitational constant. The member counts, total masses, mean masses,  and half-mass radii were adopted from the catalog of \citet{Hunt2024}.  The ratio $t_{\rm age}/t_{\rm relax}$ provides an estimate of how many relaxation times a cluster has undergone since formation, and thus how dynamically evolved it is expected to be. Among our nine clusters, NGC~2204 and NGC~6939 stand out with the highest $t_{\rm age}/t_{\rm relax}$ ratios, and these are precisely the two clusters that show statistically significant mass segregation in the KS test ($p_{\rm KS} = 0.016$ and $0.006$, respectively). This correspondence is broadly consistent with the expectation that more dynamically evolved clusters have had more time to redistribute their more massive members toward the core through dynamical friction \citep{Ferraro2012, Lanzoni2016}.

In contrast, clusters such as NGC~2533 and NGC~2192 show no statistically significant segregation ($p_{\rm KS} = 0.996$ and $0.446$, respectively), despite having non-negligible relaxation timescales. This suggests that either the clusters have not yet completed enough relaxation cycles or that the small number of BSSs in these systems limits the statistical power of the KS test. We caution that the connection between $t_{\rm relax}$ and BSS segregation is not one-to-one in our sample; other factors, such as cluster richness, the initial binary fraction, and the mass of the BSS systems, likely also play a role. These results should therefore be interpreted as broadly indicative rather than definitive.

\subsection{Binary Fraction of Host Clusters}

To provide a simple comparison with our UV-excess BSS results, we estimated the photometric binary fraction for each of the nine clusters. We followed the CMD-based approach of \citet{Milone2012}, also used by \citet{Donada2023}. MS members were selected from the catalogues of \citet{Hunt2024}. The resulting $t_{\rm relax}$ values are listed in Table~\ref{tab:cluster_params}. For each cluster, we defined a single-star sequence by fitting a fourth-degree polynomial to the median color–magnitude relation. Stars lying more than 0.75~mag above this sequence were considered binary candidates (roughly $q \gtrsim 0.6$). The binary fraction was then computed as:
\begin{equation}
    f_{\rm b} = \frac{N_{\rm bin}}{N_{\rm MS}},
\end{equation}
where $N_{\rm bin}$ is the number of binary candidates and $N_{\rm MS}$ is the number of main-sequence stars. Uncertainties were estimated using 95\% Wilson confidence intervals. The resulting values listed in Table~\ref{tab:binaryfrac} range from $f_{\rm b} \approx 0.04$ to $0.15$. These are generally on the lower side compared to the broader sample of \citet{Donada2023}, which is expected since our estimates mainly trace binaries with relatively high mass ratios. Within our sample, NGC~6939, NGC~2533, and NGC~2204 show the highest fractions, while Berkeley~29 and Berkeley~37 have the lowest.

Table~\ref{tab:binaryfrac} shows a weak indication that clusters with higher binary fractions may host more binary BSS systems, although the trend is not clear. For example, Berkeley~31 and NGC~2204, which have the largest BSS populations in our sample, also contain the highest numbers of UV-excess systems. On the other hand, Berkeley~75 and NGC~2192 show relatively high fractions of binary BSSs among their small BSS samples, consistent with the expectations that binary-rich environments can favour BSS formation via mass transfer \citep{Mathieu2009, Knigge2009}.

NGC~6939, despite having the highest binary fraction, contains only one UV-excess BSS out of five. This may be related to its more advanced dynamical state (Section~\ref{sec:radial}), where the KS test ($p_{\rm KS} = 0.006$) indicates strong mass segregation. In such dynamically evolved clusters, massive systems are expected to concentrate toward the core, where crowding can make UV detections more difficult or where past interactions may have altered some binaries \citep{2011MNRAS.415.3771L, Ferraro2012}. Overall, these patterns should be interpreted with caution. The number of binary BSSs per cluster is small, and our photometric binary fractions mainly trace systems with $q \gtrsim 0.6$ \citep{Donada2023}. A larger and more homogeneous sample would be needed to clarify whether a real connection exists \citep{Childs2025, Mikhnevich2026a, Malhotra2026}.

\begin{table}
\centering
\caption{Photometric binary fractions estimated from the Gaia CMD for the nine host clusters. $N_{\rm BSS}$ is the total number of BSSs, $N_{\rm BSS,bin}$ is the number of UV-excess binary BSS systems, and the uncertainties represent 95\% Wilson binomial confidence intervals.}
\label{tab:binaryfrac}
\begin{tabular}{lccc}
\hline\hline
Cluster & $N_{\rm BSS}$ & $f_b$ & $N_{\rm BSS,bin}$ \\
\hline
Berkeley~31 & 6 & $0.09^{+0.06}_{-0.04}$ & 3 \\
Berkeley~29 & 1 & $0.04^{+0.17}_{-0.04}$ & - \\
Berkeley~37 & 1 & $0.04^{+0.05}_{-0.02}$ & - \\
Berkeley~75 & 3 & $0.08^{+0.08}_{-0.04}$ & 3 \\
NGC~2192 & 2 & $0.08^{+0.06}_{-0.04}$ & 2 \\
NGC~2204 & 6 & $0.12^{+0.03}_{-0.03}$ & 4 \\
NGC~2360 & 1 & $0.06^{+0.03}_{-0.02}$ & 1 \\
NGC~2533 & 1 & $0.12^{+0.06}_{-0.04}$ & 1 \\
NGC~6939 & 3 & $0.15^{+0.03}_{-0.03}$ & 1 \\
\hline
\end{tabular}
\end{table}

\subsection{Dynamical Environment of the Clusters}

\begin{figure}[h]
\centering
    \includegraphics[width=0.8\linewidth]{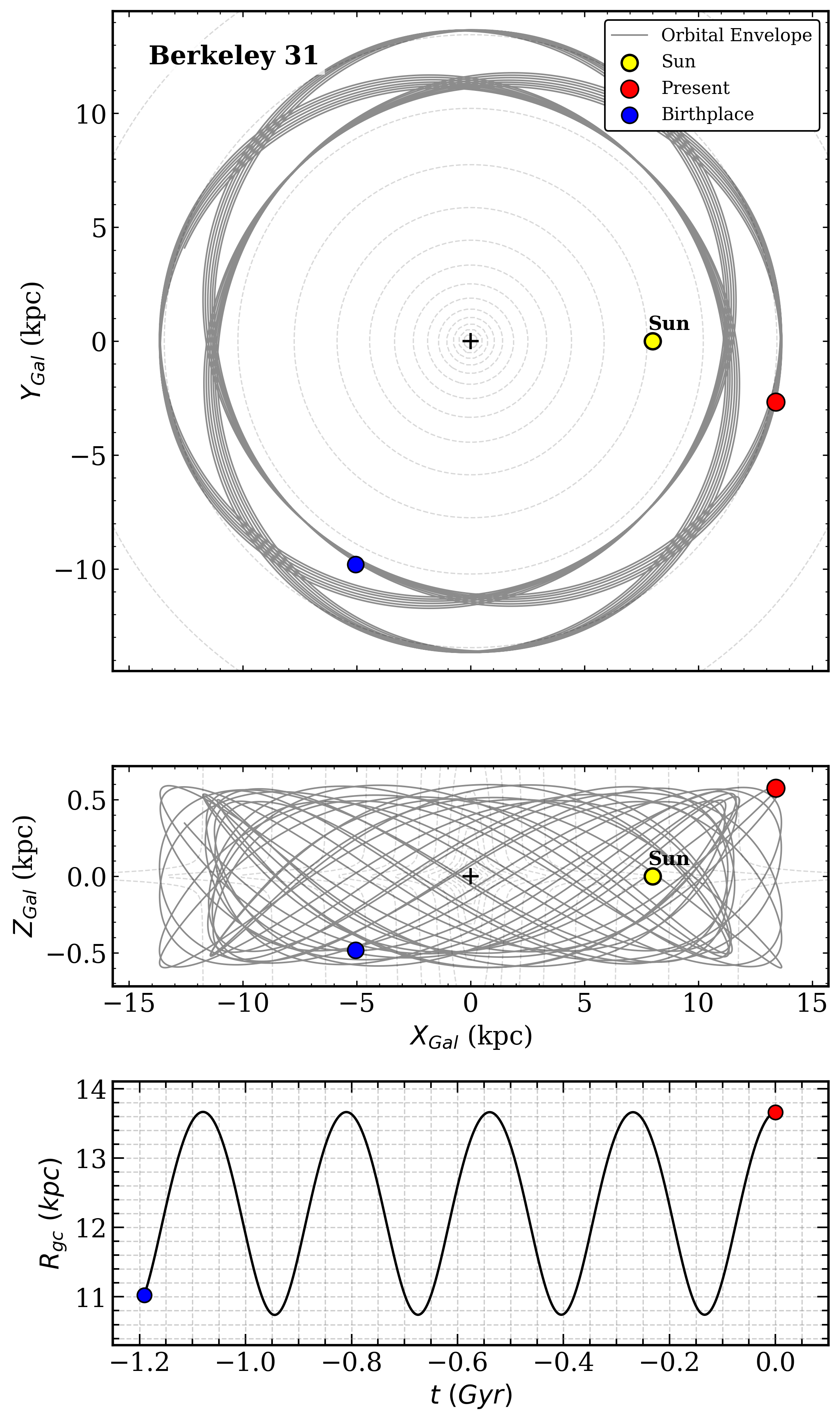}
    \caption{Galactic orbital paths for Berkeley 31. The top panel shows the orbital projection on the Galactic $X-Y$ plane, while the middle panel illustrates the motion perpendicular to the plane ($X-Z$). In both panels, the gray curves represent the integrated orbital envelope, and a yellow circle indicates the Sun. The cluster's current position and birthplace are marked with red and blue circles, respectively. The bottom panel displays the evolution of the Galactocentric distance ($R_{\mathrm{gc}}$) as a function of look-back time ($t$), where the black solid line traces the trajectory since formation. In this panel, the yellow circle represents the present-day distance, and the orange inverted triangle marks the birth distance. The grey dashed contours represent the logarithmic mass density of the Milky Way as described by the \texttt{MWPotential2014} model.}
    \label{fig:galactic_orbits}
\end{figure}

To investigate the dynamical environments of the clusters hosting BSSs with compact companions, we computed the Galactic orbits of the clusters in our sample. The orbital calculations were performed using six-dimensional (6D) phase-space information derived from Gaia DR3 astrometry, including positions, trigonometric parallaxes, proper motion components, and available radial velocities. The orbit integrations were carried out with the \texttt{galpy} Python package \citep{Bovy2015}, adopting the \texttt{MWPotential2014} Galactic potential model for the Milky Way, which consists of a Miyamoto-Nagai disk \citep{Miyamoto1975}, a Hernquist bulge \citep{Hernquist1990}, and a logarithmic halo \citep{Binney2008}. A Galactocentric distance of $R_{\rm gc}=8$ kpc, a circular rotation speed of $V_{\rm rot}=220$ km~s$^{-1}$ \citep{Bovy2012, Bovy2015}, and a vertical distance of the Sun from the Galactic plane of $27\pm 4$ pc \citep{Chen2000} were adopted. The integrations were performed over timescales corresponding to the cluster ages listed in Table~\ref{tab:cluster_params}, which were adopted from \citet{Hunt2024}. An example of the resulting orbital trajectory, corresponding to Berkeley 31, is presented in Figure~\ref{fig:galactic_orbits}. The figure shows the cluster’s motion both within the Galactic plane ($X_{\rm Gal}$–$Y_{\rm Gal}$) and in the vertical direction ($X_{\rm Gal}$–$Z_{\rm Gal}$) \citep[see also,][]{Tanik2025, Karagoz2025, Tasdemir2026}.

Because the number of clusters hosting BSSs with confirmed WD companions in our study is limited, we expanded the sample by compiling additional systems from the literature where WD companions to BSSs have been reported. In total, 14 OCs (see Table~\ref{tab:bss_lit}) from previous studies were added to our sample, resulting in a combined dataset of 23 OCs. This expanded sample allows us to explore statistical relationships between cluster structural parameters and the BSS population.

To characterize the spatial distribution of BSSs within each cluster, we calculated the $r_{50}$ parameter for the BSS population. For each BSS, the projected distance from the cluster center was determined, and the radius containing half of the BSS population was adopted as $r_{50}$. The use of $r_{50}$ provides a robust statistical description of the typical radial location of BSSs while minimizing the effects of small-number statistics. Unlike structural parameters such as the core or half-mass radius, which depend on global cluster properties and model assumptions, $r_{50}$ directly reflects the observed spatial distribution of the BSS population itself. This makes it a convenient and consistent parameter for comparing BSS distributions across clusters with different sizes and structural characteristics.

To place these measurements in a Galactic dynamical context, we compared the derived $r_{50}$ values with the early orbital radius parameter ($R_{\rm teo}$). The $R_{\rm teo}$ parameter represents the approximate early orbital configuration of a stellar system. It is calculated by tracing its orbit backward in time using the present-day phase-space coordinates and cluster age \citep{Bovy2015}. In this approach, orbital integration is performed under the assumption of a static Galactic potential, and numerous studies in the literature address this assumption \citep[e.g.,][]{Tasdemir2023,  Yontan2023, Yucel2024, 2024PARep...2....1C, Tasdemir2025}. While $R_{\rm teo}$ does not represent the exact birth radius of a cluster, it provides an estimate of its early orbital characteristics without relying on assumptions about the metallicity gradient of the Galactic disk. The $R_{\rm teo}$ values used in this work were obtained using the methodology described in \citet{Minchev2013}. Cluster ages were adopted from \citet{Hunt2024}, thereby allowing us to trace the clusters' orbital evolution back to their early dynamical configurations.

\begin{figure}
    \centering
    \includegraphics[width=1\linewidth]{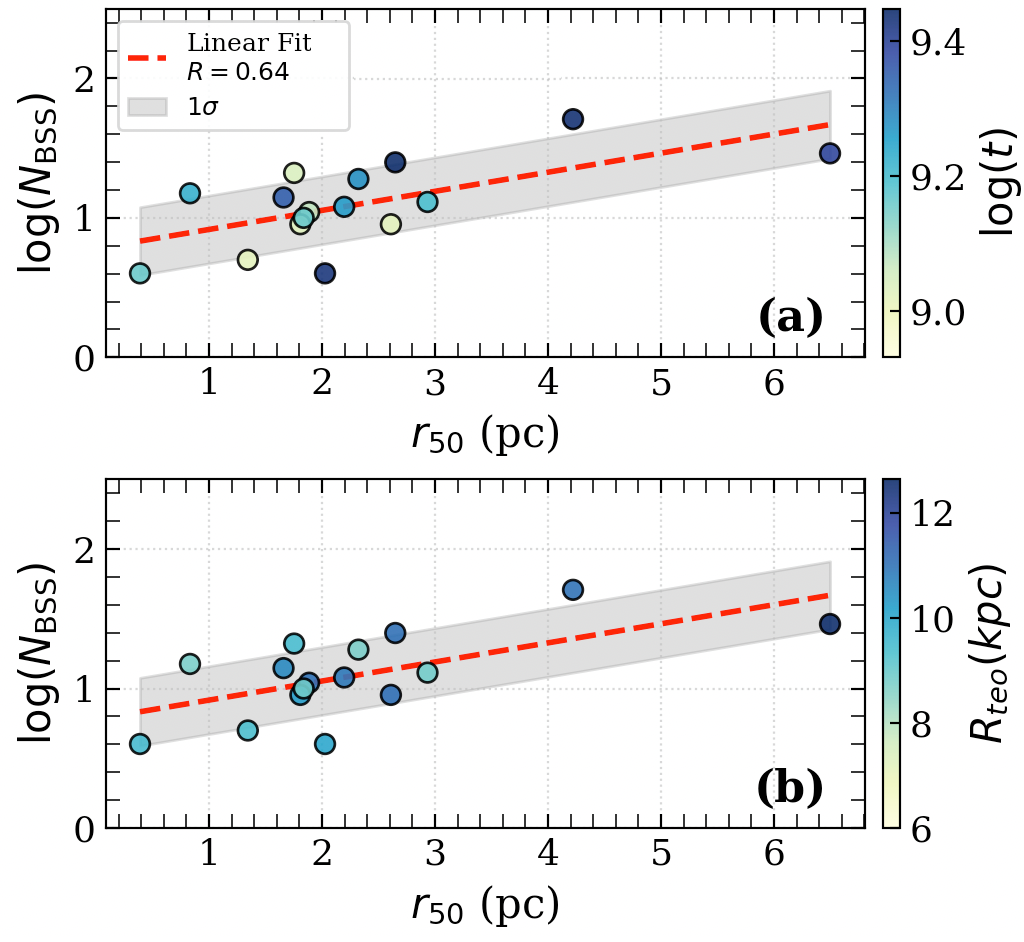}
    \caption{Correlation between the half-light radius $r_{50}$ and the number of BBSs ($\log N_{\mathrm{BSS}}$) in OCs. Color coding represents cluster age ($\log t$) in panel (a) and theoretical birth radius ($R_{\mathrm{teo}}$) in panel (b). The dashed red line indicates the linear fit, while the shaded gray area denotes the $1\sigma$.}
    \label{fig:cluster_dynamics}
\end{figure}

Figure~\ref{fig:cluster_dynamics} shows the relationship between the half-number radius of BSSs ($r_{50}$) and the logarithm of the number of BSSs ($\log N_{\rm BSS}$) for the cluster sample. Panel (a) shows the distribution color-coded by cluster age, while panel (b) displays the same relation color-coded by the early orbital radius $R_{\rm teo}$. A positive correlation between $r_{50}$ and $\log N_{\rm BSS}$ is visible, with the linear fit yielding $\log N = 0.14 \times r_{\rm 50} + 0.78$ and a moderate correlation coefficient ($R = 0.64$). 

This trend suggests that clusters with larger characteristic BSS radii tend to host larger BSS populations. Such behavior may reflect the influence of cluster structural properties and dynamical evolution on the formation and survival of BSS systems. Older clusters or those with larger orbital radii in the Galactic disk tend to occupy relatively quiescent dynamical environments, where tidal perturbations are weaker \citep[e.g.,][]{Carraro2007, Carraro2012, Angelo2020, Angelo2023, Bilir2026}. At the same time, it should be noted that cluster survival is mass-dependent: more massive clusters are generally more likely to persist to older ages, while lower-mass clusters are more easily disrupted. Therefore, the larger number of BSSs (and larger orbital radii) observed in older clusters may partly reflect this selection effect, in addition to the influence of weaker tidal perturbations. These conditions can allow primordial binaries to survive for long timescales and evolve through mass transfer, producing BSS+WD systems.

Although the sample size remains limited, the observed trend hints that both the internal structure of clusters and their Galactic dynamical history may influence the observed BSS population. All calculated parameters, including the literature clusters and the derived $r_{50}$ and $R_{\rm teo}$ values, were included in the final analysis.

\section{Summary and Conclusion}
\label{sec:conclusion}

In this study, we have conducted a systematic multi-wavelength characterization of 35 candidate stars distributed across nine OCs. By anchoring our dataset with high-sensitivity NUV photometry from \textit{Swift}/UVOT and complementing it with optical-to-IR data from Gaia DR3, Pan-STARRS, and 2MASS/WISE, we constructed robust SEDs to probe for the presence of hot, subluminous companions. We further contextualized these photometric findings by reconstructing the Galactic orbits of the host clusters, thereby linking the internal binary evolution of the BSSs to their external dynamical environments.

The main findings are summarized as follows:

\begin{itemize}
    \item Our SED analysis revealed that 20 systems are consistent with single-component atmosphere models or possess companions that remain unresolved at UV wavelengths. In contrast, 15 BSSs exhibited significant UV excesses requiring two-component SED fits, providing evidence for hot, compact companions and a binary origin \citep{Gosnell2015, 2022MNRAS.511.2274V, Chand2024}. 

    \item Among the binary systems, our analysis suggests the presence of hot companions likely representing distinct post-mass-transfer stages, specifically young, hot WDs and pre-ELM WD candidates. The coexistence of these binary systems alongside single-fit BSSs points to the heterogeneous nature of the BSS population in intermediate-age OCs, supporting the scenario that mass transfer is a major formation pathway, consistent with previous studies \citep[e.g.,][]{Subramaniam2016, 2024BSRSL..93..250V}. Component masses derived from evolutionary-track interpolation suggest BSS primaries in the range $\sim$1.2--2.1\,$M_\odot$ and WD companions spanning $\sim$0.21--0.54\,$M_\odot$, consistent with intermediate-mass binary  progenitors undergoing mass transfer in OC environments.
    
    \item Radial spatial distributions confirmed that the identified BSS candidates predominantly inhabit the central regions of their host clusters, which is a direct signature of mass segregation driven by dynamical friction. Distinct segregation within clusters such as NGC~6939 indicates a dynamically relaxed state. Conversely, a lack of strong segregation in clusters such as NGC~2192 and NGC~2360 suggests these environments may be dynamically younger or that their BSS populations formed too recently to have migrated inward.

    \item We estimated the photometric binary fractions for the nine host clusters using a CMD-based method \citep{Milone2012}. The derived values range from $f_{\rm b} \approx 0.04$ to $0.15$, with NGC~6939, NGC~2533, and NGC~2204 having the highest fractions. There is a tentative indication that clusters with higher binary fractions may host more UV-excess BSS systems; however, this trend is not statistically robust due to the limited sample size. For example, NGC~6939 contains only one UV-excess BSS, possibly due to its advanced dynamical state and observational limitations in crowded regions. A larger and more homogeneous sample is needed to test this relation more reliably \citep{Childs2025, Mikhnevich2026a, Malhotra2026}.
    
   \item Orbital solutions derived from Gaia DR3 astrometry show that clusters hosting BSSs with hot companions follow stable, mildly eccentric orbits confined to the Galactic disk. This kinematic configuration is typical for OCs and implies that they evolve in relatively quiescent dynamical environments, co-rotating with the disk and experiencing only weak tidal perturbations. As a result, these systems are largely shielded from strong tidal shocks or disruptive encounters with giant molecular clouds \citep{1987degc.book.....S, 2006MNRAS.371..793G, 2012MNRAS.426.3008K}. This suggests that dynamical stellar collisions are unlikely to play a significant role, and that the observed BSS systems are more consistently explained by long-term binary evolution through stable mass transfer \citep{2015ASSL..413...29M}.
    
    \item Our analysis reveals a positive correlation ($R = 0.64$) between the half-number radius of the BSS population ($r_{50}$) and the logarithm of the number of BSSs ($\log N_{\rm BSS}$). When examined alongside cluster age and early orbital radius ($R_{\rm teo}$), this trend suggests that clusters with larger characteristic BSS radii tend to host larger BSS populations.
\end{itemize}

The detection of both hot WDs and pre-ELM candidates highlights the diverse range of interaction products present in OCs. Future spectroscopic follow-up is strongly encouraged to determine radial velocities, orbital periods, and mass ratios for these systems, thereby providing definitive constraints on binary evolution models.

\begin{acknowledgments}
We sincerely thank the anonymous referee for a thorough review and constructive suggestions that greatly improved the clarity and quality of the manuscript. This study is a part of the PhD Thesis of Deniz Cennet Çınar. This work uses data from the European Space Agency's Gaia mission, processed by the Gaia Data Processing and Analysis Consortium (DPAC). Ultraviolet data were obtained from the Neil Gehrels Swift Observatory using calibrated Level-3 UVOT point-source catalogs. Optical photometry was taken from Pan-STARRS1 and the SkyMapper Southern Survey. Near- and mid-infrared data were obtained from 2MASS and WISE. NASA’s Astrophysics Data System, as well as the VizieR and SIMBAD databases operated by CDS in Strasbourg, France, were used. This work has also used data from the European Space Agency (ESA) mission Gaia (https://www.cosmos.esa.int/gaia), processed by the Gaia Data Processing and Analysis Consortium (DPAC; https://www.cosmos.esa.int/web/gaia/dpac/consortium). Funding for the DPAC is provided by national institutions, in particular those participating in the Gaia Multilateral Agreement. This publication makes use of VOSA, developed under the Spanish Virtual Observatory (https://svo.cab.inta-csic.es) project funded by MCIN/AEI/10.13039/501100011033/ through grant PID2020-112949GB-I00. VOSA has been partially updated with funding from the European Union's Horizon 2020 Research and Innovation Programme under Grant Agreement No. 776403 (EXOPLANETS-A).
\end{acknowledgments}

\facilities{Swift(UVOT), Gaia, Pan-STARRS, SkyMapper, 2MASS, WISE}

\software{
\texttt{VOSA} \citep{Bayo2008},
\texttt{galpy} \citep{Bovy2015},
\texttt{Aladin}
}
\appendix
\section{Single-Component SED Fits of Studied BSSs}\label{AppendixA}

\begin{figure}
    \centering
    \includegraphics[width=0.23\linewidth]{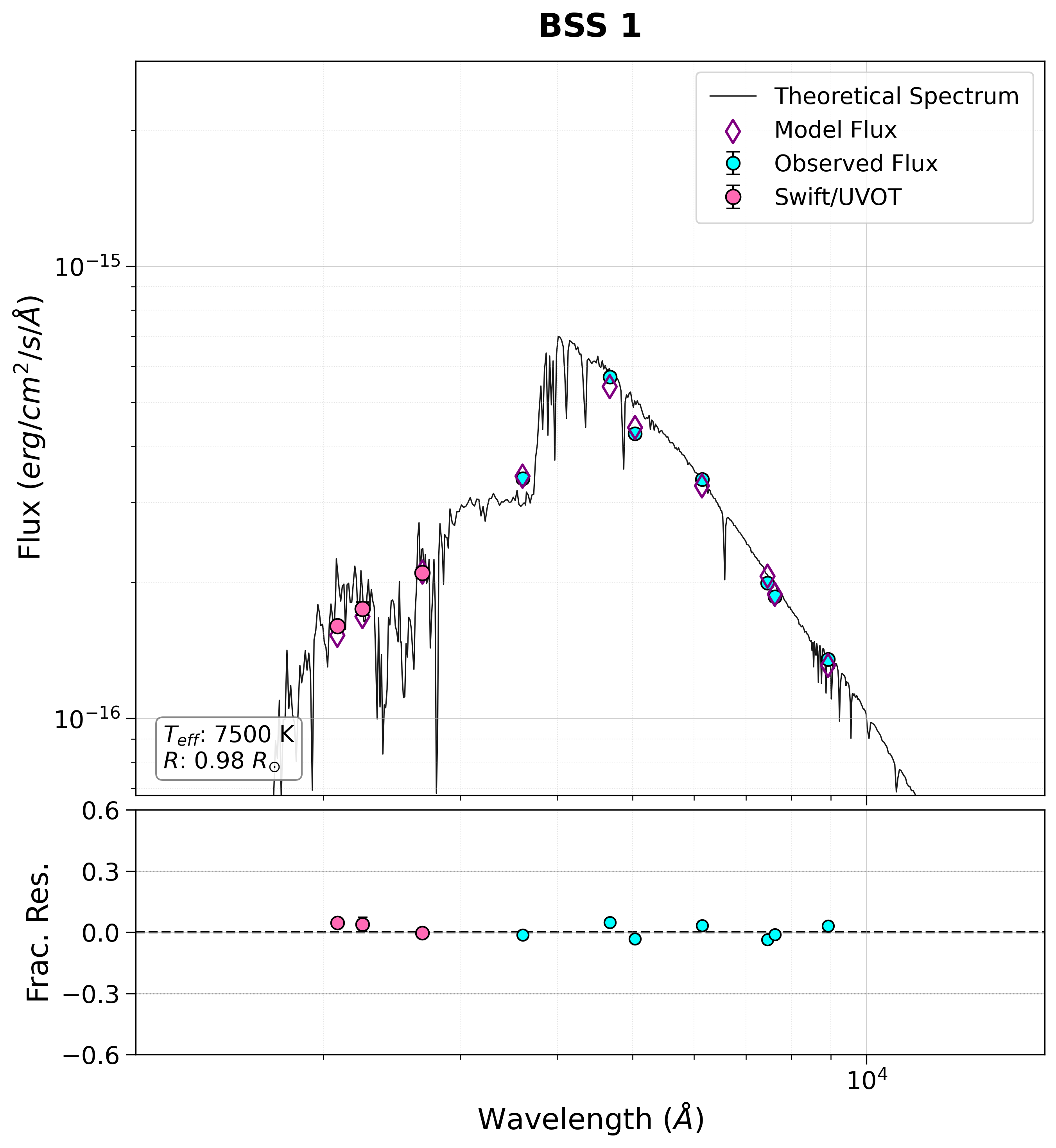}
    \includegraphics[width=0.23\linewidth]{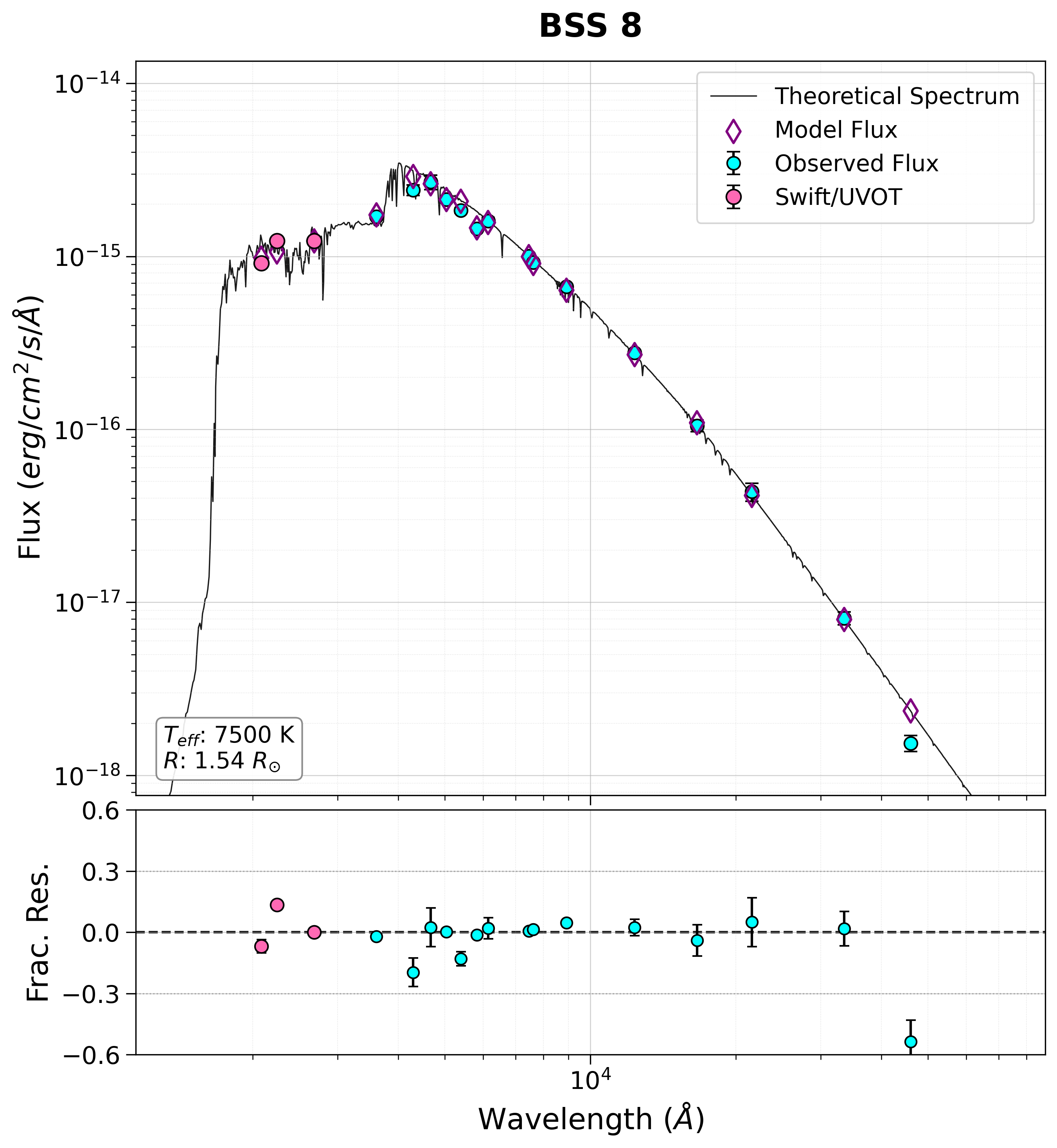}
    \includegraphics[width=0.23\linewidth]{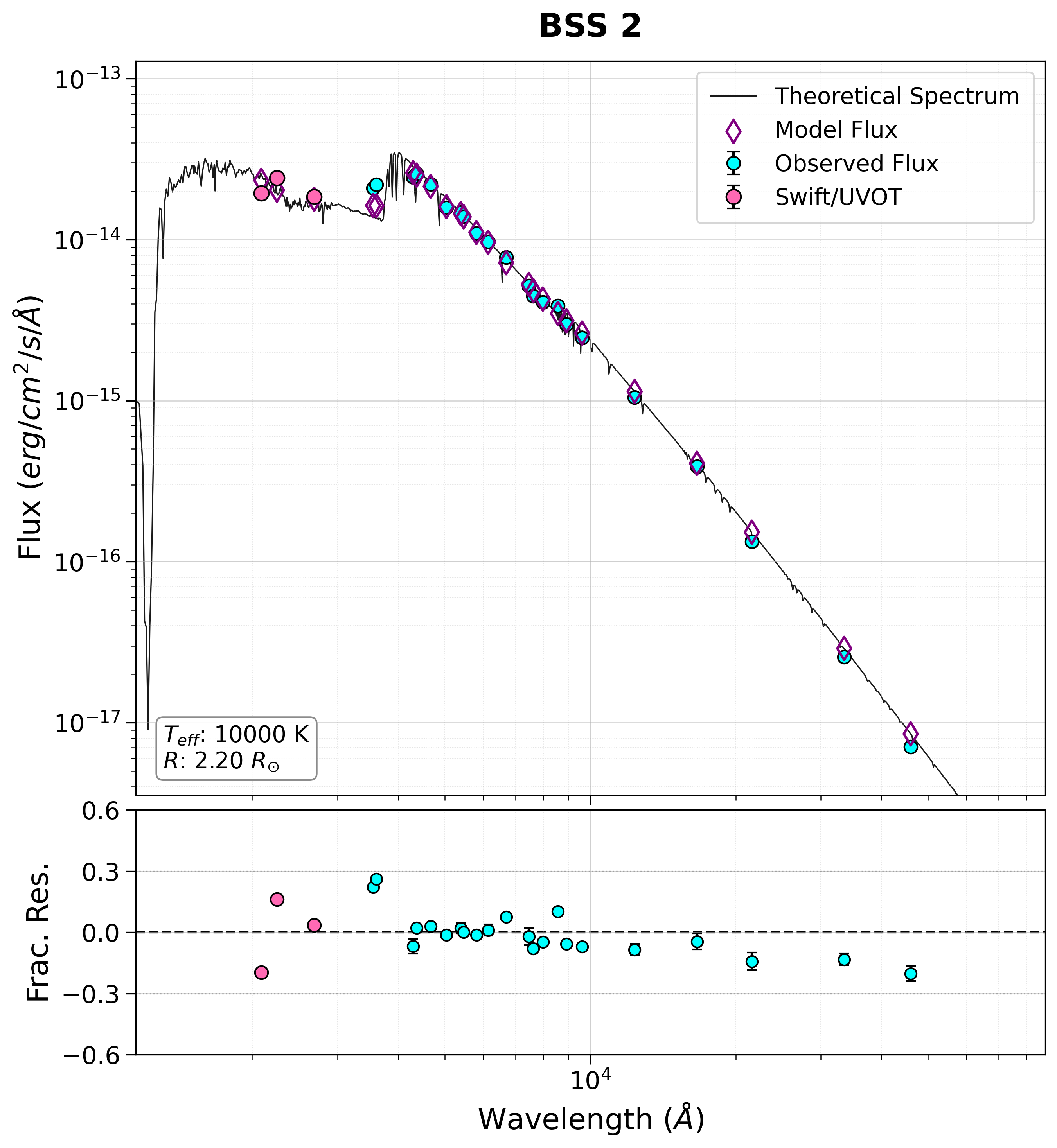}
    \includegraphics[width=0.23\linewidth]{BSS_3.png}\\

    \includegraphics[width=0.23\linewidth]{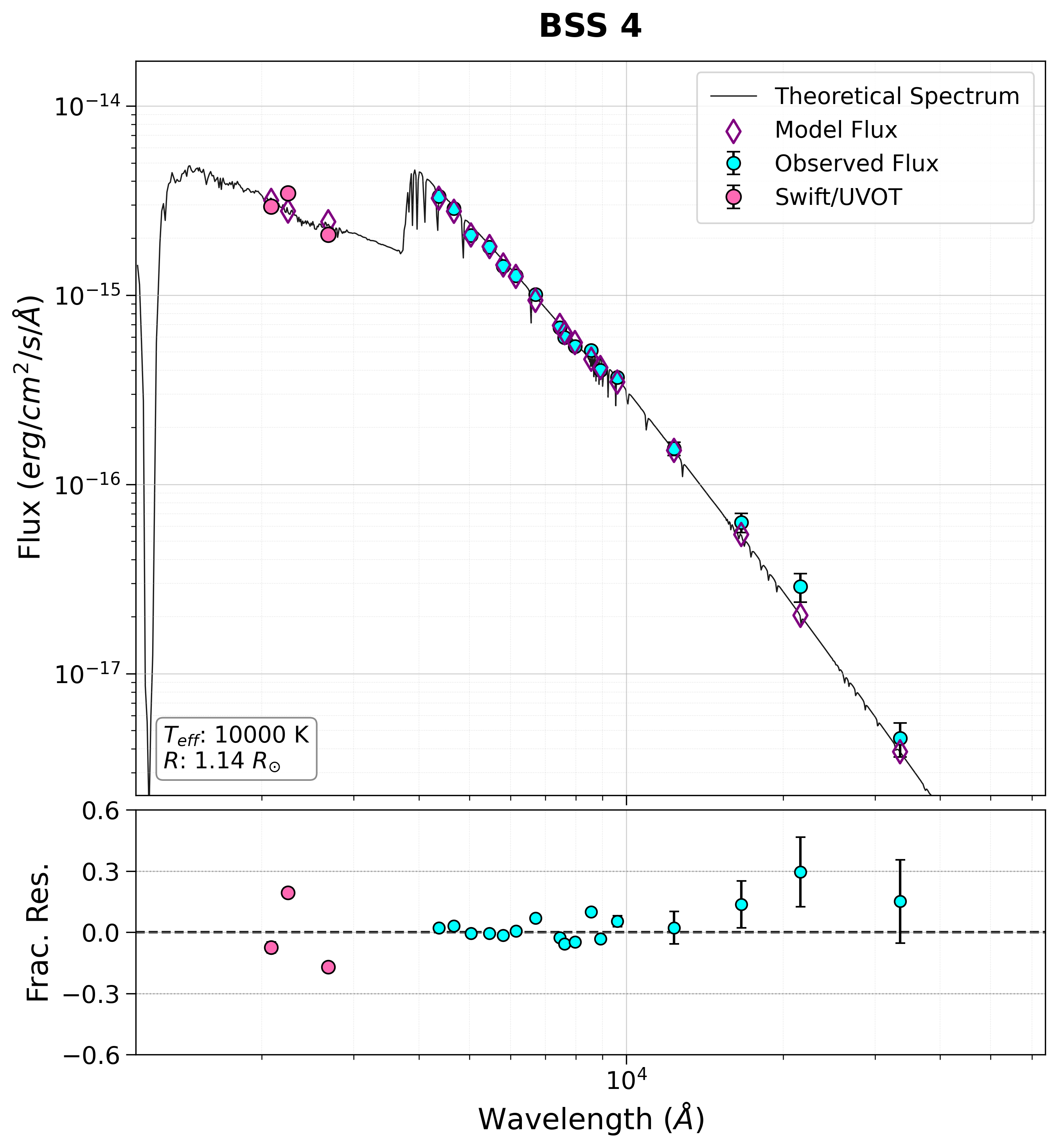}
    \includegraphics[width=0.23\linewidth]{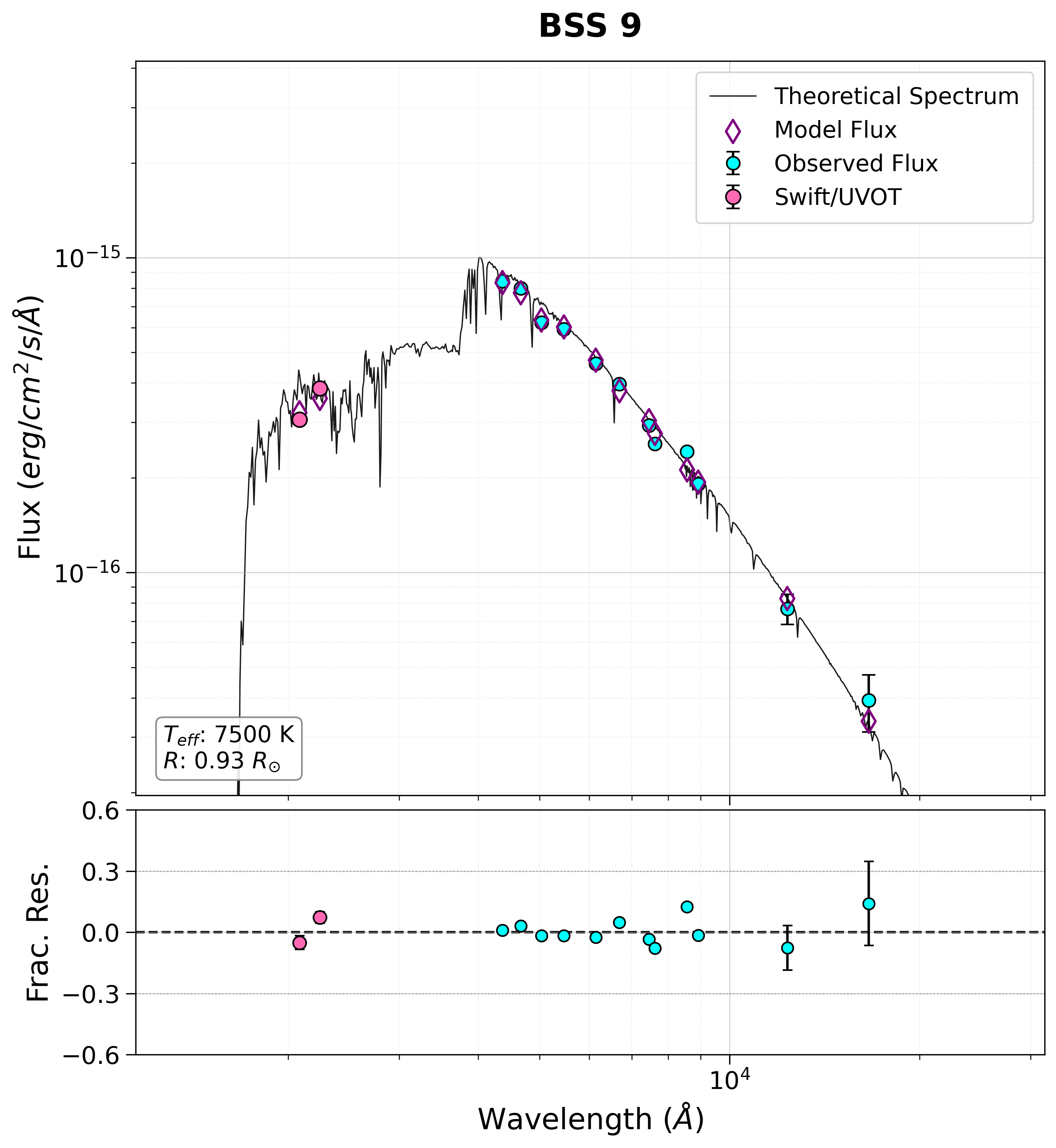}
    \includegraphics[width=0.23\linewidth]{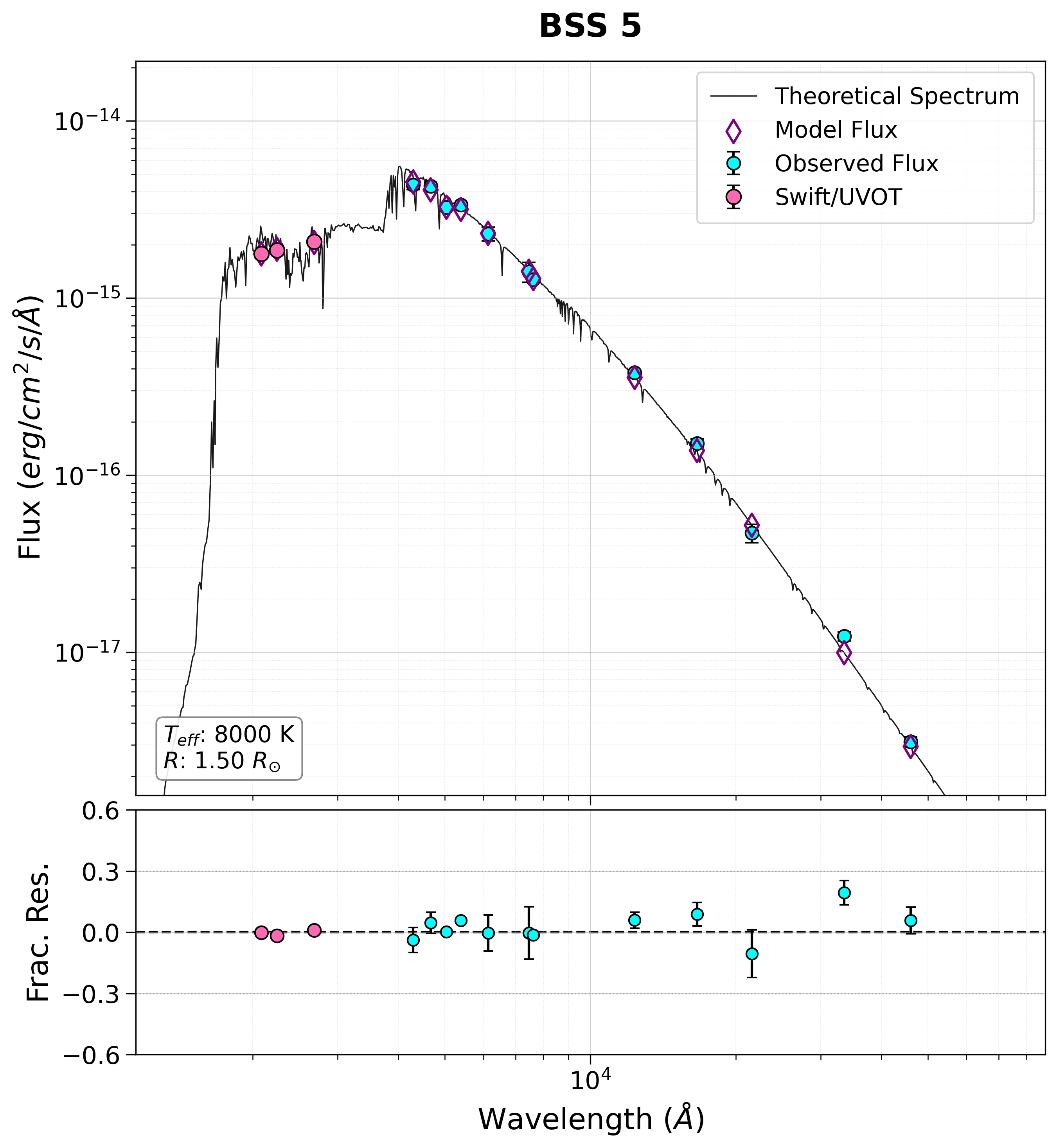}
    \includegraphics[width=0.23\linewidth]{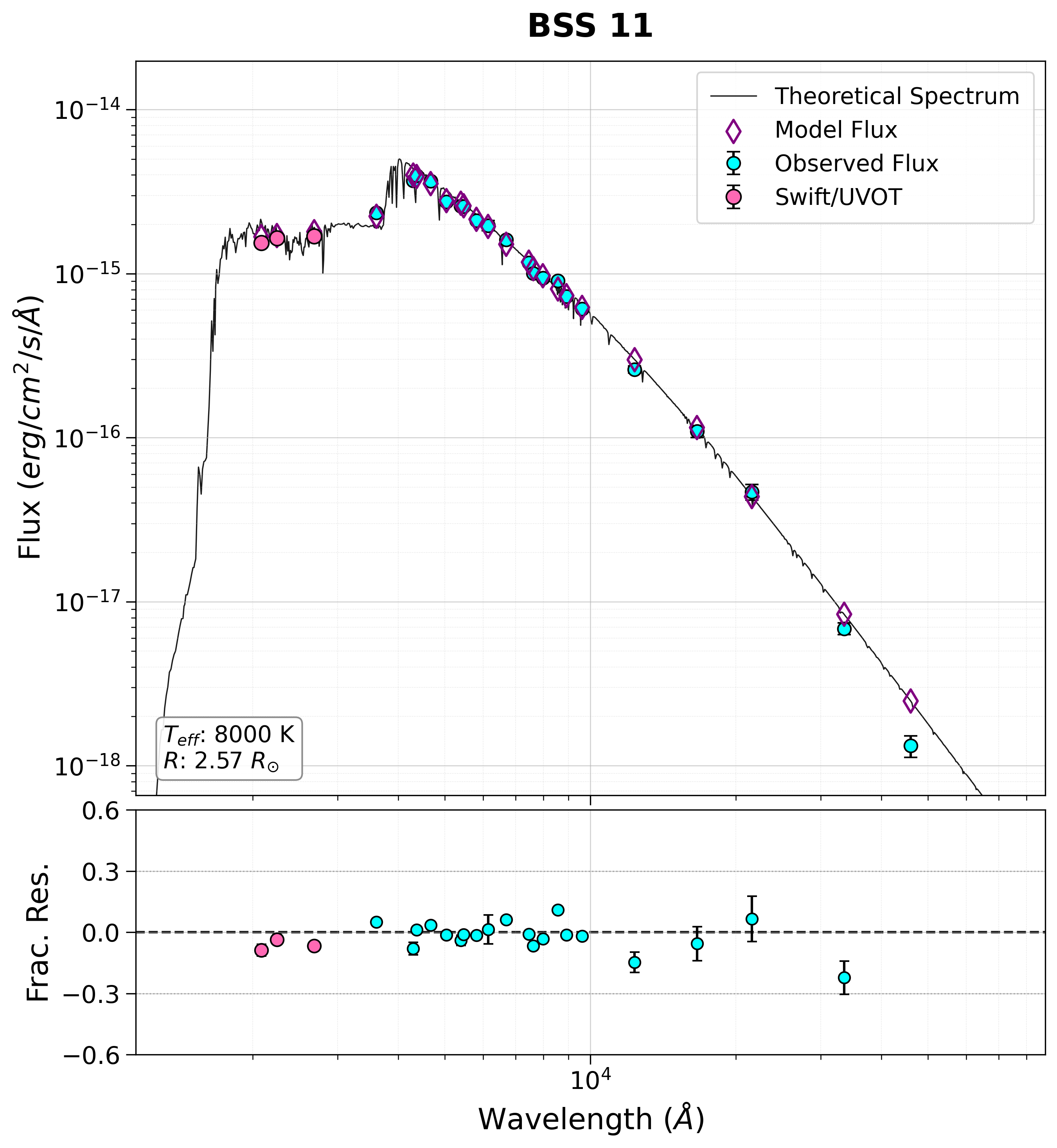}
    
    \includegraphics[width=0.23\linewidth]{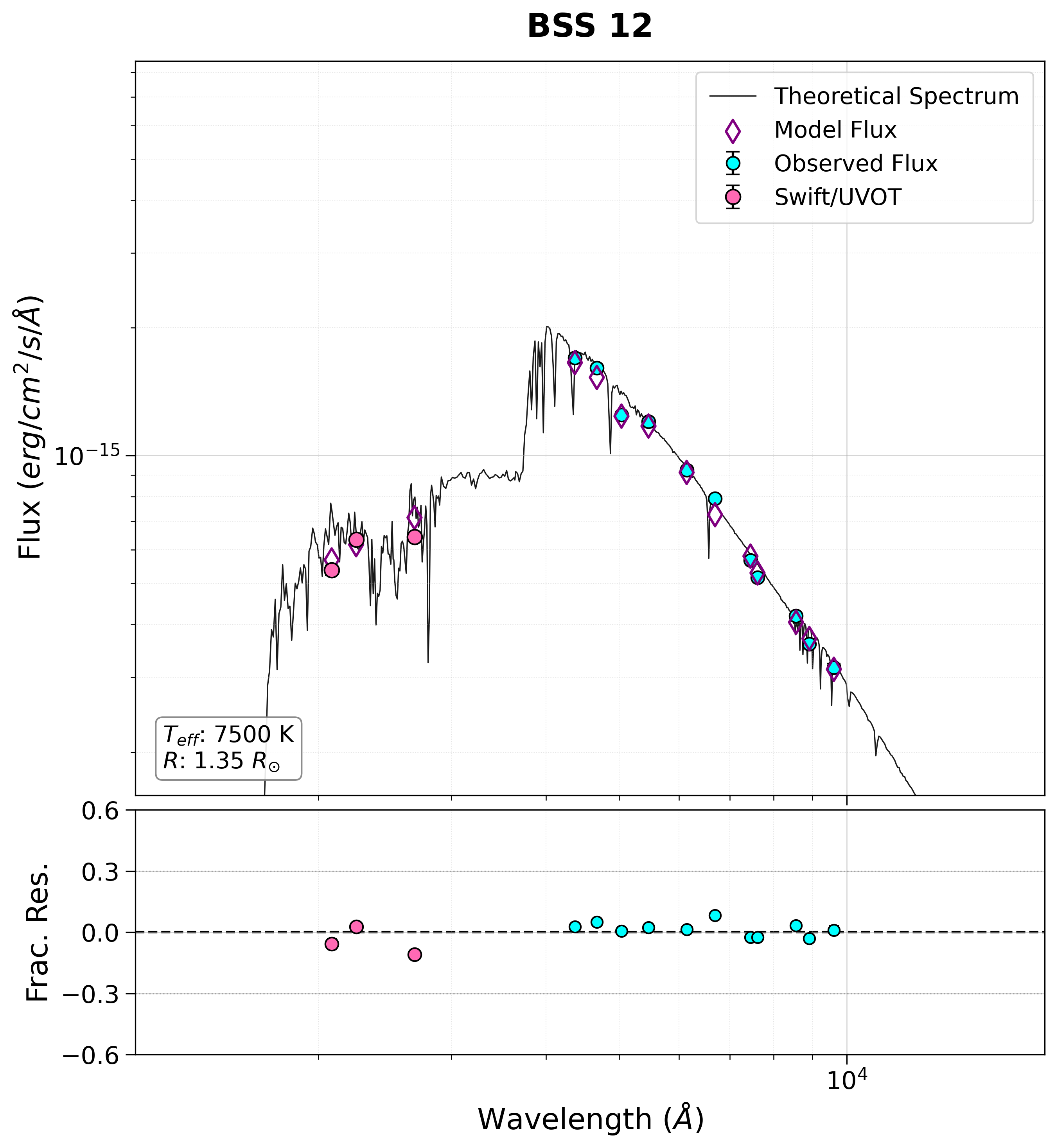}
    \includegraphics[width=0.23\linewidth]{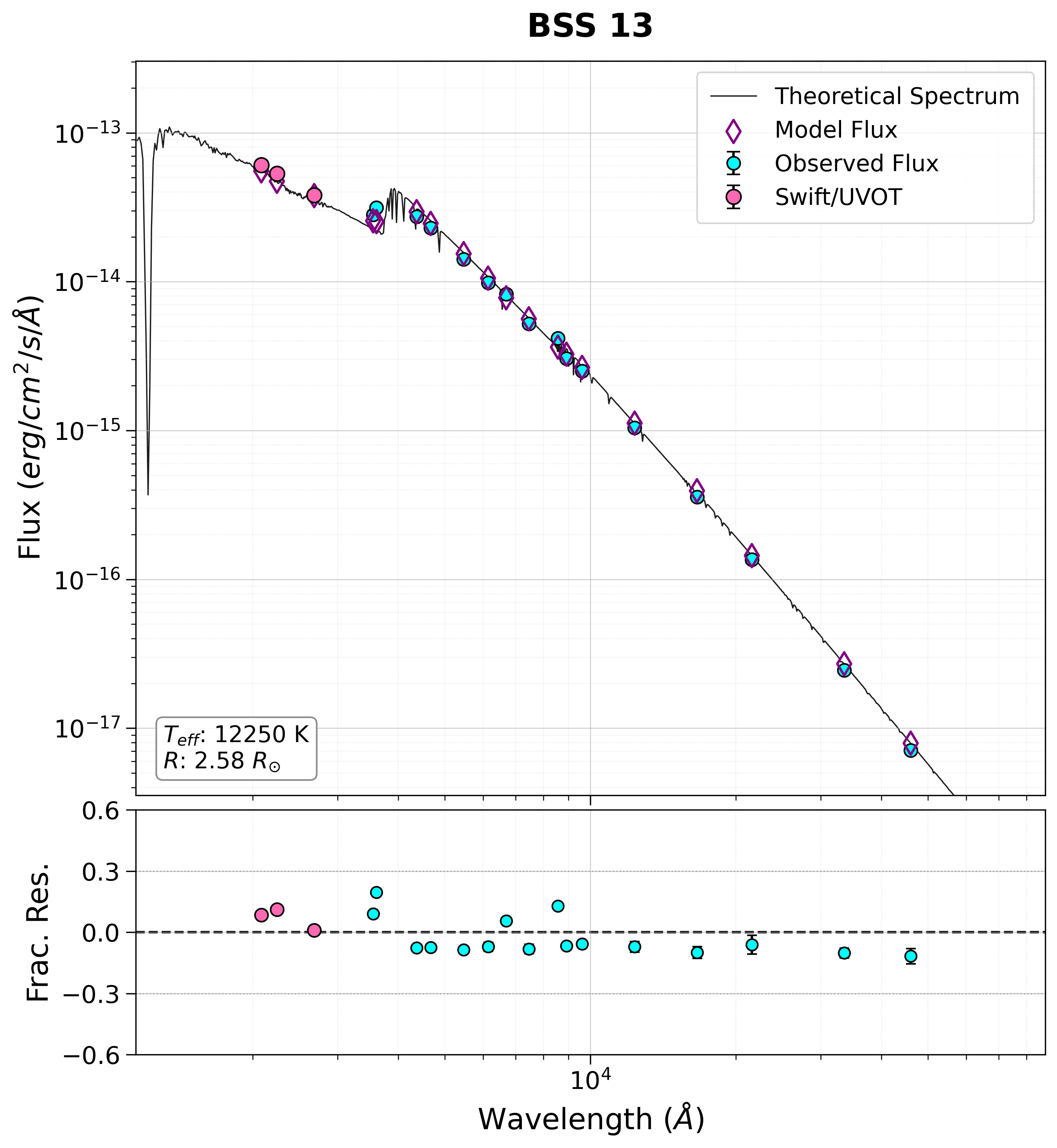}
    \includegraphics[width=0.23\linewidth]{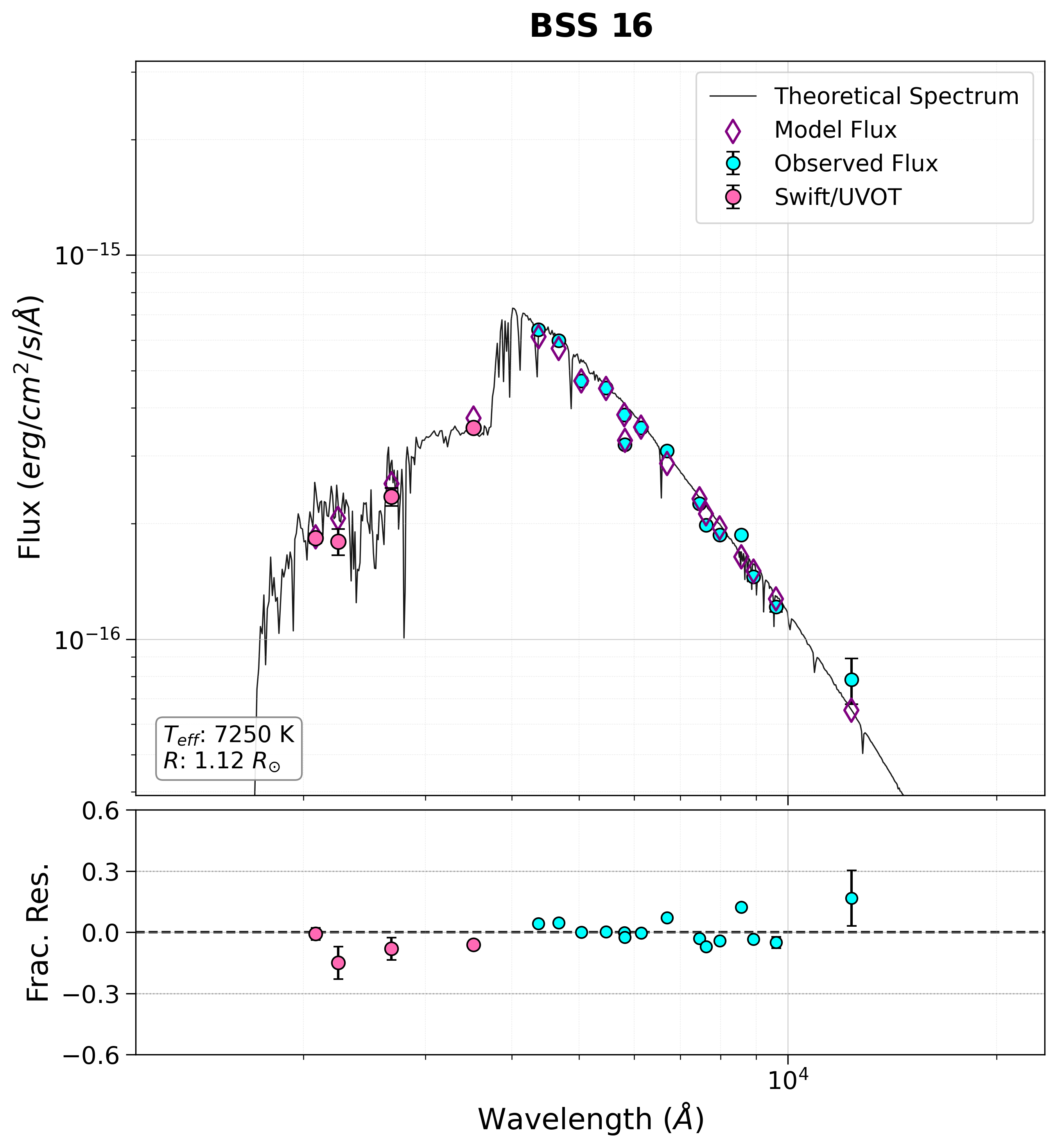}
    \includegraphics[width=0.23\linewidth]{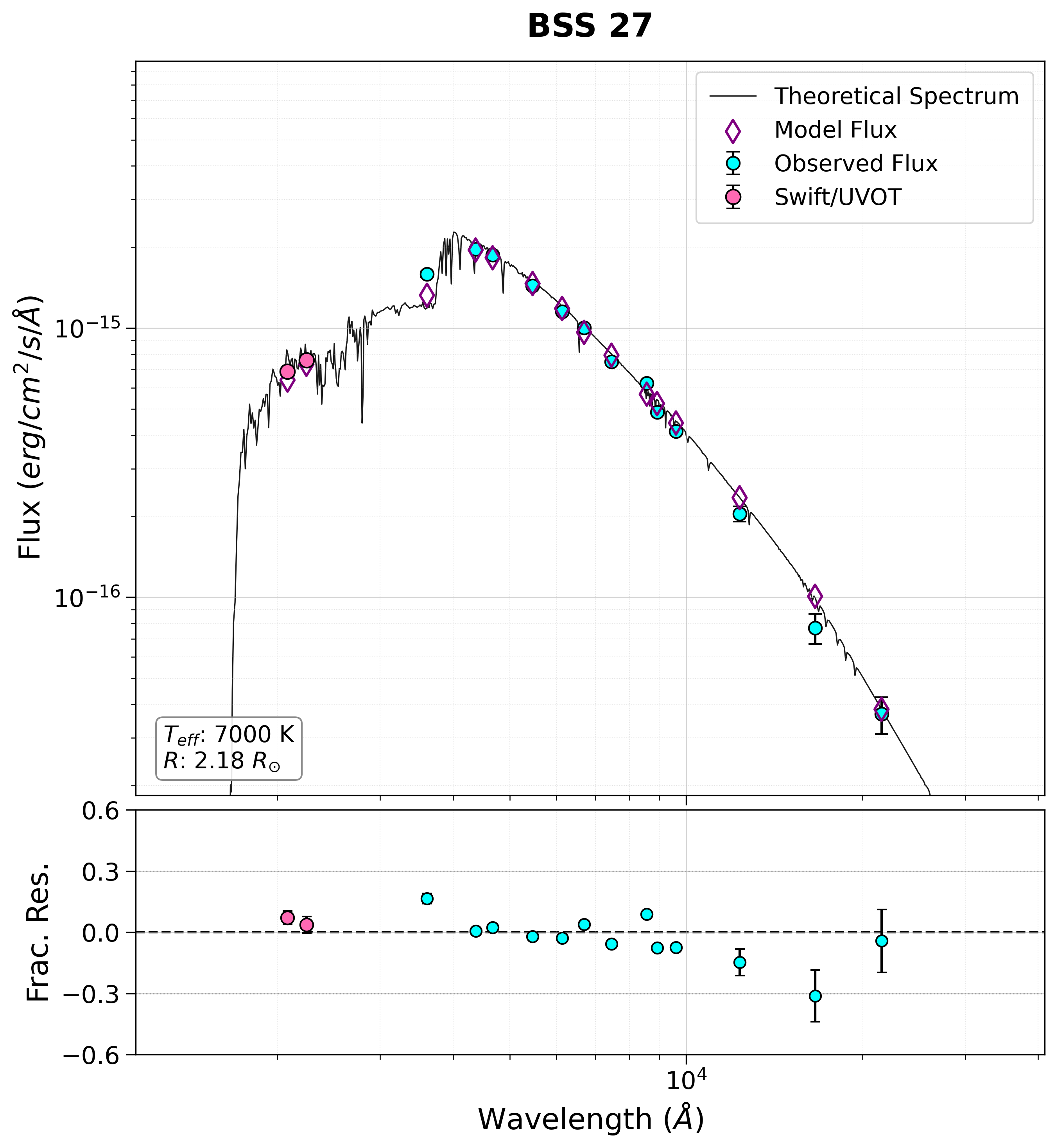}\\
    
    \includegraphics[width=0.23\linewidth]{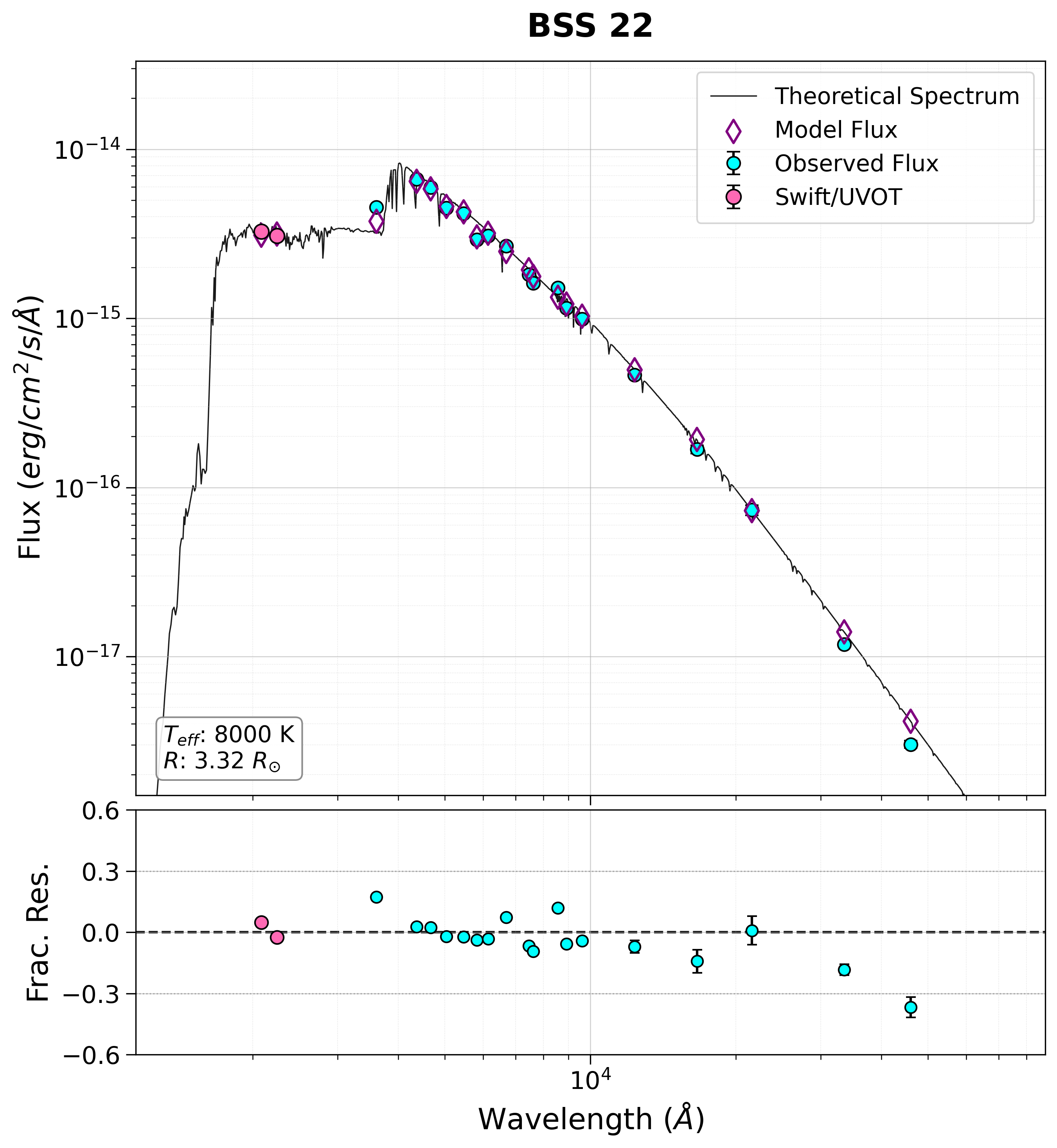}
    \includegraphics[width=0.23\linewidth]{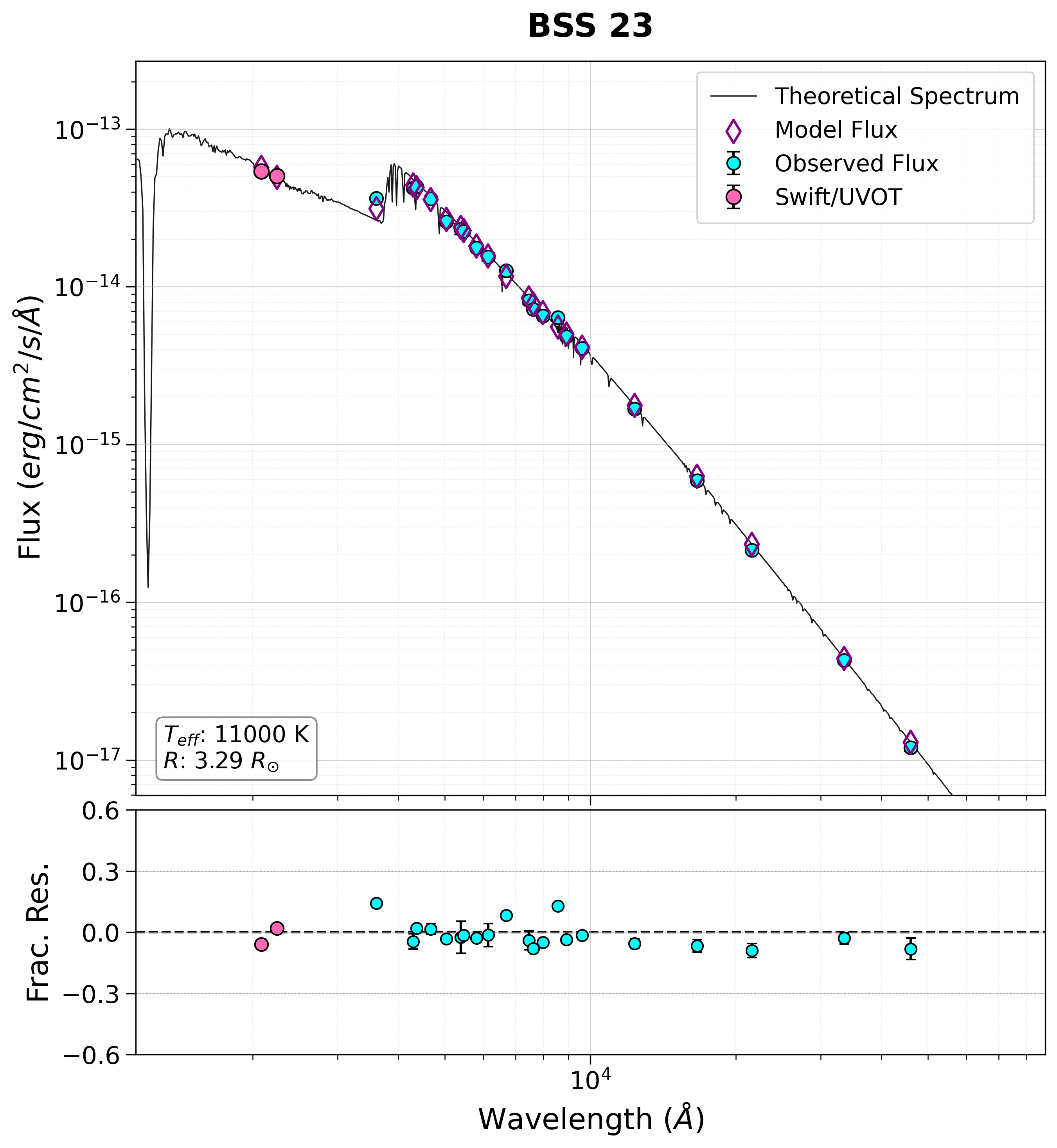}
    \includegraphics[width=0.23\linewidth]{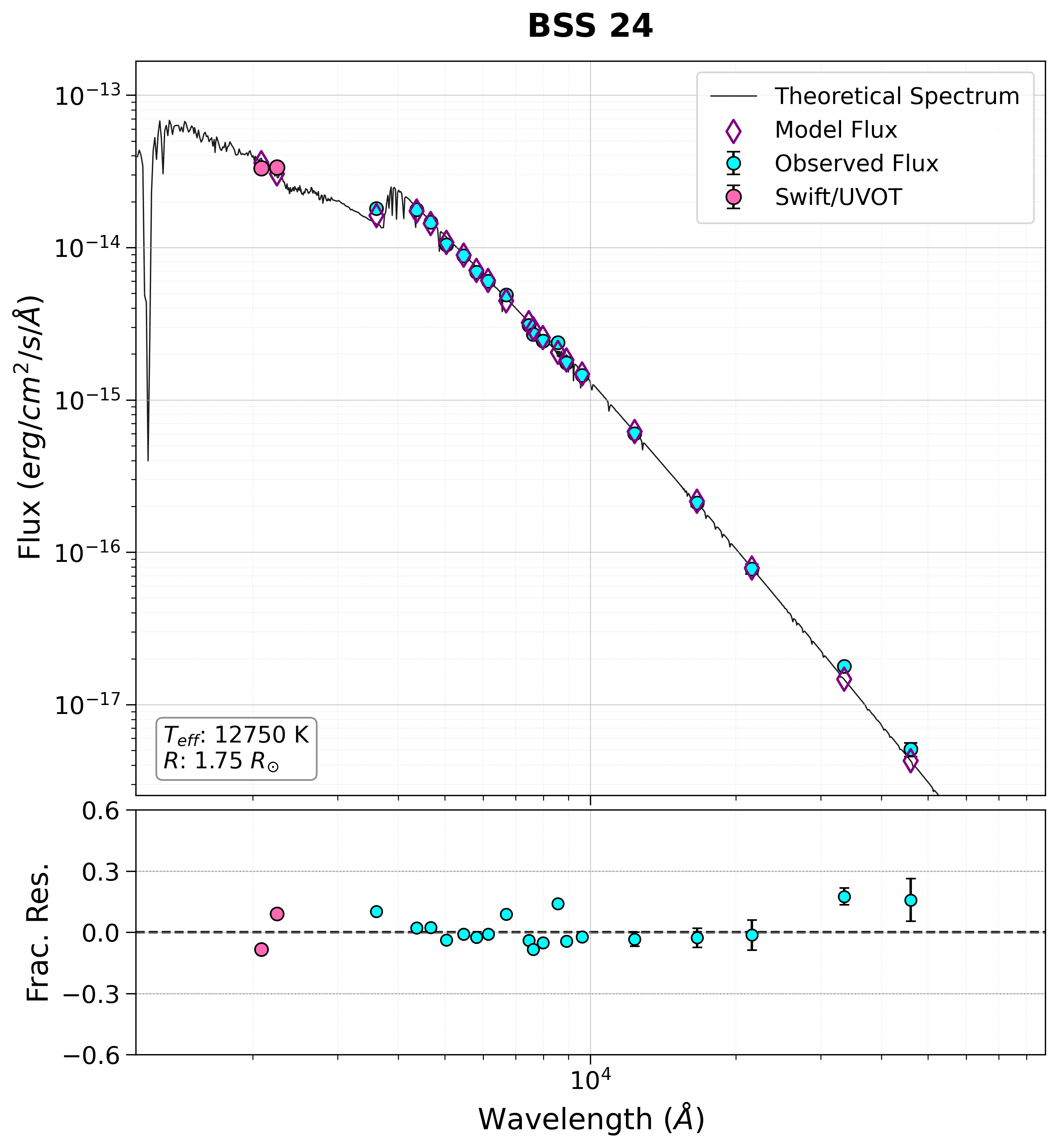}
    \includegraphics[width=0.23\linewidth]{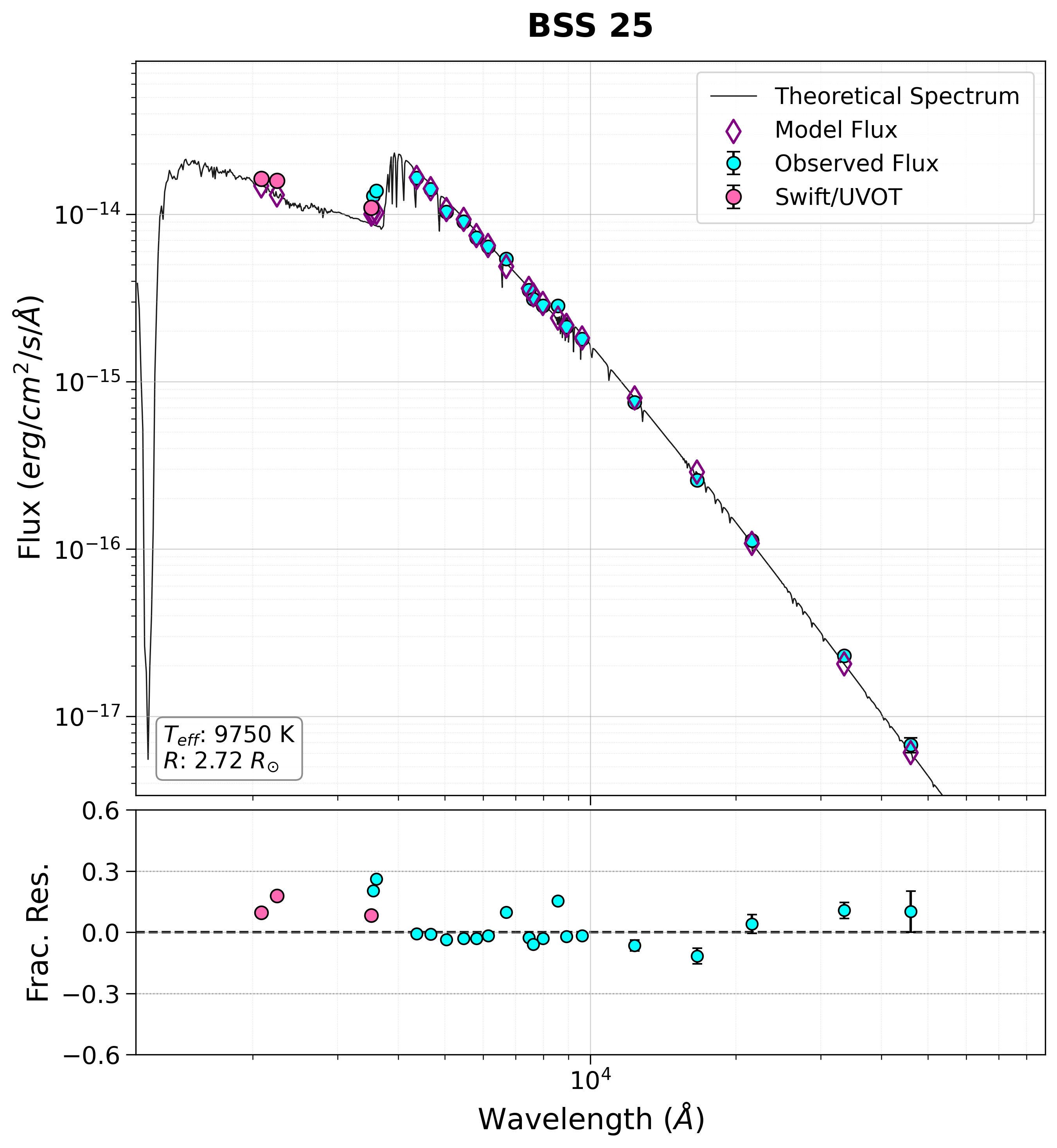}\\

    \includegraphics[width=0.23\linewidth]{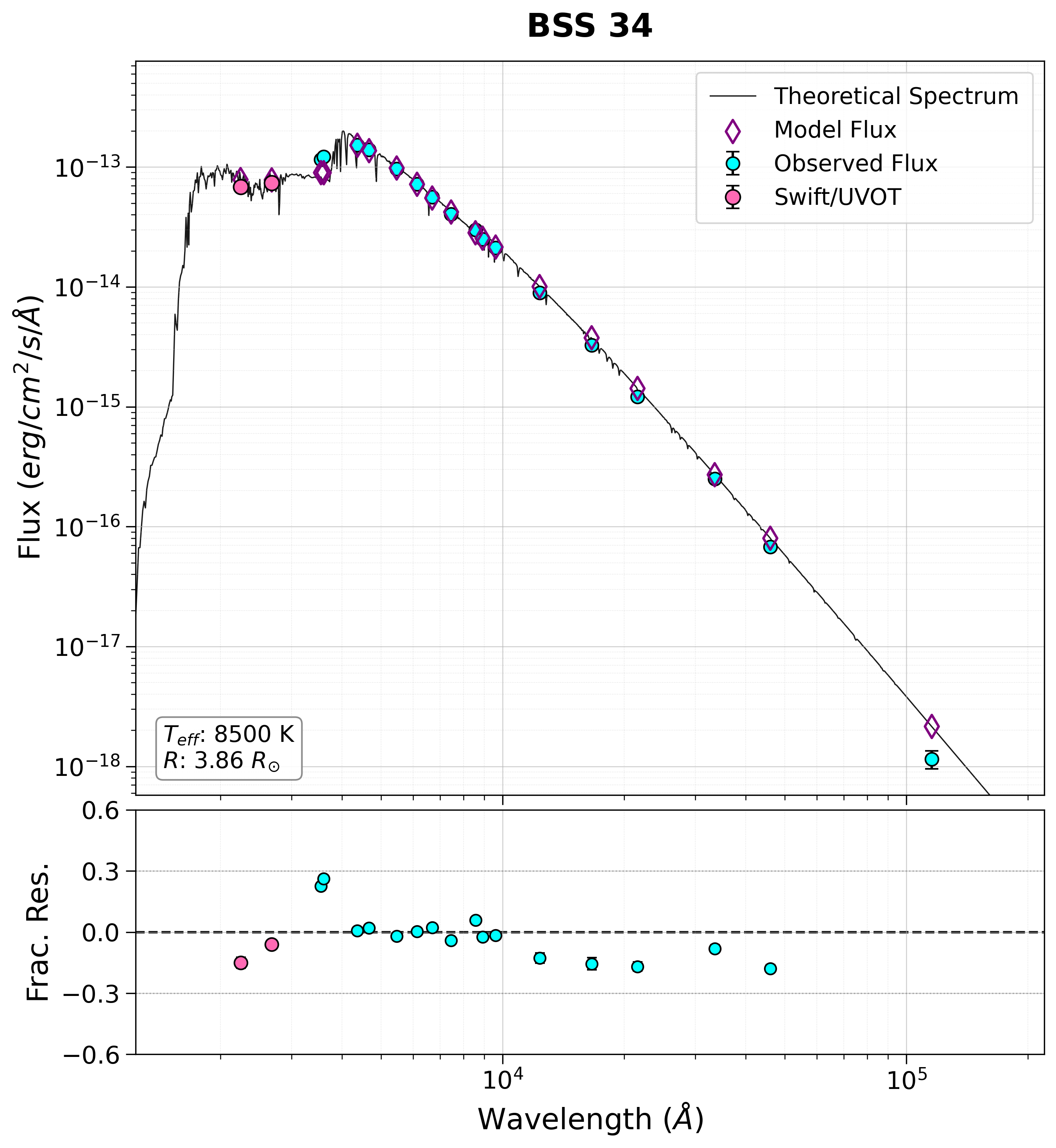}
    \includegraphics[width=0.23\linewidth]{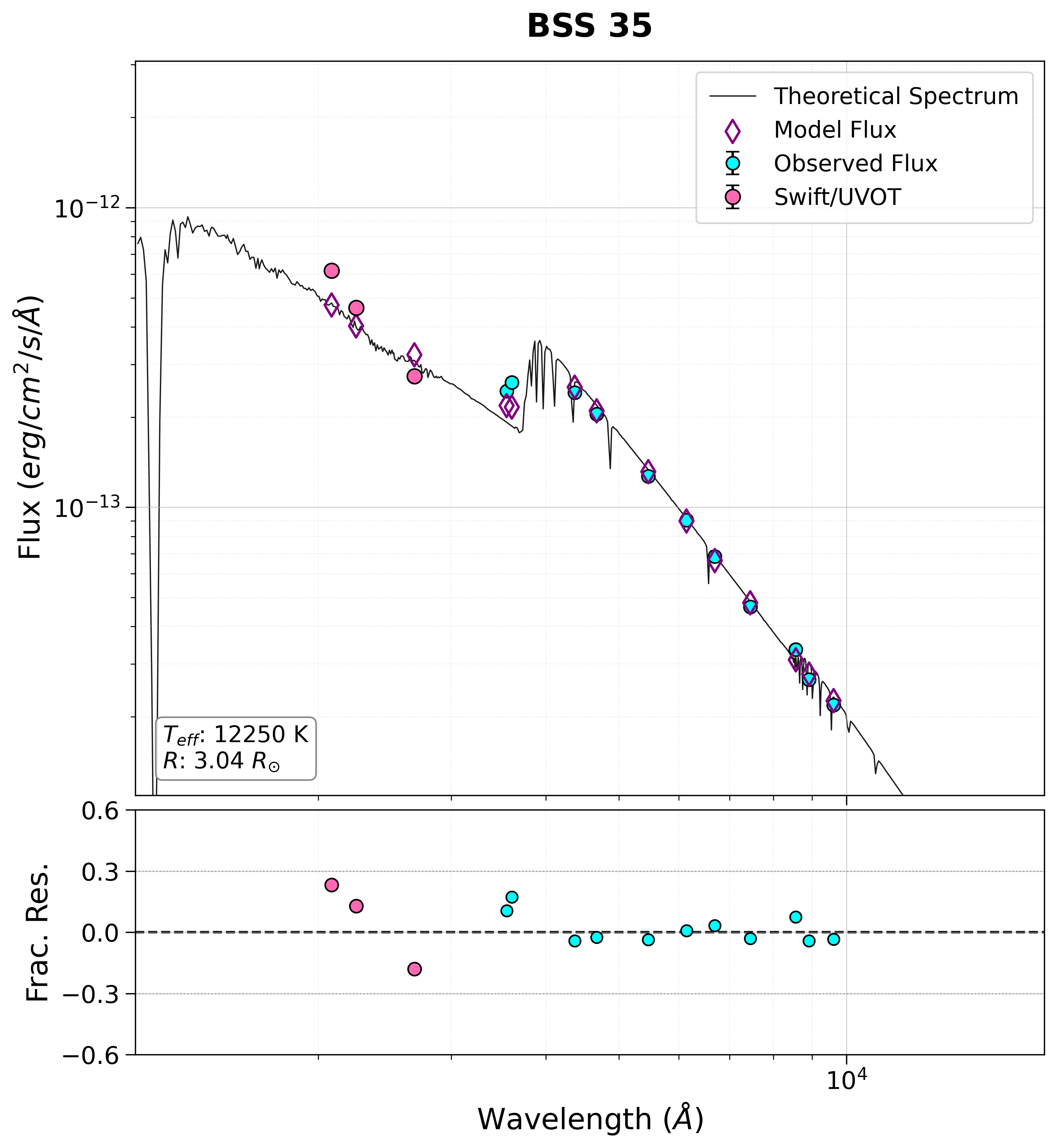}
    \includegraphics[width=0.23\linewidth]{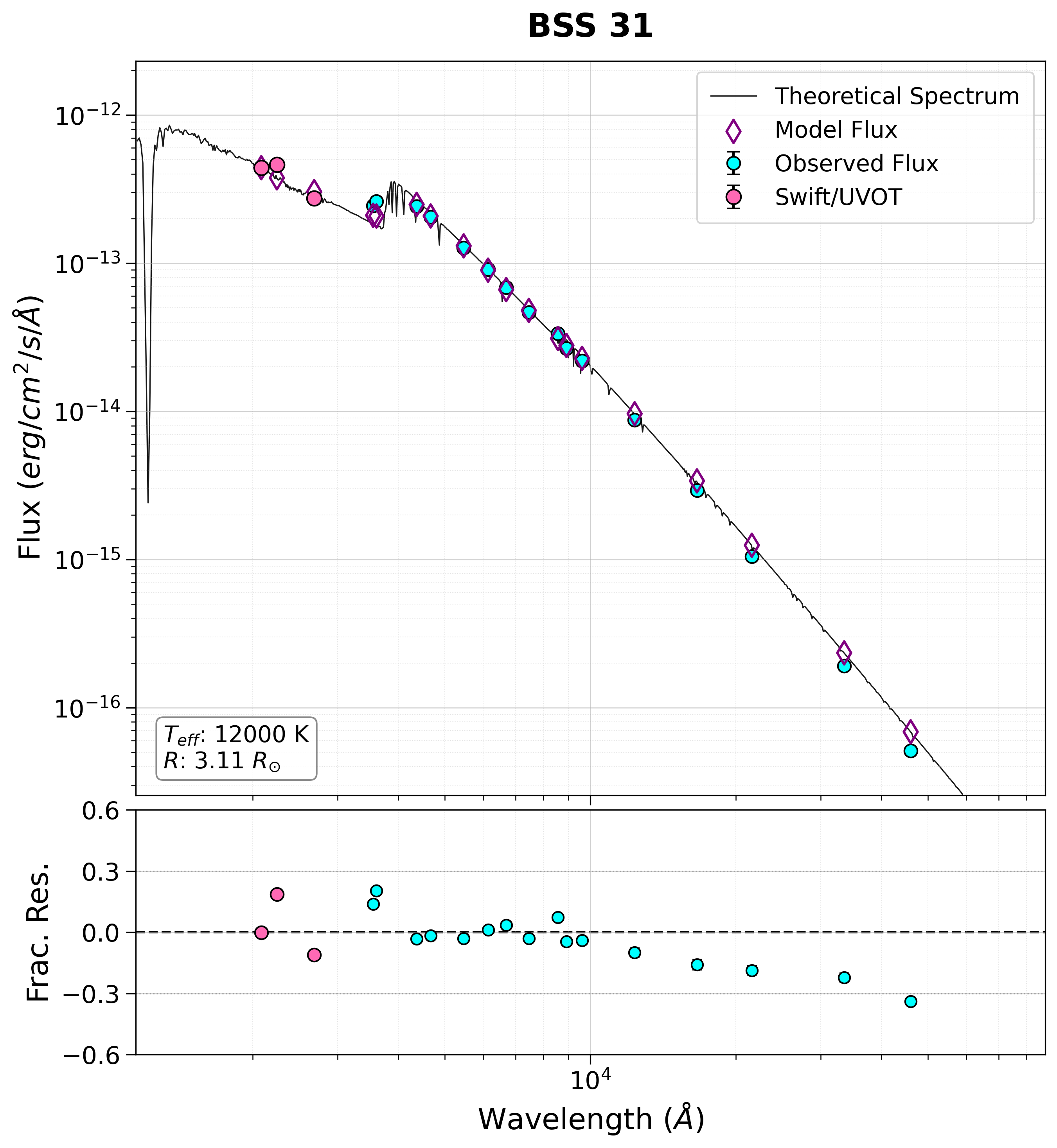}
    \includegraphics[width=0.23\linewidth]{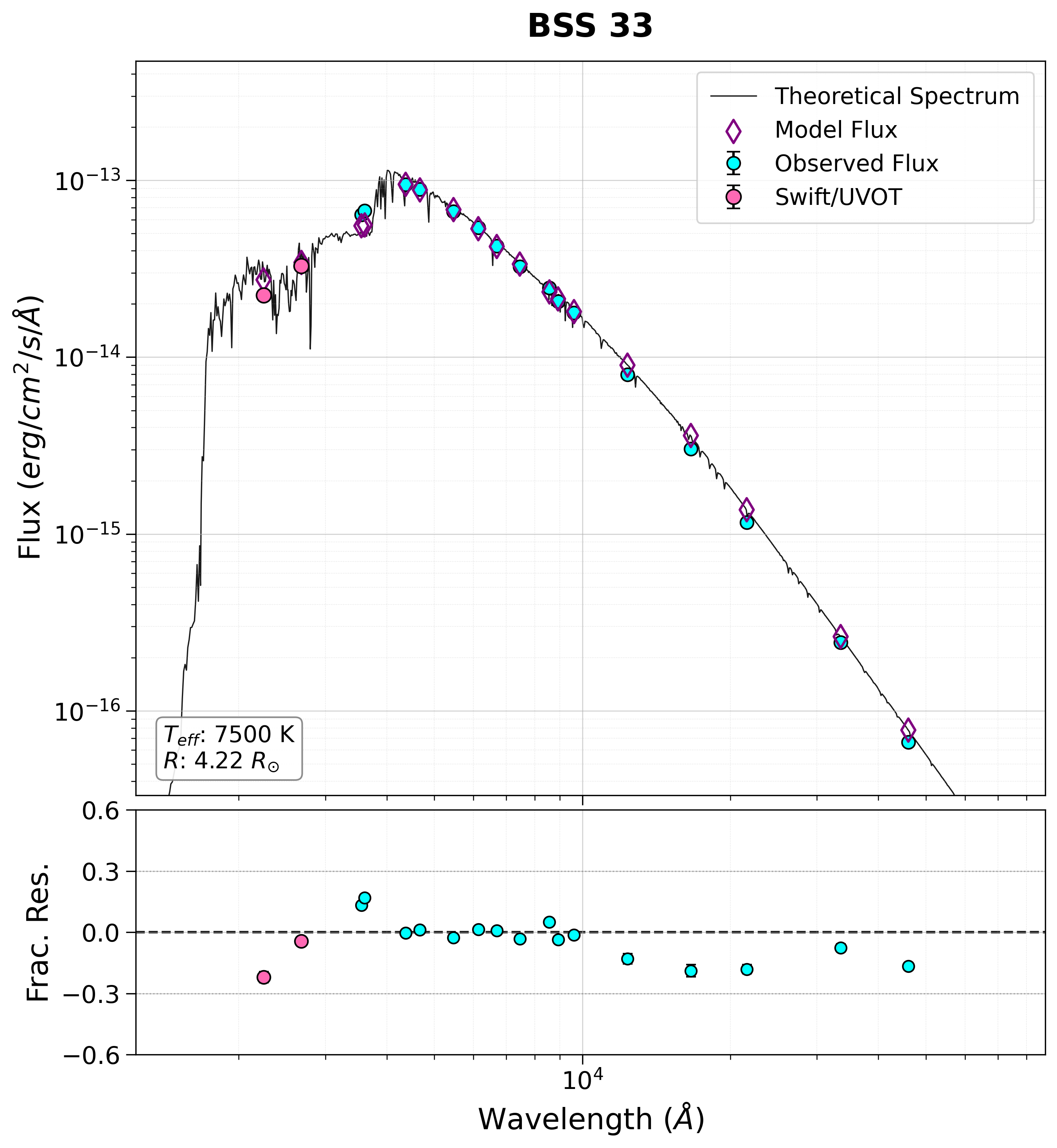}
    
    \caption{SED fits for the selected BSS stars.}
    \label{fig:bss_grid}
\end{figure}

In this appendix, we present the full set of single-component SED fits for the BSS candidates that are well reproduced by single-star atmosphere models. The fitting procedure follows the methodology described in Section~\ref{sec:section3.2}, where observed multi-wavelength fluxes were modeled using Kurucz ODFNEW/NOVER grids \citep{Castelli1997, Castelli2003}. within the VOSA framework. The derived fundamental stellar parameters for the entire sample, including $T_{\rm eff}$, radius, and luminosity, are listed in Table~\ref{tab:single_sed_results}. The corresponding SED fits are shown in Figure~\ref{fig:bss_grid}, where each panel displays the observed photometric data points along with the best-fitting theoretical model.

\begin{table*}
\centering
\footnotesize
\caption{Results of the single SED fitting analysis, showing the derived parameters for the BSS candidates. The columns list the host cluster, \textit{Gaia} DR3 source identifier, and equatorial coordinates ($\alpha$, $\delta$). For each target star, the derived fundamental parameters are provided: effective temperature ($T_{\rm eff}$), radius ($R$), and luminosity ($L$). Finally, the scaling factors, the number of photometric points used in the fit ($N_{\rm fit}$), reduced chi-square ($\chi^2_\nu$), and the goodness-of-fit metric (vgf and vgf$_b$) are presented. \label{tab:single_sed_results}}
\renewcommand{\arraystretch}{1}
\setlength{\tabcolsep}{4pt}
\begin{tabular}{llcccccccrrc}
\hline
No    & Cluster     & $\alpha$ (hh:mm:ss) & $\delta$ (dd:mm:ss) & $T_{\rm eff}$ (K) & $R$ ($R_\odot$) & $L$ ($L_\odot$)  & Scaling Factor & $N_{\rm fit}$ & $\chi^2_\nu$   & vgf   & vgf$_b$ \\
\hline
BSS 01 & Berkeley 29 & 06:53:13.59         & +16:55:24.26        & 7500$\pm$172      & 0.98$\pm$0.05   & 2.81$\pm$0.50    & 1.57E-23       & 11            & 162.25 & 41.49 & 1.67    \\
BSS 02 & Berkeley 31 & 06:57:17.42         & +08:18:24.37        & 10000$\pm$274     & 2.20$\pm$0.11   & 44.04$\pm$8.78   & 2.16E-22       & 25            & 11.19  & 7.15  & 1.48    \\
BSS 03 & Berkeley 31 & 06:57:36.93         & +08:15:25.49        & 8000$\pm$167      & 1.25$\pm$0.06   & 5.74$\pm$1.10    & 3.39E-23       & 13            & 169.71 & 29.05 & 0.56    \\
BSS 04 & Berkeley 31 & 06:57:43.62         & +08:16:51.60        & 10000$\pm$301     & 1.14$\pm$0.06   & 11.80$\pm$1.89   & 2.93E-23       & 20            & 4.94   & 4.88  & 0.93    \\
BSS 05 & Berkeley 31 & 06:57:36.17         & +08:16:29.10        & 8000$\pm$174      & 1.50$\pm$0.07   & 8.35$\pm$1.25    & 9.08E-23       & 15            & 36.04  & 14.16 & 0.62    \\
BSS 08 & Berkeley 31 & 06:57:48.73         & +08:16:02.50        & 7500$\pm$125      & 1.54$\pm$0.08   & 6.70$\pm$0.70    & 7.84E-23       & 18            & 22.48  & 7.99  & 2.31    \\
BSS 09 & Berkeley 31 & 06:57:38.82         & +08:16:26.62        & 7500$\pm$212      & 0.93$\pm$0.05   & 2.48$\pm$0.47    & 2.40E-23       & 14            & 2.53   & 2.52  & 0.45    \\
BSS 11 & Berkeley 31 & 06:57:39.96         & +08:18:13.68        & 8000$\pm$125      & 2.57$\pm$0.13   & 23.92$\pm$2.40   & 7.77E-23       & 24            & 44.96  & 7.15  & 2.26    \\
BSS 12 & Berkeley 31 & 06:57:33.78         & +08:18:12.09        & 7500$\pm$159      & 1.35$\pm$0.06   & 5.18$\pm$0.81    & 4.56E-23       & 15            & 76.70   & 12.38 & 0.51    \\
BSS 13 & Berkeley 37 & 07:20:16.87         & -01:01:29.51        & 12250$\pm$125     & 2.58$\pm$0.13   & 138.26$\pm$14.43 & 1.70E-22       & 19            & 365.99 & 22.12 & 1.04    \\
BSS 16 & Berkeley 75 & 06:48:52.55         & -23:59:09.63        & 7250$\pm$125      & 1.12$\pm$0.06   & 3.14$\pm$0.32    & 2.01E-23       & 19            & 385.61 & 11.61 & 0.53    \\
BSS 22 & NGC 2204    & 06:15:39.17         & -18:40:04.98        & 8000$\pm$173      & 3.32$\pm$0.17   & 40.62$\pm$6.84   & 1.30E-22       & 20            & 53.42  & 6.43  & 1.69    \\
BSS 23 & NGC 2204    & 06:15:35.42         & -18:39:50.76        & 11000$\pm$164     & 3.29$\pm$0.16   & 142.38$\pm$19.99 & 3.06E-22       & 23            & 105.08 & 11.26 & 0.46    \\
BSS 24 & NGC 2204    & 06:15:47.65         & -18:38:10.18        & 12750$\pm$285     & 1.75$\pm$0.08   & 72.88$\pm$11.58  & 8.64E-23       & 21            & 319.73 & 18.88 & 0.74    \\
BSS 25 & NGC 2204    & 06:15:31.42         & -18:38:51.68        & 9750$\pm$125      & 2.72$\pm$0.14   & 62.35$\pm$6.69   & 1.60E-22       & 23            & 410.58 & 9.13  & 1.2     \\
BSS 27 & NGC 2204    & 06:15:36.18         & -18:42:19.48        & 7000$\pm$196      & 2.18$\pm$0.11   & 10.40$\pm$1.97   & 7.85E-23       & 15            & 655.64 & 26.20  & 1.29    \\
BSS 31 & NGC 6939    & 20:31:31.89         & +60:38:15.81        & 12000$\pm$125     & 3.11$\pm$0.16   & 184.26$\pm$21.72 & 1.50E-21       & 19            & 260.49 & 39.34 & 2.36    \\
BSS 33 & NGC 6939    & 20:31:31.93         & +60:36:25.46        & 7500$\pm$298      & 4.22$\pm$0.21   & 50.73$\pm$11.74  & 2.56E-21       & 18            & 530.73 & 39.69 & 1.58    \\
BSS 34 & NGC 6939    & 20:31:31.93         & +60:36:25.46        & 8500$\pm$166      & 3.86$\pm$0.19   & 70.99$\pm$10.59  & 2.34E-21       & 19            & 376.99 & 55.47 & 3.42    \\
BSS 35 & NGC 6939    & 20:31:31.89         & +60:38:15.80        & 12250$\pm$125     & 3.04$\pm$0.15   & 195.58$\pm$22.68 & 1.45E-21       & 14            & 208.46 & 29.87 & 1.59     \\ \hline
\end{tabular}%
\end{table*}

\section{Data of Literature-Studied Hot Component Stars}\label{AppendixB}

We compile a homogeneous list of literature-studied hot component stars whose atmospheric and evolutionary parameters have been previously reported. For a subset of objects lacking mass estimates in the literature, we derived approximate masses by placing them on the H-R diagram and comparing their positions with the evolutionary tracks of \citet{Istrate2016}, following the same approach adopted for our sample. The full set of compiled parameters is presented in Table~\ref{tab:lit_stellar_properties}. These objects are used as a comparison sample in the H–R diagram shown in Figure~\ref{fig:hr_diagram_elm}, where they are plotted together with our pre-ELM candidates.

\begin{table}
\centering
\footnotesize
\caption{Fundamental properties of literature-studied hot component stars. Coordinates are given in degrees, $T_\mathrm{eff}$ in Kelvin, and luminosity, radius, and mass ($L$, $R$, $M$) in solar units. The ``Classification'' column denotes stellar type, where BSS stands for ``Blue Straggler Star'' and BL stands for ``Blue Lurker''.}
\label{tab:lit_stellar_properties}
\begin{tabular}{llllccccccc}
\hline
ID & Classification & Cluster & $Gaia$ DR3 Source ID & $\alpha$ & $\delta$ & $T_\mathrm{eff}$ & $L$ & $R$ & $M$ & Ref. \\
   &      &        &              & (deg) & (deg) & (K) & ($L_\odot$) & ($R_\odot$) & ($M_\odot$) &  \\
\hline
01  & BSS & Berkeley 39 & 3056678372883126528 & 116.7569 & -4.7046  & 12500 & 0.06   & 0.05 & 0.186* $\pm$ 0.092   & 01 \\
02  & BSS & King 2      & 424416887106018688  & 12.7354  & 58.1857  & 22000 & 3.10   & 0.12 & 0.222 $\pm$ 0.024    & 02 \\
03  & BSS & King 2      & 424415606039931520  & 12.8517  & 58.1522  & 24000 & 16.40  & 0.23 & 0.219 $\pm$ 0.055    & 02 \\
04  & BSS & King 2      & 424418398934484480  & 12.7482  & 58.2077  & 24000 & 2.70   & 0.10 & 0.214 $\pm$ 0.030    & 02 \\
05  & BSS & King 2      & 424418364574755328  & 12.7418  & 58.1967  & 26000 & 3.30   & 0.09 & 0.256 $\pm$ 0.029    & 02 \\
06  & BSS & King 2      & 424416234271059072  & 12.6768  & 58.1142  & 14000 & 1.90   & 0.23 & 0.206 $\pm$ 0.072    & 02 \\
07  & BSS & King 2      & 425074017100947968  & 12.4975  & 58.1352  & 19000 & 8.60   & 0.27 & 0.213 $\pm$ 0.029    & 02 \\
08  & BSS & M67         & 604917560635575808  & 132.8465 & 11.8028  & 11000 & 0.05   & 0.06 & 0.282 $\pm$ 0.057    & 03 \\
09  & BSS & Melotte 66  & 5507234395259864448 & 111.5865 & -47.6962 & 38000 & 2.99   & 0.04 & 0.378 $\pm$ 0.102    & 04 \\
11 & BSS & NGC 2243    & 2893942062036322688 & 97.4233  & -31.2615 & 19000 & 0.55   & 0.07 & 0.186* $\pm$ 0.023   & 05 \\
12 & BSS & NGC 2420    & 865405334273841536  & 114.7737 & 21.6450  & 11500 & 0.23   & 0.12 & 0.171 $\pm$ 0.051    & 06 \\
13 & BSS & NGC 2420    & 865401520342930560  & 114.6090 & 21.5778  & 10250 & 1.61   & 0.40 & 0.186* $\pm$ 0.008   & 06 \\
14 & BSS & NGC 2506    & 3038042784664224000 & 120.1130 & -10.7665 & 19000 & 0.24   & 0.05 & 0.265 $\pm$ 0.022    & 07 \\
15 & BSS & NGC 2506    & 3038046358077125248 & 119.9990 & -10.7541 & 15000 & 0.45   & 0.10 & 0.186* $\pm$ 0.037   & 07 \\
16 & BSS & NGC 2506    & 3038044880608396672 & 119.9921 & -10.7650 & 13250 & 0.44   & 0.13 & 0.186* $\pm$ 0.021   & 07 \\
17 & BSS & NGC 2627    & 5504787948259881344 & 129.2610 & -29.8309 & 14500 & 0.26   & 0.08 & 0.216 $\pm$ 0.033    & 08 \\
25 & BSS & NGC 7142    & 2217943209267285248 & 326.3131 & 65.8230  & 19750 & 0.10   & 0.03 & 0.319 $\pm$ 0.028    & 09  \\
26 & BSS & NGC 7142    & 2217989766712690432 & 325.9511 & 65.8598  & 15000 & 0.08   & 0.04 & 0.213 $\pm$ 0.003    & 09  \\
27 & BSS & NGC 7142    & 2218036534571068288 & 326.3476 & 65.8043  & 14000 & 0.09   & 0.05 & 0.202 $\pm$ 0.080    & 09  \\
27 & BSS & NGC 7142    & 2218036534571068288 & 326.3476 & 65.8043  & 14000 & 0.09   & 0.05 & 0.202 $\pm$ 0.080 & 09  \\
28 & BL  & NGC 752     & 342917133876421760  & 29.2828  & 37.8243  & 14250 & 1.30   & 0.19 & 0.202 $\pm$ 0.020 & 10 \\
29 & BL  & NGC 752     & 342918645704887040  & 29.4973  & 37.9149  & 13750 & 38.70  & 1.10 & 0.285 $\pm$ 0.024 & 10 \\
30 & BL  & NGC 752     & 342864597836478080  & 29.5476  & 37.6592  & 10750 & 120.30 & 3.17 & 0.265 $\pm$ 0.023 & 10 \\
31 & BL  & NGC 752     & 342893803614055168  & 29.6243  & 37.8603  & 9500  & 18.40  & 1.59 & 0.242 $\pm$ 0.045 & 10 \\
32 & BL  & NGC 752     & 342921738081340928  & 29.2633  & 37.9290  & 14250 & 8.70   & 0.48 & 0.228 $\pm$ 0.006 & 10 \\
33 & BL  & NGC 752     & 342914556896034176  & 29.2211  & 37.8692  & 15250 & 20.80  & 0.65 & 0.228 $\pm$ 0.006 & 10 \\
34 & BL  & NGC 752     & 342915411593428736  & 29.1635  & 37.8614  & 15250 & 41.50  & 0.92 & 0.305 $\pm$ 0.092 & 10 \\
35 & BL  & NGC 752     & 342866144024819840  & 29.6538  & 37.7529  & 13750 & 95.00  & 1.72 & 0.233 $\pm$ 0.032 & 10 \\
36 & BL  & NGC 752     & 342917610616684032  & 29.3364  & 37.8619  & 12500 & 55.30  & 1.59 & 0.287 $\pm$ 0.073 & 10 \\
37 & BL  & NGC 752     & 342919058020364544  & 29.3827  & 37.8945  & 12500 & 18.30  & 0.91 & 0.219 $\pm$ 0.029 & 10 \\
38 & BL  & NGC 752     & 342869236401143808  & 29.4123  & 37.7700  & 12750 & 1.10   & 0.22 & 0.222 $\pm$ 0.020 & 10 \\
39 & BL  & NGC 752     & 342918233388035328  & 29.4144  & 37.8737  & 11250 & 234.00 & 4.04 & 0.265 $\pm$ 0.006 & 10 \\
40 & BL  & NGC 752     & 342870640854339200  & 29.4908  & 37.8061  & 11500 & 36.90  & 1.53 & 0.265 $\pm$ 0.015 & 10 \\
41 & BL  & NGC 752     & 342864666555954048  & 29.5321  & 37.6658  & 11750 & 124.40 & 2.70 & 0.287 $\pm$ 0.026 & 10 \\
42 & BL  & NGC 752     & 342920020094927104  & 29.4364  & 37.9884  & 15000 & 21.30  & 0.68 & 0.219 $\pm$ 0.005 & 10 \\
43 & BSS & NGC 7789    & 1995060313854772736 & 359.1755 & 56.7368  & 12500 & 1.29   & 0.24 & 0.202 $\pm$ 0.016 & 11\\
44 & BSS & NGC 7789    & 1995020211733162112 & 359.5219 & 56.7841  & 15250 & 1.56   & 0.18 & 0.208 $\pm$ 0.004 & 11\\
45 & BSS & NGC 7789    & 1995014237443527680 & 359.4574 & 56.7436  & 15000 & 0.26   & 0.08 & 0.219 $\pm$ 0.028 & 11\\
46 & BSS & NGC 7789    & 1995010973268848512 & 359.4831 & 56.7121  & 15500 & 0.25   & 0.07 & 0.202 $\pm$ 0.072 & 11\\
47 & BSS & NGC 7789    & 1995010801470149504 & 359.3789 & 56.6636  & 11750 & 0.67   & 0.20 & 0.222 $\pm$ 0.042 & 11\\
\hline
\end{tabular}

\vspace{1.5ex}
\begin{minipage}{16cm}
\noindent 01: \citet{2024AJ....168..278C}, 02: \citet{2021JApA...42...89J}, 03: \citet{2021MNRAS.507.2373P}, 04: \citet{2022MNRAS.516.2444R}, 05: \citet{2024AJ....168..274S}, 06: \citet{2024ApJ...961..251Y}, 07: \citet{2022MNRAS.516.5318P}, 08: \citet{2024AJ....168...97S}, 09: \citet{2024MNRAS.527.8325P}, 10: \citet{2024AA...688A.152J}, 11: \citet{2022MNRAS.511.2274V}. \\
\textbf{*}: Indicates that the mass value is adopted from the literature.
\end{minipage}
\end{table}


\bibliography{References}
\bibliographystyle{aasjournal}

\end{document}